\documentclass[a4paper,11pt]{article}
\usepackage{a4wide}
\usepackage{amsmath,amssymb,amsthm,mathtools,array,booktabs}
\usepackage{graphicx,hyperref}
\usepackage{xcolor}
\usepackage{cancel}
\usepackage{tikz}
\usepackage{amscd}
\usepackage{latexsym,cite}
\usepackage{subcaption}
\usepackage{ytableau}
\usepackage{scalerel}
\usepackage{calc}
\input xy
\xyoption{all}

\hypersetup{
  unicode,
  bookmarksnumbered,
  linktoc = all,
  pdfborderstyle = {/S/U/W 0.5}
}

\def\cO{\mathcal{O}}

\let\Re\relax
\let\Im\relax
\DeclareMathOperator{\Re}{Re}
\DeclareMathOperator{\Im}{Im}

\DeclareMathOperator{\Sym}{Sym}

\numberwithin{equation}{section}

\begin{document}
\thispagestyle{empty}
\begin{flushright}
preprint
\end{flushright}
\vspace{1cm}
\begin{center}
{\LARGE\bf A GLSM view on Homological Projective Duality} 
\end{center}
\vspace{8mm}
\begin{center}
{\large Zhuo Chen\footnote{{\tt zhuochen2019@mail.tsinghua.edu.cn }}, Jirui Guo\footnote{{\tt jrguo@mail.tsinghua.edu.cn}}  and Mauricio Romo\footnote{{\tt mromoj@tsinghua.edu.cn}}}
\end{center}
\vspace{6mm}
\begin{center}
 Yau Mathematical Sciences Center, Tsinghua University, Beijing, 100084, China
\end{center}
\vspace{15mm}

\begin{abstract}
\noindent
Given a gauged linear sigma model (GLSM) $\mathcal{T}_{X}$ realizing a projective variety $X$ in one of its phases, i.e. its quantum K\"ahler moduli has a geometric point, we propose an \emph{extended} GLSM $\mathcal{T}_{\mathcal{X}}$ realizing the homological projective dual category $\mathcal{C}$ to $D^{b}Coh(X)$ as the category of B-branes of the Higgs branch of one of its phases. In most of the cases, the models $\mathcal{T}_{X}$ and $\mathcal{T}_{\mathcal{X}}$ are anomalous and the analysis of their Coulomb and mixed Coulomb-Higgs branches gives information on the semiorthogonal/Lefschetz decompositions of $\mathcal{C}$ and $D^{b}Coh(X)$. We also study the models $\mathcal{T}_{X_{L}}$ and  $\mathcal{T}_{\mathcal{X}_{L}}$ that correspond to homological projective duality of linear sections $X_{L}$ of $X$. This explains why, in many cases, two phases of a GLSM are related by homological projective duality. We study mostly abelian examples: linear and
Veronese embeddings of $\mathbb{P}^{n}$ and Fano complete intersections in $\mathbb{P}^{n}$. In such cases, we are able to reproduce known results as well as produce some new conjectures. In addition, we comment on the construction of the HPD to a nonabelian GLSM for the Pl\"ucker embedding of the Grassmannian $G(k,N)$. 
\end{abstract}
\newpage
\setcounter{tocdepth}{3}
\tableofcontents
\setcounter{footnote}{0}

\section{Introduction}

B-branes on $\mathcal{N}=(2,2)$ supersymmetric theories can be defined in general terms, regardless of the fact that such theory possesses or not an IR superconformal fixed point. This is achieved by simply studying boundary conditions preserving an appropriate subset $\mathbf{2}_{B}$, spanned by $\mathbf{Q}_{-}+\mathbf{Q}_{+}$ and its conjugate supercharge $\overline{\mathbf{Q}}_{-}+\overline{\mathbf{Q}}_{+}$ of the $\mathcal{N}=(2,2)$ supersymmetry algebra \cite{Hori:2000ck,Witten:1992fb,Aspinwall:2009isa}. The collection of such boundary conditions are expected to have the structure of a triangulated category. These categories are expected to be insensitive to deformations of the action of the form $\mathbf{Q}_{+}\overline{\mathbf{Q}}_{-}\mathcal{G}$ (and its adjoint) where $\overline{\mathbf{Q}}_{+}\mathcal{G}=\mathbf{Q}_{-}\mathcal{G}=0$. In the context of the gauged linear sigma model (GLSM) \cite{Witten:1993yc} these deformations can be interpreted as deformations of the twisted superpotential and the space of such deformations is called the quantum K\"ahler moduli space $\mathcal{M}_{K}$. A huge class of GLSMs have geometric points in $\mathcal{M}_{K}$. At such points, the theory in the IR can be thought of as a nonlinear sigma model (NLSM) with the target space given by a K\"ahler manifold $X$ and locally $\mathcal{M}_{K}$ is regarded as the complexified K\"ahler cone of $X$. Then the category of B-branes is equivalent to $D^{b}Coh(X)$. By moving around $\mathcal{M}_{K}$ we can find relations between $D^{b}Coh(X)$ and categories of B-branes at the different Higgs branches arising at other points in $\mathcal{M}_{K}$. When $X$ is Calabi-Yau (CY), this is known to give equivalences or autoequivalences, classified by the space of homotopy equivalent paths in $\mathcal{M}_{K}$, between $D^{b}Coh(X)$ and the aforementioned categories. For $X$ general K\"ahler manifold, the different B-brane categories generically embed into each other and paths correspond to fully faithful functors. The way the categories fit into each other depends on the structure of the Coulomb and Coulomb-Higgs branches, which are either absent or deemed singular in the CY case.  This problem has been analyzed thoroughly in the physics literature, starting with \cite{Herbst:2008jq} for nonanomalous abelian GLSMs and further developed in \cite{Clingempeel:2018iub,Hori:2013ika,hori2019notes,eager2017beijing} and in the mathematics literature \cite{segal2011equivalences,halpern2015derived,ballard2019variation}.

On the other hand, one very powerful result of the recent years regarding the interplay between derived categories of projective varieties and their semiorthogonal decompositions is Kuznetsov's homological projective duality \cite{kuznetsov2007homological}. Kuznetsov's result provides a relation between the semiorthogonal decomposition of a projective variety $X$ and its homological projective dual (HPD) variety $Z$ (or more general its HPD category $\mathcal{C}$) and linear sections of them. This result leads to a great deal of insight into equivalences of triangulated categories and their semiorthogonal decompositions (for a review see for example \cite{alex2014semiorthogonal,thomas2015notes}). Mutually HPD varieties/categories have made an appearance as B-brane categories of different phases of GLSMs \cite{Caldararu:2007tc,Hori:2016txh,Hori:2013gga,Addington:2012zv,Hori:2006dk} and more recently this point of view has been studied in a mathematical framework in \cite{ballard2017homological,rennemo2017fundamental}.

Inspired mostly by the works 
\cite{ballard2017homological,rennemo2017fundamental}, we develop a proposal for 
a GLSM construction of HPD models of a K\"ahler variety $X$, whenever a GLSM 
construction for $X$ itself is known. Let us summarize our proposal. Details 
can be found in section \ref{sec:proposal}. Starting from a GLSM  
$\mathcal{T}_{X}=(G,\rho_{m}: G\rightarrow GL(V),W,t_{\mathrm{ren}},R)$ with a 
geometric point in $\mathcal{M}_{X}$ around which we can identify the category 
of B-branes with $D^{b}Coh(X)$ and a map $f:X\rightarrow \mathbb{P}(S)$, we 
write an extension $\mathcal{T}_{\mathcal{X}}$ of $\mathcal{T}_{X}$ given by
\begin{eqnarray}
\mathcal{T}_{\mathcal{X}}=(\widehat{G}=G\times U(1)_{s+1},\hat{\rho}_{m}: \widehat{G}\rightarrow GL(V\oplus V'),\widehat{W},\widehat{R}),
\end{eqnarray}
where\footnote{Here and in the following, $\mathbb{C}_{(\chi,w)}$ denotes the one-dimensional representation of $\widehat{G}$ determined by the $G$-character $\chi$ and the $U(1)_{s+1}$-weight $w$.} $V'=\mathbb{C}_{(\chi^{-1},-1)} \oplus S^\vee$ is a representation of $\widehat{G}$ with  
$\chi\in\mathrm{Hom}(G,\mathbb{C}^{*})$ the character of $G$ defined by \eqref{chi} and $S^\vee \cong \mathbb{C}^{\oplus{n+1}}_{(1,1)}$. By an appropriate choice of basis, we 
can write $\mathbb{C}_{\chi^{-1}}$ as a representation of weight $Q$ of a 
subgroup $U(1)_{l}\subset G$. Denoting the coordinates of $V'$ as 
$(P,S_{0},\ldots,S_{n})$, the superpotential $\widehat{W}$ is given by
\begin{eqnarray}
\widehat{W}=W+P\sum_{j=0}^{n}S_{j}f_{j}(X),
\end{eqnarray}
where $f_{j}(X)$ are the components of the image of the map $f$. The GLSM $\mathcal{T}_{\mathcal{X}}$ is identified with the GLSM of the universal hyperplane class $\mathcal{X}$ of $X$. 
Its more important property is that its Higgs branch deep in the second quadrant of the FI parameter of $U(1)_{s+1}\times U(1)_{l}$, upon taking the gauge decoupling limit, is a hybrid model whose category of B-branes can be identified with the HPD category $\mathcal{C}$ of $X$. This is essentially because both categories are equivalent to the subcategory $\mathcal{W}_{-}$ (small window category, see section \ref{sec:GLSM}) of GLSM B-branes. Brane transport between this phase and the geometric phase realizes the correspondence $\mathcal{C}\cong \mathcal{W}_{-} \cong D^{b}Coh(X)$. Moreover, we can take linear sections of $\mathcal{T}_{\mathcal{X}}$, essentially by restricting the chiral fields $S_{j}$ to a linear subspace $L\subset \mathbb{P}(S^{\vee})$. This gives a GLSM termed $\mathcal{T}_{\mathcal{X}_{L}}$ which has two very interesting phases, one that we can identify with the NLSM on $X_{L}$ and one whose category of B-branes can be identified with $D(Z_{L})$ (even though in some cases, the variety $Z$ itself cannot be determined, but then, $D(Z_{L})$  can be defined in terms of a hybrid model). If we take the gauge decoupling limit $e_{s+1}\rightarrow \infty$ of $U(1)_{s+1}$, on $\mathcal{T}_{\mathcal{X}}$, while keeping $\zeta_{s+1}\ll-1$, we end up with a $G\times U(1)_{l}$ gauge theory $\mathcal{T}_{\mathcal{C}'}$ in which one of its phases is related to some category $\mathcal{C}'$ which embeds in $\mathcal{C}$ (although, in some examples $\mathcal{C}'=\mathcal{C}$, a detailed explanation can be found in section \ref{sec:proposal}). The same limit for $\mathcal{T}_{\mathcal{X}_{L}}$ is more interesting since we end up with a theory $\mathcal{T}_{X_{L}}$ in which one phase is related with $X_{L}$ and the other with $\mathcal{C}_{L}$, a subcategory 'shared' between $D(X_{L})$ and $D(Z_{L})$ (a precise definition can be found in section \ref{sec:section2}). We summarize this in the following diagram: 
\[
\xymatrix{\mathcal{T}_X \ar@<0.5ex>[d]_-{\mathrm{Extension}} & \\ \mathcal{T}_{\mathcal{X}} \ar@<0.5ex>[rr]^-{\mathrm{Restriction}} \ar@<0.5ex>[d]_-{-\zeta_{s+1},e_{s+1} \rightarrow \infty} && \mathcal{T}_{\mathcal{X}_L} \ar@<0.5ex>[d]_-{-\zeta_{s+1},e_{s+1} \rightarrow \infty} \\
\mathcal{T}_{\mathcal{C}'} \ar@<0.5ex>[rr]^-{\mathrm{Restriction}} && \mathcal{T}_{X_L} }
\]
We provide evidence of our proposal by computing it in several examples, reproducing known results and providing insight into new ones. More precisely we analyze:
\begin{itemize}
 \item \textbf{Linear and Veronese embeddings of $\mathbb{P}^{n}$}. In the linear embedding with the Lefschetz decomposition defined by the Beilinson collection, we reproduce the basic result that the HPD corresponds to the classical projective dual. For the double Veronese embedding, we reproduce the result of \cite{kuznetsov2008derived} and we also analyze higher Veronese embeddings finding a HPD analogous to the one from \cite{ballard2017homological}, described by a hybrid model $\mathcal{C}$. Using GLSM technology, we find an explicit functor from this hybrid model to the universal hyperplane section $\mathcal{X}$ of the Veronese embedding which allows us to write explicit generators of $\mathcal{C}$ as a subcategory of $D^{b}Coh(\mathcal{X})$.
 
 \item \textbf{Quadrics in $\mathbb{P}^{n}$}. When $n$ is even, we reproduce the HPD from \cite{kuznetsov2019homological}. When $n$ is odd, our construction induces a different Lefschetz decomposition than the one used in \cite{kuznetsov2019homological}. Therefore, the HPD changes and we are not aware of it being constructed explicitly anywhere else. We analyze again an explicit functor from objects in the hybrid model representing the HPD category to $D^{b}Coh(\mathcal{X})$.
 
 \item \textbf{Fano complete intersections in $\mathbb{P}^{n}$}. In this case, we only analyze the resulting HPD that takes the form of a hybrid model corresponding to the Lefschetz decomposition induced by $(\mathcal{T}_{X},\mathcal{T}_{\mathcal{X}})$. 
 
 \item \textbf{Pl\"ucker embedding of $G(k,N)$}. We devote section \ref{sec:nonabelian} to the study of the basic properties of $\mathcal{T}_{\mathcal{X}}$ for the Pl\"ucker embedding of $G(k,N)$. In this case we present the hybrid model representing the HPD and we compute to which Lefschetz decomposition it corresponds, by analyzing the Coulomb branch. We also comment on $\mathcal{T}_{\mathcal{X}_{L}}$ for some particular linear sections.
\end{itemize}

The paper is organized as follows. In section \ref{sec:section2} we review Kuznetsov's homological projective duality theorem and consequences as well as some examples. In section \ref{sec:GLSM} we provide a review of B-branes in GLSMs, focusing on the abelian case but not restriciting to nonanomalous models. The main goal of this section is to review how the grade restriction rule works for anomalous GLSMs. In section \ref{sec:proposal} we present our proposal for a GLSM containing the HPD of a variety $X$ and we analyze its properties in a general setting. In particular, we specify the Lefschetz decomposition induced by the pair $(\mathcal{T}_{X},\mathcal{T}_{\mathcal{X}})$ and we give the details on how to take linear sections and construct $\mathcal{T}_{\mathcal{X}_{L}}$ as well as its properties. In section \ref{sec:examples} we analyze several examples of $(\mathcal{T}_{X},\mathcal{T}_{\mathcal{X}})$: linear and Veronese embedding of projective space, quadrics in $\mathbb{P}^{n}$ and Fano complete intersections in $\mathbb{P}^{n}$. In section \ref{sec:mutually orthogonal section} we analyze $\mathcal{T}_{\mathcal{X}_{L}}$  for the previous examples. In section \ref{sec:nonabelian} we make some remarks on $\mathcal{T}_{\mathcal{X}}$ for the Pl\"ucker embedding of $G(k,N)$. Further background material and complimentary results are collected in the appendices.

\section{\label{sec:section2}Lightning review of homological projective duality}

In order to define the homological projective dual (HPD) of a projective variety $X$, we need to define a few elements of triangulated categories first. We denote by $D(X)$ the bounded derived category of coherent sheaves on $X$, $D^{b}Coh(X)$.
\\\\
\textbf{Definition}.\cite{kuznetsov2007homological} A (right) Lefschetz decomposition of $D(X)$ w.r.t. the line bundle $\mathcal{L} $ on $X$ corresponds to a semiorthgonal decomposition \cite{bondal1995semiorthogonal}
\begin{eqnarray}\label{LDX}
D(X)=\langle\mathcal{A}_{0},\mathcal{A}_{1}(1),\ldots,\mathcal{A}_{k}(k)\rangle
\end{eqnarray}
such that $\mathcal{A}_{0}\supseteq \mathcal{A}_{1}\supseteq\cdots\supseteq\mathcal{A}_{k}$ are a collection of admissible subcategories of $D(X)$ and $\mathcal{A}_{i}(i):=\mathcal{A}_{i}\otimes \mathcal{L}^{\otimes i}$. The Lefschetz decomposition is called rectangular if $\mathcal{A}_{0}=\mathcal{A}_{1}=\cdots=\mathcal{A}_{k}$.\\\\
A few remarks are in order. The Lefschetz decomposition is completely determined by its center $\mathcal{A}_{0}$ and $\mathcal{L}$ via the relation
\begin{eqnarray}
\mathcal{A}_{r}=^{\perp}\mathcal{A}_{r-1}(-r)\cap \mathcal{A}_{r-1},
\end{eqnarray}
We can always construct a dual Lefschetz decomposition by setting \cite{kuznetsov2008lefschetz}
\begin{eqnarray}
\mathcal{B}_{0}=\mathcal{A}_{0},\mathcal{B}_{i}=\mathcal{A}_{0}(i)^{\perp}\cap \mathcal{A}_{i},\qquad i=1,\ldots,k,
\end{eqnarray}
where $\mathcal{A}_{0}(i)^{\perp}$ denotes the right orthogonal of $\mathcal{A}_{0}(i)$. Then we have a left Lefschetz decomposition
\begin{eqnarray}
D(X)=\langle \mathcal{B}_{k}(-k),\ldots,\mathcal{B}_{1}(-1), \mathcal{B}_{0}\rangle.
\end{eqnarray}\\
Consider a smooth projective variety $X$ with a morphism $f:X\rightarrow \mathbb{P}(V)$ and assume we have a Lefschetz decomposition of the form (\ref{LDX}) w.r.t. the line bundle $\mathcal{L}=\mathcal{O}_{X}(1):=f^{*}\mathcal{O}_{\mathbb{P}(V)}(1)$. We define the incidence divisor $\mathcal{H}\subset \mathbb{P}(V)\times \mathbb{P}(V^{\vee})$ by
\begin{eqnarray}
\mathcal{H}=\{(u,v)\in \mathbb{P}(V)\times \mathbb{P}(V^{\vee}) : v(u)=0 \}.
\end{eqnarray}
Then the universal hyperplane section of $X$ is defined by
\begin{eqnarray}
\mathcal{X}:=X\times_{\mathbb{P}(V)}\mathcal{H}\subset X\times \mathbb{P}(V^{\vee})
\end{eqnarray}
and has the following semiorthogonal decomposition \cite{kuznetsov2007homological}:
\begin{eqnarray}
D(\mathcal{X})=\langle \mathcal{C},\mathcal{A}_{1}(1)\boxtimes D(\mathbb{P}(V^{\vee})),\ldots,\mathcal{A}_{k}(k)\boxtimes D(\mathbb{P}(V^{\vee}))\rangle.
\end{eqnarray}
Then, one can define
\\\\
\textbf{Definition}.\cite{kuznetsov2007homological} $\mathcal{C}$ is called the HPD category of $X$. In other words, $\mathcal{C}$ is the right orthogonal of $\langle\mathcal{A}_{1}(1)\boxtimes D(\mathbb{P}(V^{\vee})),\ldots,\mathcal{A}_{k}(k)\boxtimes D(\mathbb{P}(V^{\vee}))\rangle$.\\\\
If we have an algebraic variety $Z$ with a morphism $g:Z\rightarrow \mathbb{P}(V^{\vee})$ such that there is an equivalence $D(Z)\cong \mathcal{C}$, we call $Z$ the HPD to $f:X\rightarrow \mathbb{P}(V)$ w.r.t. the Lefschetz decomposition (\ref{LDX}). A very important property of $\mathcal{C}$, that will be used for consistency checks in several of our examples is the following: given the embedding map $\delta:\mathcal{X}\rightarrow X \times \mathbb{P}(V^{\vee})$, we have:
\begin{eqnarray}\label{deltamapp}
\delta_{*}: \mathcal{C}\rightarrow\mathcal{A}_{0}\boxtimes D(\mathbb{P}(V^{\vee})).
\end{eqnarray}
Indeed, one can define the $\mathcal{C}$ by the objects in $D(\mathcal{X})$ whose image under $\delta_{*}$ belongs to $\mathcal{A}_{0}\boxtimes D(\mathbb{P}(V^{\vee}))$. The main theorem of \cite{kuznetsov2007homological} is
\\\\
\textbf{Theorem}.\cite{kuznetsov2007homological} Assume that $Z$ is the HPD to $X$ as defined as above. Then, $Z$ is smooth and admits a dual Lefschetz decomposition
\begin{eqnarray}
D(Z)=\langle \mathcal{B}_{l}(-l),\ldots,\mathcal{B}_{1}(-1),\mathcal{B}_{0}\rangle\nonumber,\\
 \mathcal{B}_{l}\subseteq\cdots \subseteq \mathcal{B}_{1}\subseteq\mathcal{B}_{0}\subseteq D(Z),
\end{eqnarray}
where $\mathcal{B}_{0}\cong \mathcal{A}_{0}$. \footnote{More precisely \cite{kuznetsov2007homological}, $\mathcal{B}_{j}\cong \langle\mathfrak{a}_{0},\ldots,\mathfrak{a}_{N-j-2}\rangle$, where $N=\mathrm{dim}V$ and $\mathfrak{a}_{j}$ is the right orthogonal to $\mathcal{A}_{j+1}$ in $\mathcal{A}_{j}$. Also, is assumed that $N>k+1$.}
For any linear subspace $L\subset V^{\vee}$ satisfying
\begin{eqnarray}
\mathrm{dim}X_{L}&=&\mathrm{dim}X-r\nonumber,\\
\mathrm{dim}Z_{L}&=&\mathrm{dim}Z+r-\mathrm{dim}V,
\end{eqnarray}
where $r=\mathrm{dim}L$, we have the following Lefschetz decompositions
\begin{eqnarray}
D(X_{L})&=&\langle \mathcal{C}_{L}, \mathcal{A}_{r}(1),\ldots,\mathcal{A}_{k}(k+1-r)\rangle,\nonumber\\
D(Z_{L})&=&\langle \mathcal{B}_{l}(N-l-1-r),\ldots,\mathcal{B}_{N-r}(-1),\mathcal{C}_{L}\rangle,
\end{eqnarray}
where $X_{L}:=X\times_{\mathbb{P}(V)}\mathbb{P}(L^{\perp})$, $Z_{L}:=Z\times_{\mathbb{P}(V^{\vee})}\mathbb{P}(L)$.
\\\\
We list a few relevant examples:
\begin{itemize}
 \item \textbf{Trivial Lefschetz decomposition}. For any $f:X\rightarrow \mathbb{P}(V)$ we can always take the Lefschetz decomposition with a single component $\mathcal{A}_{0}=D(X)$. Then its HPD is $Z=\mathcal{X}\rightarrow \mathbb{P}(V^{\vee})$, where the map is just the projection and
 \begin{eqnarray}
D(Z_{L})=\langle D(X)\otimes \mathcal{O}_{\mathbb{P}(L)}(1-\mathrm{dim}L),\ldots,D(X)\otimes \mathcal{O}_{\mathbb{P}(L)}(-1),D(X_{L})\rangle.
\end{eqnarray}

\item \textbf{Quadrics}. \cite{kuznetsov2019homological} If $f:X\rightarrow \mathbb{P}(V)$ is a smooth quadric with the usual embedding and $V\cong \mathbb{C}^{2m+1}$, then we have the Lefschetz decomposition $\mathcal{A}_{0}=\langle \mathcal{S},\mathcal{O}\rangle$, $\mathcal{A}_{i}=\langle \mathcal{O}\rangle$, $i=1,\ldots 2m-2$ and the HPD of $X$ is $g:Z\rightarrow \mathbb{P}(V^{\vee})$, a double cover of $\mathbb{P}(V^{\vee})$ branched along the dual quadric $X^{\vee}$. On the other hand, if $V\cong \mathbb{C}^{2m+2}$ then, $\mathcal{A}_{0}=\mathcal{A}_{1}=\langle \mathcal{S}_{\pm},\mathcal{O}\rangle$, $\mathcal{A}_{i}=\langle \mathcal{O}\rangle$, $i=2,\ldots 2m-1$ and the HPD of $X$ is $g:Z\rightarrow \mathbb{P}(V^{\vee})$, where $Z=X^{\vee}$ with the usual embedding.

\item \textbf{Double Veronese embedding}. \cite{kuznetsov2008derived} If $f:X=\mathbb{P}(V)\rightarrow \mathbb{P}(S^{2}V)$ is the double Veronese embedding of $\mathbb{P}(V)$ (where $S^{2}V:=\mathrm{Sym}^{2}V$) and $X\cong \mathbb{P}^{2l-\varepsilon}$, $\varepsilon\in \{0,1\}$, we have the Lefschetz decomposition $\mathcal{A}_{0}=\ldots=\mathcal{A}_{l-2}=\langle \mathcal{O}_{X}(-1),\mathcal{O}_{X}\rangle$, and
\begin{eqnarray}
\mathcal{A}_{l-1}=\begin{cases}
                   \langle \mathcal{O}_{X}(-1)\rangle, & \text{if } \varepsilon=1\\
                   \langle \mathcal{O}_{X}(-1),\mathcal{O}_{X}\rangle, &  \text{if } \varepsilon=0
                  \end{cases}
\end{eqnarray}
The HPD category of $X$ is $D(\mathbb{P}(S^{2}V^{\vee}),\mathcal{C}l_{0})$, which consists of the coherent sheaves of modules over the even part of the universal Clifford algebra $\mathcal{C}l_0$ on $\mathbb{P}(S^{2} V^{\vee})$.

\end{itemize}

\section{\label{sec:GLSM}Lightning review of window categories in (anomalous) GLSM}
In this section we review the general definition of a GLSM \cite{Witten:1993yc}, in order to fix the notation for the following sections. We define a GLSM by specifying the following data.
\begin{itemize}
  \item \textbf{Gauge group}: a compact Lie group $G$.
  \item \textbf{Chiral matter fields}: a faithful unitary representation $\rho_{m}: G\rightarrow GL(V)$ of $G$ on some complex vector space $V\cong \mathbb{C}^{N}$. This determines the representation of the chiral $(2,2)$ fields.
  \item \textbf{Superpotential}: a holomorphic, $G$-invariant polynomial $W: V\rightarrow \mathbb{C}$, namely $W\in \mathrm{Sym}(V^{\vee})^G$.
  \item \textbf{Fayet-Illiopolous (FI)-theta parameters}: a set of complex parameters $t$ such that
  \begin{eqnarray}
  \exp(t)\in \mathrm{Hom}(\pi_{1}(G),\mathbb{C}^{*})^{\pi_{0}(G)}
  \end{eqnarray}
  i.e., $\exp(t)$ is a group homomorphism from $\pi_{1}(G)$ to $\mathbb{C}^{*}$ that is invariant under the adjoint action of $G$. It is customary to write $t=\zeta-i\theta$, therefore \cite{hori2019notes}
  \begin{eqnarray}
  t\in \left(\frac{\mathfrak{t}^{*}_{\mathbb{C}}}{2\pi i \mathrm{P}}\right)^{W_{G}}\cong\frac{\mathfrak{z}^{*}_{\mathbb{C}}}{2\pi i \mathrm{P}^{W_{G}}},
  \end{eqnarray}
  where $\mathrm{P}$ is the weight lattice, $W_{G}$ is the Weyl subgroup of $G$, 
$\mathfrak{t}$ is the Cartan subalgebra of $\mathfrak{g}=\mathrm{Lie}(G)$ and 
$\mathfrak{z}=\mathrm{Lie}(Z(G))$. We remark that the $W_{G}$ 
invariant part of $\mathfrak{t}^{*}_{\mathbb{C}}$, can be alternatively 
identified with the group of characters of $\mathfrak{g}_{\mathbb{C}}$.
  
  \item \textbf{R-symmetry}: a vector $U(1)_{V}$ and axial $U(1)_{A}$ R-symmetries that commute with the action of $G$ on $V$. In particular, we denote for future use $R:U(1)_{V}\rightarrow GL(V)$.
    To (classically) preserve the $U(1)_{V}$ symmetry the superpotential must have weight $2$ under it:
    \begin{eqnarray}
    W(R(\lambda)\cdot\phi)=\lambda^{2}W(\phi),
    \end{eqnarray}
    where $\phi$ denotes the coordinates in $V$.
  \end{itemize}
  For a GLSM with gauge group $G$, and $n$ chirals in representations $R_{i}$, $i=1,\ldots,n$, the condition for the cancellation of $U(1)_{A}$ anomaly takes the following form:
\begin{eqnarray}
  \sum_{i=1}^{n} \mathrm{tr}_{R_i}(T) = 0 \text{ \ for all \ }T \in \mathfrak{g}.
  \end{eqnarray}
Consider the case $G=U(1)^{s}$, hence $\mathfrak{g}\cong \mathbb{R}^{s}$. Denote $Q^{j}\in \mathfrak{g}^{*}$, $j=1,\ldots,N$, the weights of $\rho_{m}$. Then, the  anomaly condition reads
\begin{eqnarray}
  Q^{\mathrm{tot}}:=\sum_{j=1}^{N}Q_{j} = 0.
  \end{eqnarray}
  We will be mainly interested in models where $Q^{\mathrm{tot}}\neq 0$. Then, the FI-theta parameter gets a 1-loop correction that depends logarithmically in the energy scale:
\begin{eqnarray}\label{zetaren}
  \zeta_{\mathrm{ren}}(\mu):=\zeta_{\mathrm{bare}}+Q^{\mathrm{tot}}\log\left(\frac{\mu}{\Lambda}\right),
  \end{eqnarray}
we denote $t_{\mathrm{ren}}=\zeta_{\mathrm{ren}}-i\theta$ \footnote{The parameter $\theta$ do not receive quantum corrections, however, for anomalous models, its component along $Q^{\mathrm{tot}}$ can be redefined by a $U(1)_{A}$ rotation.}. The parameters $t_{\mathrm{ren}}$, at fixed $\frac{\mu}{\Lambda}$, span what we will refer to as the quantum K\"ahler moduli space $\mathcal{M}_{K}$. When $Q^{\mathrm{tot}}=0$, $\mathcal{M}_{K}$ takes the form $(\mathbb{C}^{*})^{s}\setminus \Delta$ where $\Delta$ is a complex codimension $1$ discriminant. For the anomalous case $Q^{\mathrm{tot}}\neq0$, the structure of $\mathcal{M}_{K}$ can be more complicated because of the existence of Coulomb and mixed Coulomb-Higgs branches that are not singular. Then a case-by-case analysis is required, but the classical phase space spanned by $t_{\mathrm{ren}}(\mu)$ gives a great deal of information about the Higgs branches, as we will review below. The Coulomb and mixed Coulomb-Higgs branches are determined by the twisted effective superpotential\footnote{The factor of $i$ inside the $\log$ term is necessary in order to match the energy scale $\Lambda$ with the one appearing in (\ref{zetaren}) as shown in \cite{hori2019notes}}:
\begin{eqnarray}\label{effpotsigma}
 \widetilde{W}_{\mathrm{eff}}(\sigma)=-t(\sigma)+2\pi i \rho_{W}(\sigma)-\sum_{j}Q_{j}(\sigma)\left(\log\left(i\frac{Q_{j}(\sigma)}{\Lambda}\right)-1\right),
  \end{eqnarray}
where $ \rho_{W}$ denotes the Weyl vector of $G$, which is set to $\rho_{W}\equiv 0$ for $G$ abelian. Then, the solutions of $\exp\frac{\partial\widetilde{W}_{\mathrm{eff}}(\sigma)}{\partial\sigma}=1$ establishes the location and nature of the Coulomb and mixed Coulomb-Higgs branches, however, further analysis of the masses of the chiral fields is necessary for each solution as performed in section \ref{sec:examples}. Define the cones $C_{I}\subset \mathfrak{g}^{*}$ as
\begin{eqnarray}
  \mathrm{C}_{I}:=\mathrm{Span}_{\mathbb{R}_{\geq 0}}\left\{ Q_{j}:j\in I\right\}\subset \mathfrak{g}^{*},
  \end{eqnarray}
i.e. the positive span of the vectors $Q_{j}$ with $j\in I\subset \{1,\ldots,N\}$, then we define the classical phase boundaries as $C_{I}$'s such that $\mathrm{dim}_{\mathbb{R}}(C_{I})=s-1$. Classically, the space of solutions to the D-term equations:
\begin{eqnarray}\label{dterm}
 \sum_{j=1}^{N}Q_{j}|\phi_{j}|^{2}=\zeta,
\end{eqnarray}
which consists of points that break $G$ into a finite subgroup (i.e. points whose stabilizer is a finite subgroup of $G$) whenever $\zeta$ is in the interior of a phase. When $\zeta$ is generic but constrained to the interior of a phase boundary, $\zeta\in C_{I}$, then solutions to (\ref{dterm}) break $G$ to $U(1)\times \Gamma$ where $\Gamma$ is a finite group\footnote{In principle, $\Gamma$ can depend on the point we take as solution to (\ref{dterm}), but we will not encounter this situation in this work, hence we ignore it}. In such a case, we can integrate out $\{\phi_{j}\}_{j\in I}$ and we are left with a $U(1)\times \Gamma$ gauge theory with chiral matter $\{\phi_{j}\}_{j\not\in I}$. The superpotential in this local model must be induced from $W$, upon integrating out $\{\phi_{j}\}_{j\in I}$, but actually details of the resulting superpotential are irrelevant for our purposes. Denote $u\in \mathfrak{g}$ as the integral generator of $U(1)\subset G$, i.e., $u$ satisfies $\exp(2\pi i n u)=1\in G$ for any $n\in\mathbb{Z}$. We then define what we call a local GLSM at the phase boundary $C_{I}$, which is given by the following data
\begin{flushleft}
\begin{eqnarray}\label{localGLSM}
&&\textbf{matter \ \ } \{\phi_{j}\}_{j\not\in I}\nonumber\\
&&\textbf{gauge group \ \ }U(1)\times \Gamma, \qquad \mathfrak{u}(1)=\mathbb{R}u\nonumber\\
&&\textbf{FI-theta parameter \ \ }  t_{\mathrm{ren}}(u)\nonumber\\
\end{eqnarray}
\end{flushleft}
To a GLSM, for fixed values of $t_{\mathrm{ren}}$ and $W$, we can assign a category of B-branes. Such category, denoted by $MF_{G}(W)$, is described by pairs of objects $(\mathcal{B},L_{t})$ which we will describe in the following. Let us start with what we will call the \emph{algebraic data} consisting of a quadruple $\mathcal{B}=(M,\rho_{M},R_{M},\mathbf{T})$:
\begin{itemize}
\item \textbf{Chan-Paton vector space}: a $\mathbb{Z}_{2}$-graded, finite dimensional free $\Sym(V^\vee)$ module denoted by $M=M_{0}\oplus M_{1}$.
\item \textbf{Boundary gauge and (vector) R-charge representation}: $\rho_{M}: G\rightarrow GL(M)$, and $R_{M}:U(1)_{V}\rightarrow GL(M)$ commuting and even representations, where the weights of $R_{M}$ are allowed to be rational.
\item \textbf{Matrix factorization of $W$}: Also known as the tachyon profile, a $\mathbb{Z}_2$-odd endomorphism $\mathbf{T} \in \mathrm{End}^{1}_{\Sym(V^{\vee})}(M)$ satisfying $\mathbf{T}^{2}=W\cdot\mathrm{id}_{M}$.
\end{itemize}
The group actions $\rho_{M}$ and $R_{M}$ must be compatible with $\rho_{m}$ and $R$, i.e.
  for all $\lambda\in U(1)_{V}$ and $g\in G$,
  \begin{equation}\label{rhodef}
    \begin{aligned}
      R_{M}(\lambda)\mathbf{T}(R(\lambda)\phi)R_{M}(\lambda)^{-1} & = \lambda \mathbf{T}(\phi) , \\
      \rho_{M}(g)^{-1}\mathbf{T}(\rho_{m}(g)\cdot \phi)\rho_{M}(g) & = \mathbf{T}(\phi) .
    \end{aligned}
  \end{equation}
The other piece of data that we need, termed $L_{t}$, is a profile for the vector multiplet scalar.
This data consists of a gauge-invariant middle-dimensional subvariety $L_{t}\subset \mathfrak{g}_{\mathbb{C}}$ of the complexified Lie algebra of $G$, or equivalently its intersection $L\subset\mathfrak{t}_{\mathbb{C}}$ with the (complexified) Cartan subalgebra $\mathfrak{t}$, which we refer to as the contour. An \textbf{admissible contour} is a gauge invariant, middle dimensional $L_{t}$ that is a continuous deformation of the real contour $L_{\mathbb{R}}:=\{\Im\sigma=0\}$, where we denote by $\sigma\in \mathfrak{t}_{\mathbb{C}}$ a point in $\mathfrak{t}_{\mathbb{C}}$, such that the imaginary part of the boundary effective twisted superpotential $\widetilde{W}_{\text{eff},q}:\mathfrak{t}_{\mathbb{C}}\rightarrow \mathbb{C}$
\begin{equation}\label{twistedbdry}
  \widetilde{W}_{\text{eff},q}(\sigma):= \left(\sum_{\alpha>0}\pm i\pi\,\alpha\cdot\sigma\right)-\left(\sum_j (Q_j(\sigma))\left(\log\left(\frac{iQ_j(\sigma)}{\Lambda}\right)-1\right)\right)-t(\sigma)+2\pi i q(\sigma)
\end{equation}
approaches $+\infty$ in all asymptotic directions of $L_{t}$ and for all the weights $q\in \mathfrak{t}^{*}$ of $\rho_{M}$.
Signs in the sum over positive roots $\alpha$ of $G$ depend on the Weyl chamber in which $\Re\sigma$ lies; this sum is absent in abelian GLSMs. When $\zeta_{\mathrm{ren}}$ is deep inside a phase i.e. we assume $|\zeta_{\mathrm{ren}}|\gg 1$, there always exists a unique, up to homotopy, admissible $L_{t}$ given a quadruple $\mathcal{B}$ and so, we can forget this information and regard the image of $\mathcal{B}$ in the gauge decoupling limit, i.e. where we take the renormalized gauge coupling  $g_{\mathrm{ren}}(\mu)$ to infinity while keeping $t_{\mathrm{ren}}(\mu)$ fixed, as an object $\pi_{\zeta_{\mathrm{ren}}}(\mathcal{B})$ in a category which we denote $\mathcal{D}_{\zeta_{\mathrm{ren}}}$. In general, this category has a decomposition of the form
\begin{eqnarray}
\mathcal{D}_{\zeta_{\mathrm{ren}}}=\langle D(Y_{\zeta_{\mathrm{ren}}},W|_{\zeta_{\mathrm{ren}}}),E_{1},\ldots,E_{k} \rangle,
\end{eqnarray}
where $D(Y_{\zeta_{\mathrm{ren}}},W|_{\zeta_{\mathrm{ren}}})$ represents the B-brane category of the Higgs branch. More precisely, denoting $\mu: V\rightarrow\mathfrak{g}^{*}$ the moment map associated to $\rho_{m}$, we identify $Y_{\zeta_{\mathrm{ren}}}=\mu^{-1}(\zeta_{\mathrm{ren}})/G$ and $W|_{\zeta_{\mathrm{ren}}}$ with $W$ restricted to $Y_{\zeta_{\mathrm{ren}}}$. When the stabilizer on $G$ of all points in $Y_{\zeta_{\mathrm{ren}}}\cap dW^{-1}(0)$ is finite, $D(Y_{\zeta_{\mathrm{ren}}},W|_{\zeta_{\mathrm{ren}}})$ denotes the category of B-branes of the hybrid model\footnote{We use the term hybrid model, to refer to Landau-Ginzburg (LG) orbifold models with nontrivial target space. In mathematics literature they are usually called just LG models.} described by the pair $(Y_{\zeta_{\mathrm{ren}}},W|_{\zeta_{\mathrm{ren}}})$ (mathematically this corresponds to coherent matrix factorizations \cite{buchweitz1987maximal,orlov2004triangulated,efimovcoherent,ballard2012resolutions}). We write $E_{1},\ldots,E_{k}$ for Coulomb or mixed Coulomb-Higgs vacua that may arise in the chosen phase. In the case of nonanomalous GLSMs, the $E_{1},\ldots,E_{k}$ pieces are absent in every phase. The map $\pi_{\zeta_{\mathrm{ren}}}$, induced by RG flow, in general, is a projection $\pi_{\zeta_{\mathrm{ren}}}:MF_{G}(W)\rightarrow \mathcal{D}_{\zeta_{\mathrm{ren}}}$ and there is no natural lift. A natural question is then, how $\mathcal{D}_{\zeta_{\mathrm{ren}}}$ and $\mathcal{D}_{\zeta'_{\mathrm{ren}}}$ are related if $\zeta_{\mathrm{ren}},\zeta'_{\mathrm{ren}}$ belong to distinct cones in $\mathcal{M}_{K}$ (if they belong to the same cone, then the categories are trivially equivalent). This problem was solved, for $G$ abelian in \cite{Herbst:2008jq} and in the mathematics literature in \cite{segal2011equivalences,halpern2015derived,ballard2019variation} for general $G$. A physics perspective on the case of general $G$ and anomalous GLSMs can be found in \cite{Clingempeel:2018iub,Hori:2013ika,hori2019notes,eager2017beijing}. Here, we mostly follow \cite{Clingempeel:2018iub,hori2019notes}. Suppose that $\zeta_{\mathrm{ren}},\zeta'_{\mathrm{ren}}$ belong to opposite sides of a phase boundary $C_{I}$. We then can define two families of subcategories $\mathcal{W}_{\pm,l}\subset MF_{G}(W)$, which we call window categories, labeled by $l\in \mathbb{Z}$. These categories are defined as the objects of $MF_{G}(W)$ such that the weights $q$ of $\rho_{M}|_{U(1)}$, for the $U(1)\subset G$ generated by $u$, previously defined, are constrained, to be more precise we define:
\begin{eqnarray}\label{constrwin}
&&\textbf{Small window}:\qquad|\theta(u)+2\pi q(u)|<\pi \mathrm{min}(N_{u,\pm})\nonumber\\
&&\textbf{Big window}:\qquad|\theta(u)+2\pi q(u)|<\pi \mathrm{max}(N_{u,\pm})
\end{eqnarray}
where $N_{u,\pm}:=\sum_{j}(Q_{j}(u))^{\pm}$ and $(x)^{\pm}:=(|x|\pm x)/2$. Therefore we have the definition of the window subcategories by the constraints (\ref{constrwin}): $\mathcal{W}_{+,l}$ (resp. $\mathcal{W}_{-,l}$) corresponds to the objects $\mathcal{B}$ such that the weights of $\rho_{M}$ satisfy the big (resp. small) window constraint for $l=\lfloor\frac{\theta(u)}{2\pi}\rfloor$. WLOG assume we fix a coordinate system such that the wall $C_{I}$ is at $\zeta_{\mathrm{ren}}(u)=0$ and $Q^{\mathrm{tot}}(u)\geq 0$. Then, we can take $\zeta_{\mathrm{ren}}\gg 1$ and $\zeta'_{\mathrm{ren}}\ll -1$. We then expect that there is an equivalence of categories associated with each side of the wall:
\begin{eqnarray}\label{derwin}
\mathcal{W}_{+,l}\cong D(Y_{\zeta_{\mathrm{ren}}(u)},W_{\zeta_{\mathrm{ren}}(u)})\cong \langle D(Y_{\zeta'_{\mathrm{ren}}(u)},W_{\zeta'_{\mathrm{ren}}(u)}),E_{1},\ldots,E_{k} \rangle
\end{eqnarray}
for all $l$ where $k=|N_{u,+}-N_{u,-}|$. The objects $E_{i}$ represent massive vacua from the Coulomb or mixed Coulomb/Higgs branches. When the model is nonanomalous $\mathcal{W}_{+,l}=\mathcal{W}_{-,l}$ and (\ref{derwin}) gives an equivalence of Higgs branch categories. For anomalous models there is a fully faithful map $\mathcal{W}_{-,l}\rightarrow \mathcal{W}_{+,l}$ and we have a refinement of (\ref{derwin}):
\begin{eqnarray}\label{derwinsmall}
\mathcal{W}_{-,l}\cong D(Y_{\zeta'_{\mathrm{ren}}(u)},W_{\zeta'_{\mathrm{ren}}(u)}).
\end{eqnarray}
We remark that the classical phase boundaries defined by the cones $C_{I}$ are not the whole phase space. When taking into account quantum corrections, new walls can appear and some classical ones can be lifted. However, the equivalences (\ref{derwin}) and (\ref{derwinsmall}) are always valid when taking the gauge decoupling limit for $|\zeta_{\mathrm{ren}}(u)|$ sufficiently large. Note that the definition of the window categories comes precisely from the study of B-branes in the local GLSM (\ref{localGLSM}). Because of this, and the previous remark on quantum corrected phase boundaries, the vacua $E_{j}$ in (\ref{derwin}) can be further decomposed into smaller vacua as $\zeta'_{\mathrm{ren}}(u)$ becomes more and more negative, but the simple analysis of the local GLSM will be not enough to see this effect and one has to analyze the global behavior of the GLSM. We will see examples of such a situation in section \ref{sec:examples}.

\section{\label{sec:proposal}Proposal for GLSM for HPD of projective varieties}

\begin{figure}
\centering
 \begin{tikzpicture}[scale=0.70]
  	\draw[thin,->] (0,0) -- (4,0) node[anchor=north west]{$\zeta_l$};
	\draw[thin,->] (0,0) -- (0,4) node[anchor=south east]{$\zeta'$};
	\draw [ultra thick] (0,0) -- (0,3.5);
	\draw [ultra thick] (0,0) -- (3.5,0);
	\draw [ultra thick] (0,0) -- (-3.5,-3.5);
	\draw [ultra thick, dashed] (0,0) -- (-4,0);
	\draw [thick, dotted, ->] (0,0) -- (-1,-4);
    \node at (2.5,2.5) {$D(\mathcal{X})$};
    \node at (-2.5,2.5) {${\cal C}$};
    \node at (-2.5,-1.25) {${\cal C}'$};
    \node at (1,-2.5) {$\varnothing$};
 \end{tikzpicture}
 \caption{Higgs branches of the GLSM $\mathcal{T}_{\mathcal{X}}$}\label{fig:phase}
{\footnotesize \begin{flushleft} The theory $\mathcal{T}_{\mathcal{X}}$ has a geometric phase realizing the universal hyperplane section $\mathcal{X}$ and a LG phase realizing the HPD category $\mathcal{C}$ (assuming that the Lefschetz decomposition is nontrivial). When $D(Y_{\zeta_{l}\ll -1},W_{\zeta_{l}\ll -1})$ is empty, $\mathcal{C}' \cong \mathcal{C}$, otherwise $\mathcal{C}'$ is a subcategory of $\mathcal{C}$. The dashed arrow shows the direction of RG flow. \end{flushleft} }
 \end{figure}
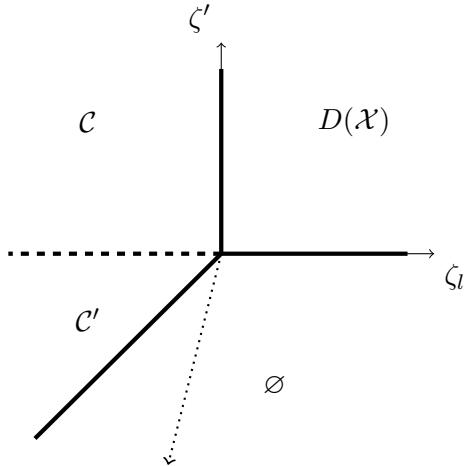

We start with a GLSM\footnote{For all of our general analysis, the R-charge 
assignment is irrelevant, so it will be left unspecified.} 
$\mathcal{T}_{X}=(G,\rho_{m}: G\rightarrow GL(V),W,t_{\mathrm{ren}},R)$ and 
assume it has a large volume point in $\mathcal{M}_{K}$ where we can identify a 
geometric phase whose category of B-branes at low energies is equivalent to 
$D(X)$. In the following, we will omit the subscript `$\mathrm{ren}$' from all 
the FI parameters to avoid cluttering. Consider a map $f:X\rightarrow 
\mathbb{P}(S)$. In order for this map to be well defined in all of $X$ it must 
be base point free, i.e. $\{x\in X : f(x)=0\}=\emptyset$. 
Since there is a $G$-action on $S$, one can see that there exists a map 
$\alpha:\mathrm{Hom}(G,\mathbb{C}^{*})\rightarrow \mathrm{Pic}(X)$ 
induced by the $G$-equivariant line bundle $V \times \mathbb{C}_{\mu}$ over $V$ for any $\mu \in \mathrm{Hom}(G,\mathbb{C}^{*})$,
and hence there 
exists a character $\chi$ of $G$ such that
\begin{equation}\label{chi}
\alpha(\chi)= f^{*}\mathcal{O}_{\mathbb{P}(S)}(1). 
\end{equation}
The character $\chi$ defines a one-dimensional representation $\mathbb{C}_\chi$ of $G$. In the context of the GLSM $\mathcal{T}_{X}$, we can write the image of 
$f$ as $[f_{0}(X),\ldots,f_{n}(X)]$, where $n+1=\mathrm{dim}(S)$ and each $f_{j}(X)$ 
is a $\chi$-invariant function, i.e. $g\in G$ acts on $f_j(x)$ as multiplication 
by $\chi(g)$. 
For later use, by fixing a basis for the 
generators of $G$, each $f_j$ becomes a $G/U(1)_{l}$-invariant function for 
some $U(1)_{l}\subseteq G$ and transforms under $U(1)_{l}$ with the weight\footnote{We can consider also $f_{j}(x)$'s with different weights. This will 
be interpreted as a map from $X$ to a weighted projective space 
$\mathbb{WP}(S)$, but in this work we only consider the case $\mathbb{P}(S)$.} $Q$ determined by $\chi$. The functions $f_{j}(X)$ depend on some fields $X_{a}$, representing the coordinates of $X$, however they are not 
necessarily fundamental fields, in general they will be polynomials in the 
coordinates $\phi_{\alpha}$, $\alpha=1,\ldots, N=\mathrm{dim}(V)$ of $V$. 
Define $\zeta_{l}:=\zeta(h)$ where $h$ is the integral generator of 
$\mathfrak{u}(1)_{l}$ and assume WLOG that the large volume phase is located at 
a region $\zeta\gg 1$, in particular with $\zeta_{l}\gg 1$. We will require that 
our GLSM $\mathcal{T}_{X}$ satisfies the condition:
\begin{eqnarray}\label{cdt1}
Y_{\zeta\gg 1}\cap \{\phi_{\alpha}\in V^{\vee} : f(X(\phi))=0\}=\emptyset.
\end{eqnarray}
Therefore we write\footnote{Even though we assume the large volume phase is weakly coupled, our arguments should carry on for cases where the NLSM on $X$ is realized nonperturbatively such as in \cite{Hori:2006dk}.}
\begin{eqnarray}
X_{\zeta_{l}\gg 1}:=Y_{\zeta\gg 1}\cap dW^{-1}(0),
\end{eqnarray}
where we identify $X$ with $X_{\zeta_{l}\gg 1}$, at the point in the K\"ahler moduli determined by $t$, with $\zeta=\Re(t)\gg 1$. Then, we fix all the FI parameters corresponding to $G/U(1)_{l}$ and we write:
\begin{eqnarray}\label{Xcats}
D(X_{\zeta_{l}\gg 1})\cong \langle D(Y_{\zeta_{l}\ll -1},W_{\zeta_{l}\ll -1}),E_{1},\ldots,E_{k} \rangle
\end{eqnarray}
where $E_{i}$ denote the Coulomb/mixed vacua deep in the phase where $\zeta_{l}\ll -1$. We construct the following GLSM
\begin{eqnarray}
\mathcal{T}_{\mathcal{X}}=(\widehat{G}=G\times U(1)_{s+1},\hat{\rho}_{m}: \widehat{G}\rightarrow GL(V\oplus V'),\widehat{W},\widehat{R}),
\end{eqnarray}
where $s=\mathrm{dim}(\mathfrak{z})$ and the representation $V'$ of $ 
\widehat{G}$ is given by 
$V'=\mathbb{C}_{(\chi^{-1},-1)}\oplus S^\vee$ 
where $\chi$ is defined by \eqref{chi}, $U(1)_{s+1}$ acts on $S^\vee$ with weight $1$ and $G$ acts on $S^\vee$ trivially. We denote the 
coordinates of $V'$ as $(P,S_{0},\ldots,S_{n})$. The superpotential 
$\widehat{W}$ is given by
\begin{eqnarray}
\widehat{W}=W+P\sum_{j=0}^{n}S_{j}f_{j}(X).
\end{eqnarray}
Here we denoted $f_{j}(X)$ for the components of the image of the map $f:X\rightarrow \mathbb{P}(S)$. The GLSM  $\mathcal{T}_{\mathcal{X}}$ is quite analogous to the model studied in \cite{ballard2017homological,ballard2014derived}. We interpret $\mathcal{T}_{\mathcal{X}}$ as the GLSM of the universal hyperplane section of $X$, to be more precise, $\mathcal{T}_{\mathcal{X}}$, by construction will have a large volume phase at $\boldsymbol{\zeta}\gg 1$, $\boldsymbol{\zeta}:=(\zeta_{l},\zeta')$ (and fix the other FI parameters to the large volume phase of $X$), where $\zeta'$ is the FI parameter corresponding to $U(1)_{s+1}$. The F-term equations
\begin{eqnarray}
\frac{\partial\widehat{W}}{\partial S_{j}}=0
\end{eqnarray}
implies that $f_{j}(X)=0$ for all $j$ if $P\neq 0$,. Then, the moment map equation for $\zeta_{l}\gg 1$ and the condition (\ref{cdt1}) implies this is not possible and $P$ must vanish. Therefore, we must have $P=0$ on the Higgs branch $\boldsymbol{\zeta}\gg 1$, then it is easy to see that $\widehat{Y}\cap d\widehat{W}^{-1}(0)$ reduces to $\mathcal{X}$. Therefore we identify this phase with the NLSM with target $\mathcal{X}_{\boldsymbol{\zeta}\gg 1}$. Consider now the phase $(\zeta_{l}\ll -1,\zeta'\gg 1)$. Then we have the following correspondence of B-brane categories
\begin{eqnarray}
D(\mathcal{X}_{\boldsymbol{\zeta}\gg 1})\cong \langle D(\widehat{Y}_{\zeta_{l}\ll -1},\widehat{W}_{\zeta_{l}\ll -1}),C_{1},\ldots,C_{k'} \rangle,
\end{eqnarray}
where $C_{i}$ denote the Coulomb/mixed vacua deep in the phase $(\zeta_{l}\ll -1,\zeta'\gg 1)$ and $\widehat{Y}_{\zeta_{l}\ll -1}$ is the appropriate symplectic quotient associated with the D-terms of the GLSM $\mathcal{T}_{\mathcal{X}}$. The vacua $C_{i}$, $i=1,\ldots,k'$ can be computed using the local model (\ref{localGLSM}) at the phase boundary $(\zeta_{l}=0,\zeta'\in \mathbb{R}_{\geq 0})$. Then, we claim
\begin{eqnarray}
\mathcal{C}\cong  D(\widehat{Y}_{\zeta_{l}\ll -1},\widehat{W}_{\zeta_{l}\ll -1}),
\end{eqnarray}
where $\mathcal{C}$ is the HPD category of $X$ with respect to $\mathcal{L}$ and the Lefschetz deocomposition described below. In general, in our examples, we will be able to find an explicit description of $D(\widehat{Y}_{\zeta_{l}\ll -1},\widehat{W}_{\zeta_{l}\ll -1})$ in terms of a hybrid model. Denoting by $\mathcal{W}^{U(1)_{l}}_{r,\pm}$ the big/small window categories associated with the $U(1)_{l}$ subgroup of $\widehat{G}$ on $\mathcal{T}_{\mathcal{X}}$, we can rephrase this result in terms of them
\begin{eqnarray}
\mathcal{C}\cong \mathcal{W}^{U(1)_{l}}_{r,-}\hookrightarrow \mathcal{W}^{U(1)_{l}}_{r,+}\cong D(\mathcal{X}_{\zeta_{l}\gg 1})\qquad \text{ \ for all \ }r\in \mathbb{Z},
\end{eqnarray}
where $r$ corresponds to the choice of $\theta_{l}=\theta(h)$. The Lefschetz decomposition induced by our construction, we claim, is the one determined by the center
\begin{eqnarray}\label{lefscenter}
\mathcal{A}_{0}= \langle D(X_{\zeta_{l}\ll -1},W_{\zeta_{l}\ll -1}),E_{1},\ldots,E_{k-k'} \rangle,
\end{eqnarray}
therefore a Lefschetz decomposition is imposed on us, by construction. This is exactly the same case in \cite{ballard2017homological,ballard2014derived}. Generically the phase space will look like figure \ref{fig:phase}.
We remark that we have two more phases that we can distinguish in the (classical) $\boldsymbol{\zeta}$ space. One is the phase $(Q\zeta'-\zeta_{l}\ll -1,\zeta'\ll -1)$ which has an empty Higgs branch (since $P\neq0$), and $(Q\zeta'-\zeta_{l}\gg 1,\zeta'\ll -1)$ whose B-brane category embeds into $\mathcal{C}$.  If the category $D(Y_{\zeta_{l}\ll -1},W_{\zeta_{l}\ll -1})$ in (\ref{Xcats}) is empty, the phase boundary $(\zeta_{l},\zeta')=(0,\mathbb{R}_{\leq 0})$ will be lifted by quantum effects or not present at all (for an explanation see footnote \ref{footbdry}) and therefore the Higgs branch of this latter phase is trivially equivalent to $\mathcal{C}$. We will see both cases occurring in our examples. The analysis of the Coulomb and mixed Coulomb-Higgs vacua is more involved and a classical analysis is not enough, therefore is not possible to provide a full picture at this level of generality. We provide a full detailed analysis for each family of examples in section \ref{sec:examples}.
\subsection{\label{sec:linearsec} Taking linear sections}

An important feature of HPD is the behavior of $\mathcal{C}$ under the intersection of $X$ with linear sections $L\subset S^{\vee}$. We can just modify the GLSM $\mathcal{T}_{\mathcal{X}}$ accordingly. From the matter structure of $\mathcal{T}_{\mathcal{X}}$, is clear that, in the large volume phase $(\zeta_{l}\gg 1,\zeta'\gg 1)$ we can identify the fields $S_{0},\ldots, S_{n}$ with the homogeneous coordinates of $\mathbb{P}(S^{\vee})$, therefore, if $\mathrm{dim}L=r\leq n$, denote a basis of $L$ as $\mathbf{l}^{*}_{a}$, $a=1,\ldots,r$ and we write the rank $r$ matrix $m_{a,i}:=\mathbf{l}^{*}_{a}(\mathbf{e}_{i})$ where $\mathbf{e}_{i}$, $i=0,\ldots,n$ is a basis of $S$. Then, we propose the GLSM $\mathcal{T}_{\mathcal{X}_{L}}$ given by
\begin{eqnarray}
\mathcal{T}_{\mathcal{X}_{L}}=(\widehat{G},\hat{\rho}_{m}: \widehat{G}\rightarrow GL(V\oplus V'_{L}),\widehat{W}_{L},\widehat{R}_{L}),
\end{eqnarray}
where $V'_{L}=\mathbb{C}_{(\chi^{-1},-1)}\oplus L$ with $\chi$ given by \eqref{chi}. We denote the coordinates of $V'_L$ as $(P,S_{1},\ldots,S_{r})$. The superpotential 
$\widehat{W}_{L}$ is given by
\begin{eqnarray}
\widehat{W}_{L}=W+P\sum_{a,j}S_{a}m_{a,j}f_{j}(X),
\end{eqnarray}
then the classical phase boundaries (in the $\boldsymbol\zeta$ space) of $\mathcal{T}_{\mathcal{X}_{L}}$ are the same as those of $\mathcal{T}_{\mathcal{X}}$ model previously analyzed. However, the Higgs branches are different. They become the following
\begin{figure}
\centering
 \begin{subfigure}[b]{0.5\textwidth}
 \begin{tikzpicture}[scale=0.60]
  	\draw[thin,->] (0,0) -- (4,0) node[anchor=north west]{$\zeta_l$};
	\draw[thin,->] (0,0) -- (0,4) node[anchor=south east]{$\zeta'$};
	\draw [ultra thick] (0,0) -- (0,3.5);
	\draw [ultra thick] (0,0) -- (3.5,0);
	\draw [ultra thick] (0,0) -- (-3.5,-3.5);
	\draw [ultra thick, dashed] (0,0) -- (-4,0);
    \node at (2.5,2.5) {$D(\mathcal{X}_L)$};
    \node at (-2.5,2.5) {$D(Z_{L})$};
    \node at (-2.5,-1.25) {${\cal C}'_L$};
    \node at (1,-2.5) {$D(X_L)$};
 \end{tikzpicture}
 \caption{$\mathcal{T}_{\mathcal{X}_L}$}\label{fig:phase_restricted_a}
 \end{subfigure}
\begin{subfigure}[b]{0.3\textwidth}
\begin{tikzpicture}[scale=0.60]
	\draw[thick,->] (-4,0) -- (4,0) node[anchor=north west]{$\zeta_1$};
	\draw [thick, dotted, ->] (2.5,0.5) -- (-3.,0.5);
	\node at (-4,1) {${\cal C}'_L$};
	\node at (4.75,1) {$D(X_L)$};
	\node at (0,-2.75) {$ $};
\end{tikzpicture}
 \caption{$\mathcal{T}_{X_L}$}\label{fig:phase_restricted_b}
 \end{subfigure} 
\caption{Higgs branches of the GLSM $\mathcal{T}_{\mathcal{X}_L}$}\label{fig:phase_restricted}
 {\footnotesize \begin{flushleft} (a)~The phase diagram of $\mathcal{T}_{\mathcal{X}_L}$. When restricted to a linear subspace $L \subset S^\vee$, a Higgs branch described by NLSM on $X_L$ appears. (b)~The strong coupling limit of $U(1)_{s+1}$ sees the correspondence between $X_L$ and $\mathcal{C}'_L$. The dashed arrow indicates the direction of RG flow (There is no RG flow if $X_L$ is Calabi-Yau). Note that this arrow can run in either direction, depending on the model.  \end{flushleft}}
 \end{figure}
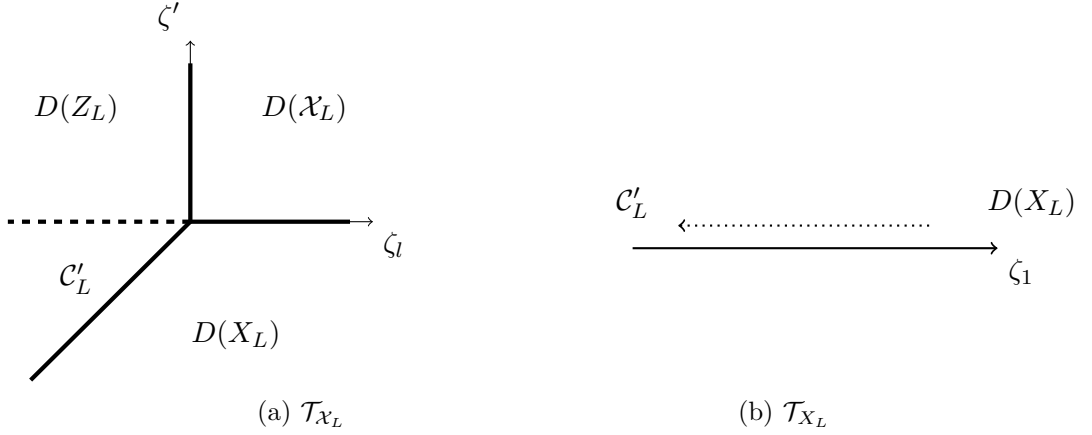
\begin{itemize}
\item $(\zeta_{l}\gg 1,\zeta'\gg 1)$: An analogous reasoning shows that this phase becomes the universal hyperplane section of $X_{L}$, denoted by $\mathcal{X}_{L}=X\times_{\mathbb{P}(V)}\mathcal{H}_{L}$ where
\begin{eqnarray}
\mathcal{H}_{L}=\{(u,v)\in \mathbb{P}(V)\times \mathbb{P}(L) : v(u)=0 \}.
\end{eqnarray}
\item $(Q\zeta'-\zeta_{l}\ll -1,\zeta'\ll -1)$: In this phase we have the condition $P\neq 0$ therefore $F_{a}:=m_{a,j}f_{j}(X)=0$ for all $a$. In this case, since the rank of $m_{a,j}$ is less than $n+1$, we can have a nonempty solution. Then, the F-term equation for $\phi_{\alpha}$ is given by
\begin{eqnarray}\label{ftermgeneral}
\frac{\partial W}{\partial\phi_{\alpha}}+P\sum_{a}S_{a}\frac{\partial F_{a}}{\partial \phi_{\alpha}}=0
\end{eqnarray}
if the Jacobian $\frac{\partial F_{a}}{\partial \phi_{\alpha}}$ is full rank in $Y_{\zeta\gg 1}$ then, the value of $S_{a}$ is completely fixed by (\ref{ftermgeneral}). This is the case if we assume transversality of $X_{L}\subset Y_{\zeta\gg 1}$. In all our examples we see that smoothness of $X_{L}\subset Y_{\zeta\gg 1}$ moreover fixes $S_{a}\equiv 0$, and we can identify the Higgs branch with $X_{L}$ and we propose that, in general, this should be the case. However, we do not have a proof for the general case. For instance, if $W\equiv 0$ this is clearly the case, or if we can identify $X$ with the zeroes of a section $s$ of a vector bundle $\mathcal{V}\rightarrow B$ over some manifold $B$, then, we can write a superpotential of the form $W=p(s)$ where $p$ belongs to the dual bundle $\mathcal{V}^{\vee}$. Then, locally $W=p_{\alpha}s^{\alpha}$, and $X$ is described by the complete intersection $s_{\alpha}(x)=0$. Transversality condition for $X_{L}$ will read 
\begin{eqnarray}
\left\{ \mathrm{rk}\left[\frac{\partial (s_{\alpha},F_{a})}{\partial x_{i}}\right]< \mathrm{codim}X_{L} \right \}\cap X_{L}=\emptyset
\end{eqnarray}
and $Y_{\zeta\gg 1}$ will be identified with $\mathcal{V}^{\vee}\rightarrow B$. Then, in such cases, (\ref{ftermgeneral}) will imply the vanishing of $P_{\alpha}$ and $S_{a}$.

\item $(\zeta_{l}\ll-1,\zeta'\gg -1)$: The Higgs branch in this phase corresponds to the hybrid model $(\widehat{Y}_{L,\zeta_{l}\ll -1},\widehat{W}_{L,\zeta_{l}\ll -1})$ and we identify $D(\widehat{Y}_{L,\zeta_{l}\ll -1},\widehat{W}_{L,\zeta_{l}\ll -1})\cong D(Z_{L})$ for the case we have a variety $Z$ which we can identify with the HPD of $X$, namely $Z$ satisfies $D(Z)\cong \mathcal{C}$ (see section \ref{sec:section2}).
\item $(Q\zeta'-\zeta_{l}\gg 1,\zeta'\ll -1)$: Generically, the category of B-branes of this phase embeds into $D(\widehat{Y}_{L,\zeta_{l}\ll -1},\widehat{W}_{L,\zeta_{l}\ll -1})$. If the category $D(Y_{\zeta_{l}\ll -1},W_{\zeta_{l}\ll -1})$ in (\ref{Xcats}) is empty, the Higgs phase will be identical to $(\widehat{Y}_{L,\zeta_{l}\ll -1},\widehat{W}_{L,\zeta_{l}\ll -1})$ i.e. the phase boundary $(\zeta_{l},\zeta')=(0,\mathbb{R}_{\leq 0})$ will be lifted by quantum effects or not present at all\footnote{\label{footbdry}This is simply because, in the case of $D(Y_{\zeta_{l}\ll -1},W_{\zeta_{l}\ll -1})=\emptyset$, the $\zeta_{l}\ll-1$ phase in $\mathcal{T}_{X}$ (with all the other FI parameters positive) will be just composed by Coulomb vacua or, more precisely, the category $\mathcal{W}_{-}$ will be empty and only $\mathcal{W}_{+}$ will prevail. This means, there exist a GLSM equivalent to $\mathcal{T}_{X}$ where all the charges $Q_{\alpha}(h)$ are positive, hence there is no phase boundary at $(\zeta_{l},\zeta')=(0,\mathbb{R}_{\leq 0})$ in $\mathcal{T}_{\mathcal{X}}$ and $\mathcal{T}_{\mathcal{X}_{L}}$.}. Let us call the category of B-branes of this phase $\mathcal{C}'_{L}$.  More remarkably, we claim that this category is equivalent to $\mathcal{C}_{L}$, defined in section \ref{sec:section2} whenever the RG flow, determined by the direction of $-Q^{\mathrm{tot}}$ points towards the $(Q\zeta'-\zeta_{l}\gg 1,\zeta'\ll -1)$ phase. We do not have a proof of this fact, but we check this on the families of examples we analyzed by computing the Witten indices of the different phases. In such a case
\begin{eqnarray}
\mathcal{C}'_{L}\cong\mathcal{C}_{L}\hookrightarrow D(X_{L}).
\end{eqnarray}
Moreover, if the phase boundary $(\zeta_{l},\zeta')=(0,\mathbb{R}_{\leq 0})$ is lifted by quantum effects or not present at all, we have that $D(Z_{L})= \mathcal{C}_{L}=\mathcal{C}'_{L}$. On the other hand, if $-Q^{\mathrm{tot}}$ points against the $(Q\zeta'-\zeta_{l}\gg 1,\zeta'\ll -1)$ phase we have 
\begin{eqnarray}
\mathcal{C}_{L}\cong D(X_{L})\hookrightarrow \mathcal{C}'_{L}.
\end{eqnarray}
Finally, if the direction $-Q^{\mathrm{tot}}$ coincides exactly with the wall $\mathbb{R}_{\geq 0}\cdot(-Q,-1)$, we have
\begin{eqnarray}
\mathcal{C}_{L}\cong D(X_{L})\cong\mathcal{C}'_{L}.
\end{eqnarray}
\end{itemize}

In conclusion, generically the phase space will look like figure \ref{fig:phase_restricted_a}. Again, we omit a detailed analysis of the Coulomb and mixed Coulomb-Higgs vacua. Interestingly, in this case, we find that two phases of the GLSM $\mathcal{T}_{\mathcal{X}_{L}}$ can be identified with $X_{L}$ and $D(Z_{L})$ respectively (even though it may be not possible to identify the variety $Z$). This is a different situation than the case when the section $L$ was trivial and is more reminiscent of the GLSM studied in \cite{rennemo2017fundamental}. So, our approach unifies the approaches of \cite{ballard2017homological,ballard2014derived} and \cite{rennemo2017fundamental}.

We can take the strong coupling limit of $U(1)_{s+1}$, while keeping $\zeta_{s+1}\ll -1$, and get a GLSM $\mathcal{T}_{X_L}$ with gauge group $G$ that realizes $D(X_L)$ and $\mathcal{C}'_L$ in two different phases (figure \ref{fig:phase_restricted_b}). Note that $\mathcal{C}'_L$ is equivalent to $D(Z_{L})$ when $D(Y_{\zeta_{l}\ll -1},W_{\zeta_{l}\ll -1})$ in (\ref{Xcats}) is empty.



\section{\label{sec:examples}Examples}



In this section, we explicitly construct and analyze the GLSM $\mathcal{T}_{\mathcal{X}}$ realizing the universal hyperplane section and HPD of three families of embeddings, namely linear embedding of projective space, Veronese embedding and complete intersections. In each case, we will show that our proposal reproduces known results such as linear embedding, double Veronese embedding and odd dimensional quadrics, and make predictions on more general cases. In all these examples we are also concerned with the Coulomb and mixed Coulomb-Higgs branches, our convention for the computations of these vacua is
\begin{equation}\label{eqssigma}
\exp\left(\frac{\partial \mu^{-1}\widetilde{W}_{\mathrm{eff}}(\sigma')}{\partial \sigma'}\right)=\exp\left(\frac{\partial \widetilde{W}_{\mathrm{eff}}(\sigma)}{\partial \sigma}\right)=1,
\end{equation}
where $\widetilde{W}_{\mathrm{eff}}(\sigma)$ is given in (\ref{effpotsigma}) and $\sigma':=\sigma/\mu$ is dimensionless. Then (\ref{eqssigma}) implies:
\begin{equation}\label{eqssigma2}
\prod_{j=1}^{N}Q_{j}(\sigma')^{Q_{j}^{\alpha}}=e^{-(t')^{\alpha}},\qquad \alpha=1,\ldots,\mathrm{dim}(\mathfrak{t}),
\end{equation}
where
\begin{equation}
t':=t_{\mathrm{ren}}+i\frac{\pi}{2}Q^{\mathrm{tot}}.
\end{equation}
In the following we will simply make the replacement 
\begin{equation}
\sigma'\rightarrow\sigma.
\end{equation}

\subsection{Linear embedding of projective space}

Let $V$ be an $n$ dimensional vector space and $E$ an $m$ dimensional subspace. Then we have the linear embedding
\begin{equation}\label{linear_embed}
\mathbb{P}(E) \hookrightarrow \mathbb{P}(V).
\end{equation}
For a fixed basis $\{e_1,e_2,\cdots,e_n\}$ of $V$, let $X_1,\cdots,X_n$ denote the corresponding coordinates. Assume that $E$ is defined by $n-m$ linear equations:
\[
L_\alpha(x) = \sum_{j=1}^n a_{\alpha,j} x_j=0, \alpha=1,\cdots,n-m.
\]
The matrix $a_{\alpha,j}$ has rank $n-m$. The embedding \eqref{linear_embed} can be realized by a $U(1)$ GLSM with the following matter content and charges (i.e. the weights under $U(1)$ representation):
\begin{equation}\label{GLSM_linear_embed}
\begin{array}{ccccccc}
X_1 & X_2 & \cdots & X_n & P_1 & \cdots & P_{n-m} \\
1 & 1 & \cdots & 1 & -1 & \cdots & -1
\end{array}
\end{equation}
together with a superpotential:
\[
W = \sum_{\alpha=1}^{n-m} P_\alpha L_\alpha(X).
\]
When the FI parameter $\zeta \gg 1$, the low energy degrees of freedom are described by NLSM on $\mathbb{P}(E)$.
When the FI parameter $\zeta \ll -1$, there is no Higgs branch, there are $m$ distinct Coulomb vacua satisfying
\[
\sigma^m = (-1)^{n-m} q,
\]
where $q = \exp(-t')$. The number of Coulomb vacua is consistent with the semiorthogonal decomposition determined by \eqref{linear_embed}
\[
D(\mathbb{P}(E)) = \langle \mathcal{A}_0, \mathcal{A}_1(1), \cdots , \mathcal{A}_{m-1}(m-1) \rangle
\]
with $\mathcal{A}_a = \mathcal{O}$ for all $a=0,\ldots,m-1$. By integrating out the massive fields, the GLSM \eqref{GLSM_linear_embed} is the same as the GLSM realizing $\mathbb{P}(E) \cong \mathbb{P}^{m-1}$, namely the $U(1)$ GLSM with the following matter content and charges:
\begin{equation}\label{GLSM_P}
\begin{array}{cccc}
X_1 & X_2 & \cdots & X_m \\
1 & 1 & \cdots & 1
\end{array}
\end{equation}
and without a superpotential.

The homological projective duality of \eqref{linear_embed} is given by (which coincides with the classical projective dual of $\mathbb{P}(E)$)
\begin{equation}\label{linear_embed_HPD}
\mathbb{P}(E^\perp) \hookrightarrow \mathbb{P}(V^\vee).
\end{equation}
Fix a basis for $E$ and denote it by $\{v_1,\cdots,v_m\}$ with $v_a = \sum_{j=1}^n b_{a,j} e_j$, where $b_{a,j}$ has full rank and $\sum_{j}a_{\alpha,j}b_{a,j}=0$ for all $a,\alpha$. Then,
as discussed in section \ref{sec:proposal}, to get the GLSM $\mathcal{T}_{\mathcal{X}}$ describing the HPD of the linear embedding \eqref{linear_embed}, we first extend the GLSM \eqref{GLSM_P} to a GLSM with gauge group $U(1) \times U(1)$ and the following matter content and charges:
\begin{equation}\label{firstuniversal}
(X_{i},P_{\alpha},P,Y_{j})\in \mathbb{C}(1,0)^{\oplus n}\oplus\mathbb{C}(-1,0)^{\oplus n-m}\oplus\mathbb{C}(-1,-1)\oplus \mathbb{C}(0,1)^{\oplus n}
\end{equation}
where $\mathbb{C}(r,s)$ denotes an irreducible representation of $U(1)\times U(1)$ of weights $(r,s)\in \mathbb{Z}^{2}$.
The superpotential is given by
\[
\widehat{W} =\sum_{j,\alpha} P_{\alpha}a_{\alpha,j}X_j+ P\sum_{j}X_{j} Y_j.
\]
The fields $Y_i$ serve as homogeneous coordinates of $\mathbb{P}(V^\vee)$. Is easy to see that the phases $(\zeta_{2}-\zeta_{1} \gg 1,\zeta_{2}\ll -1)$ and $(\zeta_{1}\ll -1,\zeta_{2}\gg 1)$, in the gauge decoupling limit, have the same Higgs branch, namely, we can set $P=1$, $P_{\alpha}\in \mathbb{C}^{n-m}\setminus \{ 0 \}$ and the $Y_{j}$ are constrained by the equation:
\begin{equation}
Y_{j}+\sum_{\alpha}P_{\alpha}a_{\alpha,j}=0.
\end{equation}
Then we identify this Higgs branch with $\mathbb{P}(E^{\perp})\hookrightarrow \mathbb{P}(V^{\vee})$, where $E^{\perp}$ is spanned by the vectors $a_{\alpha}$. Therefore the classical boundary wall $(\zeta_{1},\zeta_{2})\in (0,\mathbb{R}_{\leq 0})$ is lifted. The phase $(\zeta_{2}-\zeta_{1} \ll -1,\zeta_{2}\ll -1)$ has an empty Higgs branch and the local model at the wall $\mathbb{R}_{\leq 0}(-1,-1)$ is a $U(1)$ GLSM with matter content:
\begin{equation}\label{GLSM_linear_embed_HPD}
\begin{array}{ccccccccc}
Y_1 & \cdots & Y_n & X_1 & \cdots & X_n & P_{1} & \cdots & P_{n-m} \\
-1 & \cdots & -1 & 1 & \cdots & 1 & 1 &\cdots & 1
\end{array}
\end{equation}
From here we can read the number of Coulomb vacua, given by $n-m$, consistent with the  semiorthogonal decomposition induced by the embedding \eqref{linear_embed_HPD}, namely
\begin{equation}\label{sodperp}
D(\mathbb{P}(E^\perp)) = \langle \mathcal{B}_{n-m-1}(1+m-n), \cdots , \mathcal{B}_{1}(-1), \mathcal{B}_0 \rangle
\end{equation}
with $\mathcal{B}_i = \mathcal{O}$. We remark also, that $\mathbb{P}(E^\perp)$ is the correct HPD induced by the Lefschetz decomposition that can be computed using the prescription (\ref{lefscenter}). Namely, the local model at the phase boundary $\mathbb{R}_{\geq 0}(0,1)$ gives $N_{+}-N_{-}=m-1$ which implies $\mathcal{A}_{0}$ on the Lefschetz decomposition of $X=\mathbb{P}(E)$ has only one element: $\mathcal{A}_{0}=\mathcal{O}$. This plus the line bundle induced by \eqref{linear_embed} reproduces the expected Lefschetz decomposition.
In order to provide further consistency of our proposal in this simple model, we notice that we could have started from the GLSM \eqref{GLSM_P} as $\mathcal{T}_{X}$ and then, $\mathcal{T}_{\mathcal{X}}$ will be given by the GLSM with matter content
\begin{equation}\label{GLSM_linear_embed_universal}
\begin{array}{ccccccccc}
X_1  & \cdots & X_m & P & Y_1 & Y_2 & \cdots & Y_n \\
1 &  \cdots & 1 & -1 & 0 & 0 & \cdots & 0 \\
0 &  \cdots & 0 & -1 & 1 & 1 & \cdots & 1
\end{array}
\end{equation}
with superpotential
\[
\widehat{W} = P\sum_{a,j} X_{a} b_{a,j} Y_j.
\]
It is easy to show that the phase space of this model coincides with \eqref{firstuniversal}. In this case $\mathbb{E^{\perp}}$ is described as $\sum_{j} b_{a,j} Y_j = 0$. Taking the gauge decoupling limit in the region $\zeta_{2}\ll -1$ in \eqref{GLSM_linear_embed_universal} gives the model
\begin{equation}\label{GLSM_linear_embed_HPD2}
\begin{array}{ccccccccc}
X_1  & \cdots & X_m & Y_1 & \cdots & Y_n \\
1 &  \cdots & 1  & -1 & \cdots & -1
\end{array}
\end{equation}
with superpotential
\[
\widehat{W}' = \sum_{a,j} X_{a} b_{a,j} Y_j.
\]
Again we can see that when the FI parameter of \eqref{GLSM_linear_embed_HPD2} $\zeta \gg 1$, there is no Higgs branch, there are $n-m$ Coulomb vacua satisfying
\[
e^{-t} \sigma^{n-m} = (-1)^n,
\]
which we identify with the semiorthogonal decomposition \eqref{sodperp}.

\subsection{Veronese embedding}

\begin{figure}\label{fig:Veronese}
	\centering
	\begin{tikzpicture}
	\draw[thin,->] (0,0) -- (4,0) node[anchor=north west]{$\zeta_1$};
	\draw[thin,->] (0,0) -- (0,4)node[anchor=south east]{$\zeta_2$};
	\draw [ultra thick] (0,0) -- (0,3.5);
	\draw [ultra thick] (0,0) -- (3.5,0);
	\draw [ultra thick] (0,0) -- (-4,-2);
	\node at (-5.1,-2.2) {\footnotesize $d \zeta_2 - \zeta_1 = 0$};
	\draw [dashed, thick] (0,0) -- (-4, 1.2);
	\node [above] at (-5.75,1) {\footnotesize ${\zeta_1 + (n+1-d) \zeta_2 = 0}$};
	\draw [dashed, thick] (0,0) -- (1, -3.5) ;
	\node [below] at (1,-3.5) {\footnotesize $(N-1) \zeta_1 + d \zeta_2 =0$};
	\node at (3,2.5) {\small Geometric phase};
	\node at (-2.5,3)  {\small LG phase I };
	\node at (-3.75,-0.25) {\small LG phase II};
	\node at (-1.5, -2.75) {\small Coulomb phase};
	\node at (3, -2) {\small Mixed phase};
\end{tikzpicture}
	\quad \quad \quad \caption{Phase diagram of GLSM for HPD of degree $d$ Veroness embedding of $\mathbb{P}^n$.}\label{fig:phase_Ver}
	{\footnotesize \begin{flushleft} Here $N = {n+d \choose d}$. The geometric phase describes the universal hypersurface. LG phase I has a Higgs branch described by the LG model realizing the HPD, and $(n+1-d)$ mixed branches, each described by $\mathbb{P}^{N-1}$. LG phase II has the same Higgs branch as LG phase I but with $((n+1-d)N)$ Coulomb vacua. The mixed phase contains $(N-1)$ mixed branches, each described by $\mathbb{P}^n$. The Coulomb phase contains $((N-1)(n+1))$ Coulomb vacua. \end{flushleft}}
\end{figure}
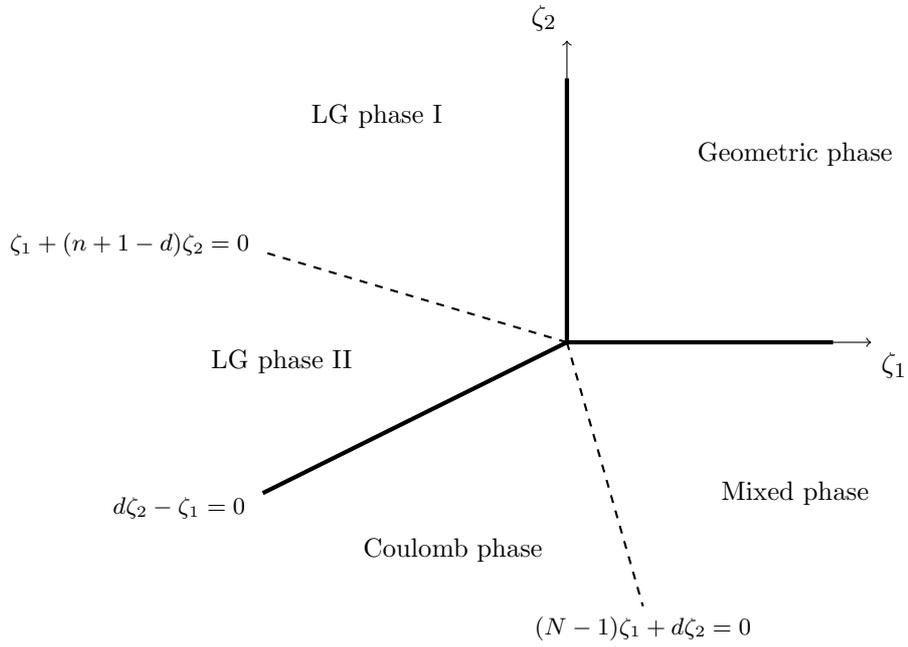

The degree-$d$ Veronese embedding $\mathbb{P}(V) \rightarrow \mathbb{P}(\mathrm{Sym}^d V)$ can be realized by abelian GLSM as discussed in \cite{Caldararu:2017usq}. Upon integrating out the massive fields, this theory is equivalent to the GLSM describing $\mathbb{P}(V)$, namely $n+1$ matter fields with charge 1 under the $U(1)$ gauge symmetry, where $\dim(V) = n+1$. Since $\dim(\mathrm{Sym}^d V) = {n+d \choose d}$, our discussion in section \ref{sec:proposal} shows that the GLSM $\mathcal{T}_{\mathcal{X}}$ describing the universal hyperplane section and HPD has gauge group $U(1) \times U(1)$ with matter content:
\begin{equation}\label{GLSM_Veronese_universal}
\begin{array}{ccccccccc}
X_0 & X_1 & \cdots & X_n & P & S_1 & S_2 & \cdots & S_{n+d \choose d} \\
1 & 1 & \cdots & 1 & -d & 0 & 0 & \cdots & 0 \\
0 & 0 & \cdots & 0 & -1 & 1 & 1 & \cdots & 1
\end{array}
\end{equation}
and superpotential
\[
W = P \sum_{a=1}^{n+d \choose d} S_a f_a(X),
\]
where $f_a(X)$ form a basis of the monomials in $X_i$ with degree $d$. Higgs branch of one of the phases gives the HPD of degree-$d$ Veronese embedding.
From the F-term and D-term constraints, one can show that the Higgs branches in different phases are as follows:\\
(i)~$\zeta_1>0, \zeta_2>0$: The Higgs branch is the universal hyperplane section of the Veronese embedding, i.e. the universal degree-$d$ hypersurface
\[
\mathcal{X} = \{ \sum_a S_a f_a(X) = 0 \} \subseteq \mathbb{P}^n \times \mathbb{P}^{{n+d \choose d}-1},
\]
where $X_i$ are homogeneous coordinates of $\mathbb{P}^n$ and $S_a$ are homogeneous coordinates of $\mathbb{P}^{{n+d \choose d}-1}$.
(ii)~$\zeta_1<0, \zeta_1<d \zeta_2$: The Higgs branch is the LG model on (the notation $\mathcal{O}(m)$ with $m\in\mathbb{Q}$ is explained in appendix \ref{app:MFgerbe})
\begin{equation}\label{LGVspace}
\mathrm{Tot}\left( \mathcal{O}\left( -\frac{1}{d} \right)^{\oplus(n+1)} \rightarrow \mathbb{P}^{{n+d \choose d}-1} \right)/\mathbb{Z}_d
\end{equation}
with superpotential
\begin{equation}\label{LGVpotential}
W_{0}=\sqrt{\frac{-\zeta_1}{d}} \sum_a S_a f_a(X).
\end{equation}
(iii)~$\zeta_1>d \zeta_2, \zeta_2<0$: There is no Higgs branch.\\
The classical phase diagram of GLSM \eqref{GLSM_Veronese_universal} is shown in Figure \ref{fig:phase_Ver}. The Coulomb/mixed branches can be determined by studying the local model associated with the phase boundaries and by analyzing the asymptotic behavior of the equations of motion on the Coulomb branch:
\[
\left\{ \begin{array}{l}
\sigma_1^{n+1} = q_1 (-d \sigma_1 - \sigma_2)^d, \\
\sigma_2^{{n+d \choose d}} = -q_2 (d \sigma_1+ \sigma_2),
\end{array} \right.
\]
where $q_a = \exp(-t'_a)$.
We provide the details of this analysis in appendix \ref{app:CoulombEOM}.
To determine which phase of the GLSM \eqref{GLSM_Veronese_universal} describes the HPD, let's consider the local model at the boundary between phase (i) and (ii) above. It is a $U(1)$ GLSM with matter content
\[
\begin{array}{ccccc}
X_0 & X_1 & \cdots & X_n & P \\
1 & 1 & \cdots & 1 & -d
\end{array}
\]
Thus we have $N_+ = n+1, N_- = d$. The small window consists of the branes satisfying
\[
\left|q+\frac{\theta}{2 \pi}\right| < \frac{1}{2} \mathrm{min} \{ d,n+1 \}.
\]
We can choose $\theta$ such that the charge $q$ of the branes in the small window satisfy
\[
0 \leqslant q \leqslant \mathrm{min} \{ d-1,n \}.
\]
For $d \leq n+1$, the corresponding Lefschetz decomposition is
\begin{equation}\label{Lef_Veronese}
D(\mathbb{P}^n) = \langle \mathcal{A}_0, \mathcal{A}_1(d), \cdots, \mathcal{A}_p(pd) \rangle,
\end{equation}
where $\mathcal{A}_0 = \mathcal{A}_1 = \cdots = \mathcal{A}_{p-1} = \langle \mathcal{O},\cdots,\mathcal{O}(d-1) \rangle$, $\mathcal{A}_p = \langle \mathcal{O},\cdots,\mathcal{O}(k) \rangle$, and $p \ge 0, 0 \le k < d$ with $n = pd+k$.
Branes in the small window have charges $q=0,1,\cdots,d-1$. These branes project to the HPD category $\mathcal{C}$ in phase (i) because they restrict to $\mathcal{A}_0$ on $\mathbb{P}^n$. From \cite{Clingempeel:2018iub}, we know that this subcategory $\mathcal{C}$ of $D(\mathcal{X})$, which is the HPD category by definition, is equivalent to the category of matrix factorizations of the LG model corresponding to LG phase I in figure \ref{fig:phase_Ver}. Therefore, we claim that the HPD of degree-$d$ Veronese embedding with respect to the Lefschetz decomposition \eqref{Lef_Veronese} is described by the LG orbifold defined by \eqref{LGVspace} and \eqref{LGVpotential}. By taking the strong coupling limit of the second $U(1)$ gauge group in \eqref{GLSM_Veronese_universal}, the HPD can also be realized by the negative phase of the $U(1)$ GLSM with matter content
\[
\begin{array}{ccccccccc}
X_0 & X_1 & \cdots & X_n & S_1 & S_2 & \cdots & S_{n+d \choose d} \\
1 & 1 & \cdots & 1 & -d & -d & \cdots & -d
\end{array}
\]
and superpotential
\[
\sum_{a=1}^{n+d \choose d} S_a f_a(X).
\]

If $d \geqslant n+1$, branes in the small window have charge $q=0,1,\cdots,n$. These branes generate $D(\mathcal{X})$ in phase (i). In this case $D(\mathcal{X}) = \mathcal{C}$ because $D(\mathbb{P}^n) = \langle \mathcal{O},\mathcal{O}(1),\cdots, \mathcal{O}(n)\rangle = \mathcal{A}_0$. Therefore, the HPD category is equivalent to $D(\mathcal{X})$.

In the special case $d = n+1$, the HPD can be described by $D(\mathcal{X})$ or the matrix factorizations of the LG orbifold \eqref{LGVspace}\eqref{LGVpotential} because these two categories are equivalent.\\

\textbf{Double Veronese Embedding}. Now let's study the GLSM more carefully in the case $d=2$, we will see that the HPD we obtained matches the mathematical result in \cite{kuznetsov2008derived}. The HPD of double Veronese embedding is the Clifford space $Y = (\mathbb{P}(\mathrm{Sym}^2 V^\vee), \mathcal{C}l_0)$ \cite{kuznetsov2008derived}, where $\mathcal{C}l_0$ denotes coherent sheaves of modules over the even part of the universal Clifford algebra determined by the corresponding quadratic forms in $\mathrm{Sym}^2 V^\vee$. Assume that $\dim V=n+1$, then $\dim \mathrm{Sym}^2 V = (n+1)(n+2)/2$. As discussed above, the HPD $Y$ can be realized by a $U(1)$ GLSM with the following matter content and charges:
\begin{equation}\label{oneHPD2V}
\begin{array}{ccccccc}
X_0 & X_1 & \cdots & X_n & S_1 & \cdots & S_{(n+1)(n+2)/2} \\
1 & 1 & \cdots & 1 & -2 & \cdots & -2
\end{array}
\end{equation}
and a superpotential $W = \sum_{i=1}^{(n+1)(n+2)/2} S_i G_i(X)$, where $G_i$'s form a complete set of quadratic monomials in $X_i$.
When the FI parameter $\zeta \gg 1$, there is no Higgs branch and there are $(n+1)^2$ Coulomb vacua satisfying
\[
2^{(n+1)(n+2)} \sigma^{(n+1)^2} q = 1,
\]
which correspond to the semiorthogonal decomposition
\[
D(\mathbb{P}(\mathrm{Sym}^2 V^\vee), \mathcal{C}l_0) = \langle \mathcal{C}l_{1-(n+1)^2}, \mathcal{C}l_{2-(n+1)^2} \cdots , \mathcal{C}l_{-1}, \mathcal{C}l_0 \rangle,
\]
where $\mathcal{C}l_1$ is the odd part of the Clifford algebra and $\mathcal{C}l_{k}$ for $k\leq 1$ are defined recursively, $\mathcal{C}l_{k-2} = \mathcal{C}l_k \otimes \mathcal{O}_{\mathbb{P}(\mathrm{Sym}^2 V^\vee)}(-1)$.

When the FI parameter $\zeta \ll -1$, the low energy theory is a hybrid model on
\[[\mathrm{Tot}(\mathcal{O}(-1/2)^{\oplus (n+1)} \rightarrow \mathbb{P}(\mathrm{Sym}^2 V^\vee))/\mathbb{Z}_2],\]
where $X_i$'s are fiber coordinates, $S_a$'s are base coordinates.
At each point $[S_1, \cdots, S_{(n+1)(n+2)/2}]$ on $\mathbb{P}(\mathrm{Sym}^2 V^\vee)$, there is a superpotential
\[ W_0 = \sum_{i=1}^{(n+1)(n+2)/2} S_i G_i(X)\] quadratic in $X_i$. Because matrix factorizations of $\mathbb{Z}_2$-orbifold of LG model with quadratic superpotential is equivalent to modules of even part of Clifford algebras \cite{Kapustin:2002bi}, the category of matrix factorizations of the above hybrid model is equivalent to the HPD category $D(\mathbb{P}(\mathrm{Sym}^2 V^\vee), \mathcal{C}l_0)$. Thus our construction recovers the mathematical result in \cite{kuznetsov2008derived}.\\

In general, for the case $d \leq n+1$ we expect that $\mathcal{C}$ has a semiorthgonal decomposition with center $\mathcal{B}_{0}\cong\mathcal{A}_{0}$ and ${n+d \choose d}-1-p$ objects, fitting the diagram
\begin{figure}[!h]
 \centering
\begin{tikzpicture}[inner sep=0in,outer sep=0in]
\node (n) {
\centering
    \begin{tabular}{r@{}l}
    \raisebox{-9.5ex}{$d\left\{\vphantom{\begin{array}{c}~\\[10ex] ~
    \end{array}}\right.$} &
    \begin{ytableau}
    \none[{\cal A}_0] & \none[{\cal A}_1]  & \none & \none[\cdots] &\none& \none[{\cal A}_{\scaleto{p-1\mathstrut}{7pt}}] & \none[{\cal A}_p]        & \none \\
    *(gray)           & *(gray)                  & \none & \none        &\none & *(gray) &  *(gray)    &       &  \none & \none         & \none &  &  \\
    *(gray)           & *(gray)                  & \none & \none         &\none& *(gray) &  *(gray)    &       &  \none  & \none         & \none &  &  \\
    *(gray)           & *(gray)                  & \none & \none [\cdots]&\none& *(gray) &  *(gray)    &       &  \none & \none[\cdots] & \none &  &  \\
    *(gray)           & *(gray)                  & \none & \none         &\none& *(gray) &             &       &  \none & \none         & \none &  &  \\
    *(gray)           & *(gray)                  & \none & \none         &\none& *(gray) &             &       &  \none & \none         & \none &  &  \\
    \none             & \none                    & \none & \none         &\none& \none   &  \none[{\cal B}_{\scaleto{l\mathstrut}{7pt}}]  & \none[{\cal B}_{\scaleto{l-1\mathstrut}{7pt}}]&\none &\none[\cdots] &\none &\none[{\cal B}_1] &\none[{\cal B}_0]\\
    \end{ytableau}
    \raisebox{-5.7ex}{$\left\}\vphantom{\begin{array}{c}~\\[5ex] ~
    \end{array}}\right.k+1$}    \end{tabular}};
\draw[thin,] (3,1.48) -- (-3.3,1.48);
\draw[thin,] (3,-1.48) -- (-3.3,-1.48);
\end{tikzpicture}
    \caption{HPD of degree-$d$ Veronese embedding}
    {\footnotesize \begin{flushleft} The gray part of the diagram corresponds to the Lefschetz decomposition of degree-$d$ Veronese embedding. The white part corresponds to Lefschetz decomposition of the dual space. The number of different components for ${\cal A}_i$ and ${\cal B}_a$ are counted by the corresponding number of rows. The number of ${\cal B}_a$'s is given by $l= {n+d \choose d} - p -1$, where $n=pd+k$.\end{flushleft}}
\end{figure}
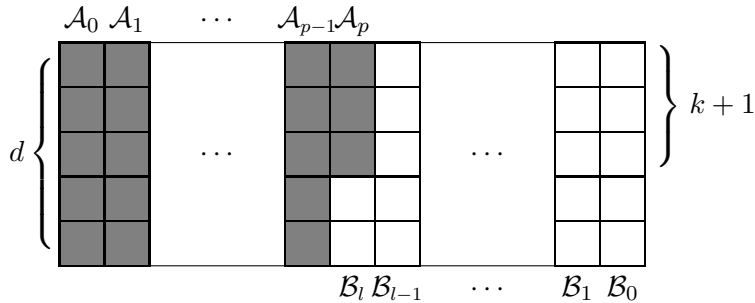

\subsubsection{B-brane transport}

Let us now consider the situation $d \leq n+1$ and denote the LG model \eqref{LGVspace}\eqref{LGVpotential} by $\mathrm{LG}_{\mathrm{HPD}}$. We want to realize the equivalence between $MF(\mathrm{LG}_{\mathrm{HPD}})$ and the HPD category $\mathcal{C} \subset D(\mathcal{X})$ through brane transport.

Remember that there is a semiorthogonal decomposition
\[
D(\mathcal{X}) = \langle \mathcal{C}, \mathcal{A}_1(1)_{\mathbb{P}(\mathrm{Sym}^2 V^\vee)},\cdots,\mathcal{A}_{m-1}(m-1)_{\mathbb{P}(\mathrm{Sym}^2 V^\vee)} \rangle,
\]
where the non-trivial component $\mathcal{C}$ is the HPD category by definition. When considering the wall crossing between the geometric phase and LG phase I in figure \ref{fig:phase_Ver}, one should expect that, under brane transport, $\mathcal{C}$ is transported to $MF(\mathrm{LG}_{\mathrm{HPD}})$ in LG phase I while the image of $\mathcal{A}_i(i)_{\mathbb{P}(\mathrm{Sym}^d V^\vee)}$ have components on Coulomb vacua or mixed branches. When transported through the small window, $MF(\mathrm{LG}_{\mathrm{HPD}})$ should have images in $\mathcal{C}$.

Explicitly, we take a set of generators of $MF(\mathrm{LG}_{\mathrm{HPD}})$ and lift them to GLSM matrix factorizations in the small window and transport them to the geometric phase, we check that the projections of these matrix factorizations are all equivalent to objects in $\mathcal{A}_0 \boxtimes D(\mathbb{P}(\mathrm{Sym}^d V^\vee))$. Objects in the HPD category $\mathcal{C}$ are precisely the objects in $D(\mathcal{X})$ that can be lifted to objects in the small window. This procedure gives a set of generators of $\mathcal{C}$. Now let us see how the brane transport works.

\textbf{Example:} $\mathbb{P}^2 \rightarrow \mathbb{P}^5$. The small window consists of branes with charges $-1, 0$, the big window consists of branes with charges $-1,0,1$. In the LG phase, non-empty branes are of the form
\begin{equation}\label{MFLG}
\xymatrix{
\mathcal{O}(m+\frac{3}{2}) \ar@<0.5ex>[r]^-{X \bar{\eta}} & \mathcal{O}(m+1)^{\oplus 3} \ar@<0.5ex>[l]^-{\frac{\partial W_{0}}{\partial X}\eta} \ar@<0.5ex>[r]^-{X\bar{\eta}} & \mathcal{O}(m+\frac{1}{2})^{\oplus 3} \ar@<0.5ex>[l]^-{\frac{\partial W_{0}}{\partial X}\eta} \ar@<0.5ex>[r]^-{X \bar{\eta}} & \mathcal{O}(m) \ar@<0.5ex>[l]^-{\frac{\partial W_{0}}{\partial X}\eta}}.
\end{equation}
If $m = n - \frac{1}{2}$ is a half-integer, then \eqref{MFLG} can be lifted to
\begin{equation}\label{liftMFLGG}
\xymatrix{
\mathfrak{W}(-2,n)_{-2} \ar@<0.5ex>[r]^-{X \bar{\eta}} & \mathfrak{W}(-1,n)_{-1}^{\oplus 3} \ar@<0.5ex>[l]^-{\frac{\partial W}{\partial X}\eta} \ar@<0.5ex>[r]^-{X\bar{\eta}} & \mathfrak{W}(0,n)_0^{\oplus 3} \ar@<0.5ex>[l]^-{\frac{\partial W}{\partial X}\eta} \ar@<0.5ex>[r]^-{X \bar{\eta}} & \mathfrak{W}(1,n)_1 \ar@<0.5ex>[l]^-{\frac{\partial W}{\partial X}\eta}}
\end{equation}
with $\theta = B_* + 2 \pi$. Here, $ \mathfrak{W}(a,b)_r$ denotes the so-called \emph{Wilson line brane} defined in \cite{Herbst:2008jq} which determines the boundary interactions. It corresponds to an irreducible representation of weights $(a,b)\in\mathbb{Z}^{2}$ under the gauge group $U(1) \times U(1)$ and R-charge $r$. The fermions $\eta,\bar{\eta}$ in the maps correspond to a representation of the Clifford algebra satisfying $\{\eta_{i},\bar{\eta}_{j}\}=\mathbf{1}\delta_{i,j}$ and all the other anticommutators vanish. In other words, the matrix factorization associated with \ref{liftMFLGG} can be written as
\begin{equation}
\mathbf{T}=X_{i}\bar{\eta}_{i}+\frac{1}{2}\frac{\partial W}{\partial X_{i}}\eta_{i}
\end{equation}
By binding with the empty branes
\[
\xymatrix{
\mathfrak{W}(1,n)_0 \ar@<0.5ex>[r]^-P & \mathfrak{W}(-1,n-1)_{-1} \ar@<0.5ex>[l]^-{G}
}
\]
and
\[
\xymatrix{
\mathfrak{W}(0,n+1)_0 \ar@<0.5ex>[r]^-P & \mathfrak{W}(-2,n)_{-1} \ar@<0.5ex>[l]^-{G},
}
\]
we get
\begin{equation}\label{MFodd}
\xymatrix{
\mathfrak{W}(0,n+1)_{0} \ar@<0.5ex>[r]^-{P X \bar{\eta}} & \mathfrak{W}(-1,n)_{-1}^{\oplus 3} \ar@<0.5ex>[l]^-{\frac{\partial G}{\partial X}\eta} \ar@<0.5ex>[r]^-{X \bar{\eta}} & \mathfrak{W}(0,n)_0^{\oplus 3} \ar@<0.5ex>[l]^-{P \frac{\partial G}{\partial X}\eta} \ar@<0.5ex>[r]^-{P X \bar{\eta}} & \mathfrak{W}(-1,n-1)_{-1} \ar@<0.5ex>[l]^-{\frac{\partial G}{\partial X}\eta}},
\end{equation}
which is in the small window. Denote by $\delta$ the embedding of $\mathcal{X}$ in $\mathbb{P}^2 \times \mathbb{P}^5$.
Assume that the projection of \eqref{MFodd} in the geometric phase is $B_{-}(n)$, then by taking $P=0$, it is easy to see
\[
\delta_*(B_{-}(n)) =
{\xymatrix{\mathcal{O}(-1,n)^{\oplus 3}_{-2} \ar@<0.5ex>[rr]^{\frac{\partial G}{\partial X}} \ar@<0.5ex>[drdr]^{X} & & \mathcal{O}(0,n+1)_{-1} \\ \oplus & & \oplus \\
\mathcal{O}(-1,n-1)_{-2} \ar@<0.5ex>[rr]_{\frac{\partial G}{\partial X}} & & \mathcal{O}(0,n)^{\oplus 3}_{-1} }}
\]
with $B=B_*$.
Thus $\delta_*(B_{-}(n)) \in \mathcal{A}_0 \boxtimes D(\mathbb{P}^5)$, and therefore $B_{-}(n) \in \mathcal{C} \subseteq D(\mathcal{X})$.
From the construction, it is easy to see that $B_-(n)$ is equivalent to
\[
\mathcal{O}_\mathcal{X}(-2,n)[1] \oplus \mathcal{O}_\mathcal{X}(-1,n-1)[1]
\]
with $2 \pi B = \theta_1 H_1 + \theta_2 H_2 + \pi (2 H_1+H_2)$, which can be checked by the hemisphere partition function.
Equivalently, $B_-(n)$ is
\[
\mathcal{O}_\mathcal{X}(0,n+1)[1] \oplus \mathcal{O}_\mathcal{X}(1,n)[1]
\]
with $2 \pi B = \theta_1 H_1 + \theta_2 H_2$.

On the other hand, if $m = n$ is an integer in \eqref{MFLG}, then it can be lifted to
\[
\xymatrix{
\mathfrak{W}(-3,n)_{-2} \ar@<0.5ex>[r]^{X \bar{\eta}} & \mathfrak{W}(-2,n)_{-1}^{\oplus 3} \ar@<0.5ex>[l]^{\frac{\partial W}{\partial X}\eta} \ar@<0.5ex>[r]^{X\bar{\eta}} & \mathfrak{W}(-1,n)_0^{\oplus 3} \ar@<0.5ex>[l]^{\frac{\partial W}{\partial X}\eta} \ar@<0.5ex>[r]^{X \bar{\eta}} & \mathfrak{W}(0,n)_1 \ar@<0.5ex>[l]^{\frac{\partial W}{\partial X}\eta}}
\]
with $\theta = B_* + 2 \pi$.
By binding with the cone of the following morphism
\[  \xymatrix{ \mathfrak{W}(-1,n+1)_0 \ar@<0.5ex>[r]^-P \ar@<0.5ex>[d]^-X & \mathfrak{W}(-3,n)_{-1} \ar@<0.5ex>[l]^-{G} \ar@<0.5ex>[d]^-X\\
\mathfrak{W}(0,n+1)^{\oplus 3}_1 \ar@<0.5ex>[r]^-P & \mathfrak{W}(-2,n)^{\oplus 3}_{0} \ar@<0.5ex>[l]^-{G}
}
\]
which is an empty brane,
we get
\begin{equation}\label{MFeven}
\xymatrix{
\mathfrak{W}(-1,n+1)_{0} \ar@<0.5ex>[r]^{X \bar{\eta}} & \mathfrak{W}(0,n+1)_{1}^{\oplus 3} \ar@<0.5ex>[l]^{P \frac{\partial G}{\partial X}\eta} \ar@<0.5ex>[r]^{P X \bar{\eta}} & \mathfrak{W}(-1,n)_0^{\oplus 3} \ar@<0.5ex>[l]^{\frac{\partial G}{\partial X}\eta} \ar@<0.5ex>[r]^{X \bar{\eta}} & \mathfrak{W}(0,n)_{1} \ar@<0.5ex>[l]^{P \frac{\partial G}{\partial X}\eta}},
\end{equation}
which is in the small window.
Assume that the projection of \eqref{MFeven} in the geometric phase is $B_{+}(n)$, then by taking $P=0$, it is easy to see
\[
\delta_*(B_{+}(n)) =
{\xymatrix{\mathcal{O}(-1,n)^{\oplus 3}_{-1} \ar@<0.5ex>[rr]^{X} \ar@<0.5ex>[drdr]^{\frac{\partial G}{\partial X}} & & \mathcal{O}(0,n)_{0} \\ \oplus & & \oplus \\
\mathcal{O}(-1,n+1)_{-1} \ar@<0.5ex>[rr]_{X} & & \mathcal{O}(0,n+1)^{\oplus 3}_{0} }}
\]
with $B = B_*$.
Thus $\delta_*(B_{+}(n)) \in \mathcal{A}_0 \boxtimes D(\mathbb{P}^5)$, and therefore $B_{+}(n) \in \mathcal{C} \subseteq D(\mathcal{X})$ as expected. Again from the construction, it is easy to see that $B_+(n)$ is equivalent to
\[
\mathcal{O}_{\mathcal{X}}(-3,n) \stackrel{X}{\rightarrow} \mathcal{O}_{\mathcal{X}}(-2,n)^{\oplus 3}
\]
with $2 \pi B = \theta_1 H_1 + \theta_2 H_2 + \pi (2 H_1+H_2)$, which can be checked by the hemisphere partition function.
Equivalently, $B_+(n)$ is
\[
\mathcal{O}_{\mathcal{X}}(-1,n+1) \stackrel{X}{\rightarrow} \mathcal{O}_{\mathcal{X}}(0,n+1)^{\oplus 3}
\]
with $2 \pi B = \theta_1 H_1 + \theta_2 H_2$.

Now let us consider the degree-$d$ Veronese embedding of $\mathbb{P}^n$. Assume that $n+1 = dk +n'$.
The GLSM branes
\begin{equation}
{\cal B}(m):
	\xymatrix{
	\mathfrak{W}(m,l)_0 \ar@<0.5ex>[r]^-{X \bar{\eta}} & \mathfrak{W}(m+1,l)_1^{\oplus (n+1)} \ar@<0.5ex>[l]^-{\frac{\partial W}{\partial X}\eta} \ar@<0.5ex>[r]^-{X \bar{\eta}} & \cdots \ar@<0.5ex>[l]^-{\frac{\partial W}{\partial X}\eta} \ar@<0.5ex>[r]^-{X \bar{\eta}} & \mathfrak{W}(m+n+1,l)_{n+1} \ar@<0.5ex>[l]^-{\frac{\partial W}{\partial X}\eta},
	} \label{cpx-d}
\end{equation}
with $m= -(n+1), -n, \cdots,  -(n+1)+d-1$, are lifts of nonempty branes
\begin{equation}
	\xymatrix{
	{\cal O}(\frac{m}{d}) \ar@<0.5ex>[r]^-{X \bar{\eta}} & {\cal O}(\frac{m+1}{d})^{\oplus (n+1)} \ar@<0.5ex>[l]^-{\frac{\partial W_{0}}{\partial X}\eta} \ar@<0.5ex>[r]^-{X \bar{\eta}} & \cdots \ar@<0.5ex>[l]^-{\frac{\partial W_{0}}{\partial X}\eta} \ar@<0.5ex>[r]^-{X \bar{\eta}} & {\cal O}(\frac{m+n+1}{d}) \ar@<0.5ex>[l]^-{\frac{\partial W_{0}}{\partial X}\eta},
	}
\end{equation}
in the LG phase I,
where ${\cal O}(\frac{i}{d})$ represents sheaves on the $\mathbb{Z}_d$-gerbe over $\mathbb{P}^{\begin{psmallmatrix} n+d \\ d \end{psmallmatrix}-1}$.

First, we need to put the branes (\ref{cpx-d}) into the small window, which means the gauge charges under the first $U(1)$ gauge group of the Wilson line branes in the complex should belong to $ \{ -d+1, \cdots,1, 0 \}$. The procedure mainly involves two kinds of cone constructions with the empty brane
\begin{equation}
	\xymatrix{
	\mathfrak{W}(m, l)_0 \ar@<0.5ex>[r]^-{P} & \mathfrak{W}(m-d,l-1)_{-1}\ar@<0.5ex>[l]^-{G}.
	}
\end{equation}in LG phase I.
By taking the cone
\begin{equation}
 \begin{split}
	\xymatrix{
	 \mathfrak{W}(m,l)_r \ar@<0.3ex>[r]^-{X \bar{\eta}} \ar@<0.3ex>[dr]^-{\mathrm{Id}} & \mathfrak{W}(m+1,l)_{r+1} \ar@<0.3ex>[l] \ar@<0.3ex>[r]^-{X \bar{\eta}}& \cdots \ar@<0.3ex>[l]\\
	 \mathfrak{W}(m+d,l+1)_{r+2} \ar@<0.3ex>[r]^-{P}& 	 \mathfrak{W}(m,l)_{r+1}  \ar@<0.3ex>[l]^-{G}
	}	 	
 \end{split},\label{cone-1}
\end{equation}
the gauge charge of the first element on the first row is raised to $(m+d,l+1)$ and the new complex is given by
\begin{equation}
	\xymatrix{
	 \mathfrak{W}(m+d,l+1)_{r+2} \ar@<0.3ex>[r]^-{PX \bar{\eta}}  & \mathfrak{W}(m+1,l)_{r+1} \ar@<0.3ex>[l] \ar@<0.3ex>[r]^-{X \bar{\eta}}& \cdots \ar@<0.3ex>[l]
	}.
\end{equation}
Notice that $R-$charge is also raised by two and the morphisms running to the left become $PX\bar{\eta}$ and $\frac{\partial G}{\partial X}\eta$. Subsequently, by taking the cone
\begingroup
\footnotesize
\begin{equation}
 \begin{split}
	\xymatrix{
	\cdots\ar@<0.3ex>[r]^-{X \bar{\eta}}  &\mathfrak{W}(m+d-1,l+1)_{r+1} \ar@<0.3ex>[l]  \ar@<0.3ex>[r]^-{PX \bar{\eta}} \ar@<0.3ex>[dr]^-{X \bar{\eta}} & \mathfrak{W}(m,l)_r \ar@<0.3ex>[l]  \ar@<0.3ex>[r]^-{X \bar{\eta}} \ar@<0.3ex>[dr]^-{\mathrm{Id}} & \mathfrak{W}(m+1,l)_{r+1} \ar@<0.3ex>[l] \ar@<0.3ex>[r]^-{X \bar{\eta}}& \cdots \ar@<0.3ex>[l]\\
	 & & \mathfrak{W}(m+d,l+1)_{r+2} \ar@<0.3ex>[r]^-{P}& 	 \mathfrak{W}(m,l)_{r+1}  \ar@<0.3ex>[l]
	}	 	
 \end{split},  \label{cone-2}
\end{equation}
\endgroup
one can raise the gauge charges of elements in the middle of a complex. The resulting complex is
\begingroup
\footnotesize
\begin{equation}
	\xymatrix{
	\cdots\ar@<0.3ex>[r]^-{X \bar{\eta}}  &\mathfrak{W}(m+d-1,l+1)_{r+1} \ar@<0.3ex>[l]  \ar@<0.3ex>[r]^-{X \bar{\eta}}  & \mathfrak{W}(m+d,l+1)_{r+2} \ar@<0.3ex>[l]  \ar@<0.3ex>[r]^-{PX \bar{\eta}}  & \mathfrak{W}(m+1,l)_{r+1} \ar@<0.3ex>[l] \ar@<0.3ex>[r]^-{X \bar{\eta}}& \cdots \ar@<0.3ex>[l]
	}.
\end{equation}
\endgroup
The Wilson line brane $\mathfrak{W}(m,l)_r$ is replaced with Wilson line brane $\mathfrak{W}(m+d,l+1)_{r+2}$. Also, the right-going morphism changes from $PX\bar{\eta}$ to $X\bar{\eta}$ and the left-going morphism changes from $X\bar{\eta}$ to $PX\bar{\eta}$. Similarly, there are two analogous cone constructions to lower the gauge charges.

Let us take the brane ${\cal B}(-n-1)$ as an example. This brane complex has $n+2$ elements and the first $n+2-d$ elements are outside the small window. In order to bring every element inside the small window, one needs to utilize the cone construction (\ref{cone-1}) for the first element and the cone construction (\ref{cone-2}) one by one for the rest $n+1-d$ elements. Consequently, gauge charge of last $d$ elements is raised into the small window. Then, repeating the procedure again for the first $n+2-2d$ elements will bring another $d$ elements inside small window. After repeating $k$ times, the entire brane complex will be in the small window and the complex becomes
\begin{equation}
\begin{split}
	\xymatrix{
	{\cal C}(0) \ar@<0.3ex>[r]^-{PX \bar{\eta}} & {\cal C}(1) \ar@<0.3ex>[r]^-{PX \bar{\eta}} \ar@<0.3ex>[l]& {\cal C}(2) \ar@<0.3ex>[r]^-{PX \bar{\eta}} \ar@<0.3ex>[l] & \cdots \ar@<0.3ex>[r]^-{PX \bar{\eta}} \ar@<0.3ex>[l] & {\cal C}(k) \ar@<0.3ex>[l]
	},
\end{split} \label{brane--n-1}
\end{equation}
where the chain complex is divided into $k+1$ small complexes with $d$ elements except for ${\cal C}(0)$ which only contains $n' +1 $ elements. The morphisms inside each small complex are $X \bar{\eta}$. The small complexes are connected with the morphism $PX \bar{\eta}$, which means morphism between the last element of ${\cal C}(i)$ and the first element of ${\cal C}(i+1)$ is $PX \bar{\eta}$. More specificly,
\begin{multline}
		{\cal C}(0) : \\
\begingroup
\footnotesize
	\xymatrix{
	\mathfrak{W}(-n', l+k)_{2k} \ar@<0.3ex>[r]^-{X \bar{\eta}}& \mathfrak{W}(-n'+1,l+k)^{\oplus (n+1)}_{2k+1} \ar@<0.3ex>[r]^-{X \bar{\eta}} \ar@<0.3ex>[l] & \cdots \ar@<0.3ex>[r]^-{X \bar{\eta}} \ar@<0.3ex>[l] &\mathfrak{W}(0,l+k)^{\oplus \begin{psmallmatrix}n+1 \\n' \end{psmallmatrix}}_{2k+n'}  \ar@<0.3ex>[l]
	},
\endgroup
\end{multline}
\begin{multline}
		{\cal C}(i) : \\
\begingroup
\footnotesize
	\xymatrix{
   \mathfrak{W}(-d+1, l+k-i)_{2(k- i)+n'+(i-1)d+1}^{\oplus \begin{psmallmatrix}n+1 \\ n' +(i-1)d+1 \end{psmallmatrix}} \ar@<0.3ex>[r]^-{X \bar{\eta}}& \mathfrak{W}(-d+2,l+k-i)^{\oplus \begin{psmallmatrix}n+1 \\ n' +(i-1)d+2 \end{psmallmatrix}}_{2(k- i)+n'+(i-1)d+2} \ar@<0.3ex>[r]^-{X \bar{\eta}} \ar@<0.3ex>[l] & \cdots \ar@<0.3ex>[l] 	
	\\
	\qquad \qquad \qquad \qquad \qquad \qquad \qquad \qquad \qquad \cdots \ar@<0.3ex>[r]^-{X \bar{\eta}}  & \mathfrak{W}(0,l+k-i)^{\oplus \begin{psmallmatrix}n+1 \\n' +id \end{psmallmatrix}}_{2(k- i)+n'+id} \,.  \ar@<0.3ex>[l]
	}
\endgroup
\end{multline}
The first element of ${\cal C}(i)$ always has charge $-d+1$ under the first $U(1)$ (except for ${\cal C}(0)$ which has charge $-n'$), and the last element has charge $0$. Notice also that if the R-charge of the last element of ${\cal C}(i)$ is $r$, then the R-charge of the first element of ${\cal C}(i+1)$ is $r-1$.

Now, denote the image of the brane (\ref{brane--n-1}) in the geometry phase by $B(-n-1)$ and denote the embedding of ${\cal X}$ into $\mathbb{P}^n \times \mathbb{P}^{\begin{psmallmatrix} n+d \\ d \end{psmallmatrix} -1}$ by $\delta$, one finds by taking $P=0$
\begin{equation}
	\delta_*(B(-n-1)) =
	\xymatrix{
	{\cal C}'(0)  & {\cal C}'(1)  \ar@<0.3ex>[l]_-{\frac{\partial G}{\partial X} \eta} & {\cal C}'(2)  \ar@<0.3ex>[l]_-{\frac{\partial G}{\partial X} \eta} & \cdots  \ar@<0.3ex>[l]_-{\frac{\partial G}{\partial X} \eta} & {\cal C}'(k) \ar@<0.3ex>[l]_-{\frac{\partial G}{\partial X} \eta}
	},
\end{equation}
with
\begin{multline}
		{\cal C}'(0) : \\
\begingroup
\footnotesize
	\xymatrix{
	{\cal O}(-n', l+k)_{2k} \ar@<0.3ex>[r]^-{X \bar{\eta}}& {\cal O}(-n'+1,l+k)^{\oplus (n+1)}_{2k+1} \ar@<0.3ex>[r]^-{X \bar{\eta}}  & \cdots \ar@<0.3ex>[r]^-{X \bar{\eta}}  & {\cal O}(0,l+k)^{\oplus \begin{psmallmatrix}n+1 \\n' \end{psmallmatrix}}_{2k+n'}
	},
\endgroup
\end{multline}
\begin{multline}
		{\cal C}'(i) : \\
\begingroup
\footnotesize
	\xymatrix{
   {\cal O} (-d+1, l+k-i)_{2(k-i)+n'+(i-1)d+1}^{\oplus \begin{psmallmatrix}n+1 \\ n' +(i-1)d+1 \end{psmallmatrix}} \ar@<0.3ex>[r]^-{X \bar{\eta}}& {\cal O}(-d+2,l+k-i)^{\oplus \begin{psmallmatrix}n+1 \\ n' +(i-1)d+2 \end{psmallmatrix}}_{2(k-i)+n'+(i-1)d+2} \ar@<0.3ex>[r]^-{X \bar{\eta}}  & \cdots  	
	\\
	\qquad \qquad \qquad \qquad \qquad \qquad \qquad \qquad \qquad \cdots \ar@<0.3ex>[r]^-{X \bar{\eta}}  & {\cal O}(0,l+k-i)^{\oplus \begin{psmallmatrix}n+1 \\n' +id \end{psmallmatrix}}_{2(k- i)+n'+id} \, ,
	}
\endgroup
\end{multline}
where all the sheaves ${\cal O}(a,b)$ is defined on the ambient space. Therefore, $\delta_*(B(-n-1)) \in {\cal A}_0 \boxtimes D(\mathbb{P}^{\begin{psmallmatrix} n+d \\ d \end{psmallmatrix} -1})$ and $B(-n-1) \in {\cal C} \subseteq D(\mathcal{X})$.

To find $B(-n-1)$, one should first write the matrix factorization as an infinite complex by the Kn{\" o}rrer map \cite{Herbst:2008jq} acting on the infinite dimensional space
\[
\widetilde{V} = V \oplus p V \oplus p^2 V \oplus p^3 V \oplus \cdots.
\]
It is easy to see that the majority part of the infinite complex is empty on the geometric phase, leaving only an finite complex, which is the $B(-n-1)$ we are looking for. The $B(-n-1)$ complex is
\begin{equation}
	\xymatrix{
	{\cal C}''(1)  & {\cal C}''(2)  \ar@<0.3ex>[l]_-{\frac{\partial G}{\partial X} \eta} & {\cal C}''(3)  \ar@<0.3ex>[l]_-{\frac{\partial G}{\partial X} \eta} & \cdots  \ar@<0.3ex>[l]_-{\frac{\partial G}{\partial X} \eta} & {\cal C}''(k) \ar@<0.3ex>[l]_-{\frac{\partial G}{\partial X} \eta}
	},
\end{equation}
with
\begin{multline}
		{\cal C}''(i) : \\
\begingroup
\footnotesize
	\xymatrix{
   {\cal O} (-d+1, l+k-i)_{2(k-i)+n'+(i-1)d+1}^{\oplus \begin{psmallmatrix}n+1 \\ n' +(i-1)d+1 \end{psmallmatrix}} \ar@<0.3ex>[r]^-{X \bar{\eta}}& {\cal O}(-d+2,l+k-i)^{\oplus \begin{psmallmatrix}n+1 \\ n' +(i-1)d+2 \end{psmallmatrix}}_{2(k-i)+n'+(i-1)d+2} \ar@<0.3ex>[r]^-{X \bar{\eta}}  & \cdots  	
	\\
	\qquad \qquad \qquad \qquad \qquad \qquad \qquad \cdots \ar@<0.3ex>[r]^-{X \bar{\eta}}  & {\cal O}\big((k-i)d,l+k-i \big)_{2(k-i)+n'+kd} \, ,
	}
\endgroup
\end{multline}
where the sheaves ${\cal O}(a,b)$ are sheaves over ${\cal X}$. Also, the morphisms between ${\cal C}''(i)$ and ${\cal C}''(i+1)$ are no longer just morphisms between two elements. Since the complexes ${\cal C}''(i)$ are much longer than ${\cal C}'(i)$ for most $i$, one needs to add back the $P$-dependent morphisms $\frac{\partial G}{\partial X} \eta$. The brane $B(-n-1)$ should be written as a double complex with ${\cal C}''(i)$ on the $i$-th row. Then, morphisms $\frac{\partial G}{\partial X} \eta$ map the elements on $(i+1)$-th row with $R$-charge $r$ to the elements on the $i$-th row with $R$-charge $r+1$.

The images of branes $\mathcal{B}(m)$ with $m = -n, -n+1, \cdots, -(n+1)+d-1$ in the geometric phase can be found similarly and they all follow the same pattern. Together they generate the HPD category ${\cal C}$.

\subsection{\label{sec:quadrics}Quadrics}

A quadric $X$ in $\mathbb{P}^n$ defined by $G(X)=0$ can be realized by the positive phase a $U(1)$ GLSM with matter content and charges
\begin{equation}\label{GLSMquadric}
\begin{array}{cccccc}
 & X_0 & X_1 & \cdots & X_n & P \\
U(1) & 1 & 1 & \cdots & 1 & -2
\end{array}
\end{equation}
and superpotential
\[
W = P \cdot G(X).
\]
The negative phase of this model is a Landau-Ginzburg model
\begin{equation}\label{LGquadric}
\mathrm{LG}([\mathbb{C}^{n+1}/\mathbb{Z}_2],\tilde{W} = \sqrt{-\zeta/2} \cdot G(X)).
\end{equation}
For a suitable choice of $\theta$-angle. The small window of \eqref{GLSMquadric} consists of branes with gauge charges $q=-1,0$. From the resolutions of the spinor bundles\footnote{Spinor bundles on quadrics are defined in \cite{ottaviani1988spinor,langer2008d,kapranov1986derived}. Here we follow the review of them included in \cite{addington2011spinor}.} $S_\pm$ on $X$
\begin{equation}\label{spinor}
\mathcal{O}^{\oplus N}_{\mathbb{P}^n}(-1)  \stackrel{\psi_{\pm}}{\rightarrow} \mathcal{O}^{\oplus N}_{\mathbb{P}^n}
\rightarrow S_{\pm},
\end{equation}
where
\[
N = \left\{ \begin{array}{cc}
2^k, & \dim X = 2k, \\
2^{k+1}, & \dim X = 2k+1,
\end{array} \right.
\]
and $\psi_+$, $\psi_-$ are morphisms such that $\psi_+ \circ \psi_- = \psi_- \circ \psi_+ = G \cdot \mathrm{id}$, we see that the lift of the spinor bundles as GLSM matrix factorizations are in the small window. But the lift of $\mathcal{O}_X(m)$ is not in the small window because its resolution reads
\[
\mathcal{O}_{\mathbb{P}^n}(m-2) \stackrel{G}{\rightarrow} \mathcal{O}_{\mathbb{P}^n}(m) \rightarrow
\mathcal{O}_X(m).
\]
Therefore, the category of matrix factorizations of \eqref{LGquadric} is equivalent to the subcategory of $D(X)$ generated by the spinor bundles. Note that $S_+ \cong S_- \equiv S$ when $\dim X$ is odd. Then we have the Lefschetz decomposition
\begin{equation}\label{Lefquadric}
D(X) = \langle \mathcal{A}_0, \mathcal{A}_1(1),\cdots, \mathcal{A}_{n-2}(n-2) \rangle,
\end{equation}
where
\[
\mathcal{A}_0 = \langle S, \mathcal{O}(-1) \rangle
\]
if $\dim X$ is odd,
\[
\mathcal{A}_0 = \langle S_-, S_+, \mathcal{O}(-1) \rangle
\]
if $\dim X$ is even, and
\[
\mathcal{A}_i = \langle \mathcal{O}(-1) \rangle
\]
for $i>0$. Note that this Lefschetz decomposition is different from the one adopted in \cite{kuznetsov2008derived,alex2014semiorthogonal,kuznetsov2019homological} when $\dim X$ is even.

The HPD of quadrics embedded in projective spaces has been studied in \cite{alex2014semiorthogonal, kuznetsov2019homological} and from a GLSM point of view, this problem was first studied in \cite{Caldararu:2007tc}. From our proposal in section \ref{sec:proposal}, for a quadric in $\mathbb{P}^n$ defined by the zero loci of a quadratic polynomial $G$, the model describing the HPD associated with Lefschetz decomposition \eqref{Lefquadric} is a $U(1) \times U(1)$ GLSM with matter content
\[
\begin{array}{ccccccccc}
X_0 & X_1 & \cdots & X_n & P_1 & P_2 & Y_0 & \cdots & Y_n \\
1 & 1 & \cdots & 1 & -2 & -1 & 0 & \cdots & 0 \\
0 & 0 & \cdots & 0 & 0 & -1 & 1 & \cdots & 1
\end{array}
\]
and superpotential
\[
\widehat{W} = P_1 G(X) + P_2 \sum_{i=0}^n Y_i X_i.
\]
This GLSM description is consistent with the mathematical description in \cite{ballard2017homological}. Again the phase with $\zeta_1>0, \zeta_2>0$ is the geometric phase describing the universal hyperplane section of the embedding, i.e. the target space is defined by $G(X)=0$ and $\sum_{i=0}^n X_i Y_i = 0$ in $\mathbb{P}^n \times \check{\mathbb{P}}^n$.
For $\zeta_2<0, \zeta_1>\zeta_2$, there is no Higgs branch. When $\zeta_1<0, \zeta_2>0$, the Higgs branch is described by the LG model on
\begin{equation}\label{LGquadric1}
\mathrm{Tot}(\mathcal{O}^{\oplus(n+1)}\oplus \mathcal{O}(-1) \rightarrow \check{\mathbb{P}}^n)/\mathbb{Z}_2
\end{equation}
with superpotential
\begin{equation}\label{LGquadric2}
 \sum_{i,j} X_i Q_{ij} X_j + P_2 \sum_{i=0}^n X_i Y_i,
\end{equation}
where we assume that $G(X) = \sum_{i,j} X_i Q_{ij} X_j$ for a $(n+1)\times (n+1)$ invertible matrix $Q$. This is the LG model description of the HPD with respect to the Lefschetz decomposition \eqref{Lefquadric}.


The first term in \eqref{LGquadric2} gives mass to all $X_i$. Upon integrating out these massive fields, we get an effective potential
\begin{equation}\label{branchquad}
W_{eff} = 2 P_2^2 \sum_{i,j} Y_i (Q^{-1})_{ij} Y_j.
\end{equation}
This suggests that the target space of the low energy theory is a double cover of $\check{\mathbb{P}}^n$ branched over the dual quadric $\sum_{i,j} Y_i (Q^{-1})_{ij} Y_j = 0$. Construction of GLSMs for branched double covers were originally proposed in \cite{Caldararu:2007tc}, and the particular case of GLSMs with superpotentials of the form (\ref{branchquad})were studied in \cite{Halverson:2013eua,Sharpe:2013bwa}. It remains to determine the $\mathbb{Z}_2$ action on this branched double cover. 

When $n$ is even, we predict that the $\mathbb{Z}_2$ action is trivial. This can be checked by comparing the Witten index of the LG model with the Euler characteristic of the branched double cover, which are both equal to $n+2$ in this case.
This prediction agrees with the result of \cite{kuznetsov2019homological}.

When $n$ is odd, our construction corresponds to the Lefschetz decomposition \eqref{Lefquadric} with
\[
\mathcal{A}_0 = \langle S_-,S_+,\mathcal{O}(-1) \rangle,
\]
which is different from the one adopted in \cite{kuznetsov2019homological}, so in this case we have a prediction for new result. In this case, the IR limit is expected to be a $\mathbb{Z}_2$ orbifold of the double cover of $\check{\mathbb{P}}^n$ branched over the dual quadric. The $\mathbb{Z}_2$ action simply exchanges the two sheets of the covering space so the dual quadric is the fixed locus. This conjecture can be checked by comparing the Witten index of the LG model \eqref{LGquadric1}\eqref{LGquadric2} with the Euler characteristic of the orbifold. In this case, the Euler characteristic of the branched double cover is $n+1$ but the Witten index of the LG model is $2n+2$. On the other hand, the Euler characteristic of the $\mathbb{Z}_2$-orbifold is
\[
(\chi(\mathbb{P}^{n+1}[2])+3 \chi(\mathbb{P}^n[2]) )/2 = 2(n+1),
\]
which is exactly the Witten index of the LG model. In the formula above, the first term in the parentheses is the contribution from the untwisted sector while the second term is the contribution from the three twisted sectors. The Witten index of the LG model \eqref{LGquadric1}\eqref{LGquadric2} can be computed directly or simply read off from the phase diagram shown in Figure \ref{fig:phase_CI}.

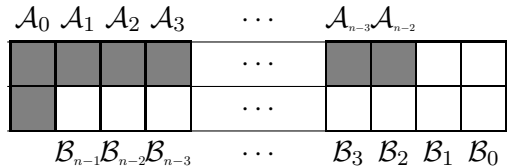
\begin{figure}[!h]
 \centering
\begin{tikzpicture}[inner sep=0in,outer sep=0in]
\node (n) {
 \centering  
    \begin{ytableau}
    \none[{\cal A}_0] & \none[{\cal A}_1]& \none[{\cal A}_2]& \none[{\cal A}_3]  & \none & \none[\cdots] &\none& \none[{\cal A}_{\scaleto{n-3\mathstrut}{4pt}}] & \none[{\cal A}_{\scaleto{n-2\mathstrut}{4pt}}]        & \none \\
    *(gray)           & *(gray) & *(gray) & *(gray)                  & \none & \none[\cdots]        &\none & *(gray) &  *(gray)    &       &     \\
       *(gray)           &      & &            & \none & \none[\cdots]  & \none        & &             &   &  \\
    \none             &\none[{\cal B}_{\scaleto{n-1\mathstrut}{5pt}}]                 &\none[{\cal B}_{\scaleto{n-2\mathstrut}{5pt}}] &\none[{\cal B}_{\scaleto{n-3\mathstrut}{5pt}}]        &\none & \none[\cdots]   &  \none   &\none[{\cal B}_3] &\none[{\cal B}_2] &\none[{\cal B}_1] &\none[{\cal B}_0] \\
    \end{ytableau}
    };
\draw[thin,] (3,.59) -- (-3.3,.59);
\draw[thin,] (3,-.59) -- (-3.3,-.59);
\draw[thin,] (3,0.0) -- (-3.3,.0);
\end{tikzpicture}
    \caption{HPD of $\mathbb{P}^n[2]$ with $n$ even}\label{evenquadricdiag}
\end{figure}

We conclude that the HPD of a quadric embedded in projective space associated with Lefschetz decomposition \eqref{Lefquadric} is the branched double cover of the dual projective space branched over the dual quadric when $n$ is even. This reproduces the result of \cite{kuznetsov2019homological}. The Lefschetz decompositions of $D(X)$ and $\mathcal{C}$ are pictured in fig. \ref{evenquadricdiag}

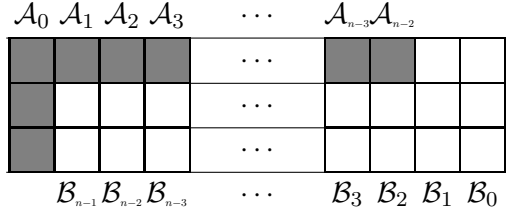
\begin{figure}[!h]
 \centering
\begin{tikzpicture}[inner sep=0in,outer sep=0in]
\node (n) {
 \centering  
    \begin{ytableau}
    \none[{\cal A}_0] & \none[{\cal A}_1]& \none[{\cal A}_2]& \none[{\cal A}_3]  & \none & \none[\cdots] &\none& \none[{\cal A}_{\scaleto{n-3\mathstrut}{4pt}}] & \none[{\cal A}_{\scaleto{n-2\mathstrut}{4pt}}]        & \none \\
    *(gray)           & *(gray) & *(gray) & *(gray)                  & \none & \none[\cdots]        &\none & *(gray) &  *(gray)    &       &     \\
    *(gray)           &     & &              & \none & \none[\cdots] & \none         & &             &        &  \\
    *(gray)           &      & &            & \none & \none[\cdots]  & \none        & &             &   &  \\
    \none             &\none[{\cal B}_{\scaleto{n-1\mathstrut}{4pt}}]                 &\none[{\cal B}_{\scaleto{n-2\mathstrut}{4pt}}] &\none[{\cal B}_{\scaleto{n-3\mathstrut}{4pt}}]        &\none & \none[\cdots]   &  \none   &\none[{\cal B}_3] &\none[{\cal B}_2] &\none[{\cal B}_1] &\none[{\cal B}_0] \\
    \end{ytableau}
    };
\draw[thin,] (3,.89) -- (-3.3,.89);
\draw[thin,] (3,-.89) -- (-3.3,-.89);
\draw[thin,] (3,.29) -- (-3,.29);
\draw[thin,] (3,-.3) -- (-3,-.3);
\end{tikzpicture}
    \caption{HPD of $\mathbb{P}^n[2]$ with $n$ odd}\label{oddquadricdiag}
\end{figure}

When $n$ is odd, we predict that the HPD associated with \eqref{Lefquadric} is the $\mathbb{Z}_2$-orbifold of the double cover of $\mathbb{P}^n$ branched over the dual quadric. We draw the diagram for the Lefschetz decompositions of $D(X)$ and $\mathcal{C}$ for this case, in fig. \ref{oddquadricdiag}

When $\zeta_1<\zeta_2<0$, the IR theory is a LG model on $(\mathbb{C}^{\oplus(n+1)} \oplus \mathbb{C}^{\oplus(n+1)})/\mathbb{Z}_2$ with coordinates $X_i$ and $Y_i$ and superpotential
\begin{equation}\label{LG2_quadric}
\sqrt{\frac{\zeta_2-\zeta_1}{2}} G(X) + \sqrt{-\zeta_2} \sum_{i=0}^n X_i Y_i.
\end{equation}
One can construct a GLSM with gauge group $U(1) \times \mathbb{Z}_2$ such that the LG model \eqref{LGquadric1}\eqref{LGquadric2} and the LG model \eqref{LG2_quadric} above describe the Higgs branches of the two phases of this GLSM, which shows that the matrix factorizations of the LG model \eqref{LG2_quadric} form a subcategory of the HPD category.

As in the case of Veronese embedding, we can apply brane transport to find image of the functor
\begin{equation}\label{functor_quadric}
F:~\mathrm{MF}(\mathrm{LG}_{\mathrm{HPD}}) \rightarrow \mathcal{C} \subset D(\mathcal{X}).
\end{equation}
To that end, we only need to lift the LG matrix factorizations to GLSM matrix factorizations in the small window corresponding to the wall crossing between the geometric phase and the HPD phase. The small window can be chosen in such a way that it consists of Wilson lines with charges $q=-2,-1,0$ under the first $U(1)$ gauge group. For example, consider the following matrix factorization of \eqref{LGquadric2}:
\begin{equation}\label{MFLG_quadric}
\xymatrix{
\mathcal{O}_0(n-1)^{\oplus N}
\ar@<0.5ex>[rr]^-{\left( \begin{array}{c} -f \\ \psi_{\pm} \end{array} \right)}
&& { \begin{array}{c} \mathcal{O}_1(n)^{\oplus N} \\ \oplus \\ \mathcal{O}_1(n-1)^{\oplus N} \end{array} }
\ar@<0.5ex>[ll]^-{(-P_2, \psi_{\mp})}
\ar@<0.5ex>[rr]^-{(\psi_{\pm},f)}
&& \mathcal{O}_0(n)^{\oplus N}
\ar@<0.5ex>[ll]^-{\left( \begin{array}{c} \psi_{\mp} \\ P_2 \end{array} \right)} },
\end{equation}
where $f = \sum_i X_i Y_i$, the subscript of $\mathcal{O}$ indicates whether it is $\mathbb{Z}_2$-even or odd, and $\psi_{\pm},N$ are defined in \eqref{spinor}. These LG matrix factorizations can be lifted to GLSM matrix factorizations in the small window as follows:
\begin{equation}\label{MFsmall_quadric}
\xymatrix{
\mathfrak{W}(-2,n-1)_{-2}^{\oplus N}
\ar@<0.5ex>[rr]^-{\left( \begin{array}{c} -f \\ \psi_{\pm} \end{array} \right)}
&& { \begin{array}{c} \mathfrak{W}(-1,n)_{-1}^{\oplus N} \\ \oplus \\ \mathfrak{W}(-1,n-1)_{-1}^{\oplus N} \end{array} }
\ar@<0.5ex>[ll]^-{(-P_2,P_1 \psi_{\mp})}
\ar@<0.5ex>[rr]^-{(\psi_{\pm},f)}
&& \mathfrak{W}(0,n)_0^{\oplus N}
\ar@<0.5ex>[ll]^-{\left( \begin{array}{c} P_1 \psi_{\mp} \\ P_2 \end{array} \right)} }
\end{equation}
which are denoted by $B_{\pm}(n)$. Thus the image of \eqref{MFLG_quadric} under the functor \eqref{functor_quadric} is $\pi_+(B_{\pm}(n))$, where $\pi_+$ is the projection functor from the category of GLSM matrix factorizations onto the derived category of the geometric phase.
Let's denote by $\delta$ the embedding of universal hyperplane section $\mathcal{X}$ in $X \times \check{\mathbb{P}}^n$, and $\iota$ the embedding of $X \times \check{\mathbb{P}}^n$ in $\mathbb{P}^n \times \check{\mathbb{P}}^n$. Then by setting $P_1=P_2=0$ in \eqref{MFsmall_quadric}, we see $\iota_* \circ \delta_* (\pi_+(B_{\pm}(n)))$ is the cone of the morphism
\[ \xymatrix{ \mathcal{O}(-2,n-1)^{\oplus N} \ar@<0.5ex>[r]^-{\psi_{\pm}} \ar@<0.5ex>[d]^-f & \mathcal{O}(-1,n-1)^{\oplus N} \ar@<0.5ex>[d]^-f \\
\mathcal{O}(-1,n)^{\oplus N} \ar@<0.5ex>[r]^-{\psi_{\pm}} & \mathcal{O}(0,n)^{\oplus N} }
\]
Therefore, from the resolution of the spinor bundles \eqref{spinor}, we get
\begin{equation}\label{universalS}
\delta_*(\pi_+(B_{\pm}(n))) = S_{\pm}(-1) \boxtimes \mathcal{O}(n-1) \stackrel{f}{\rightarrow} S_{\pm} \boxtimes \mathcal{O}(n).
\end{equation}
From the short exact sequence
\[
0 \rightarrow S_{\pm}(-1) \rightarrow \mathcal{O}_X^{\oplus N} (-1) \rightarrow S_{\mp} \rightarrow 0,
\]
we see $\delta_*(\pi_+(B_{\pm}(n)))$ is indeed an object in $\mathcal{A}_0 \boxtimes D(\check{\mathbb{P}}^n)$. \eqref{universalS} then tells us that
\[
\pi_+(B_{\pm}(n)) = (S_{\pm} \boxtimes \mathcal{O}(n))|_{\mathcal{X}}.
\]

\subsection{Complete intersection}\label{subsec:CI_HPD}

We can generalize the discussion of the previous subsection. Let's consider a complete intersection $X = \mathbb{P}^n[d_1,d_2,\cdots,d_k]$. The GLSM realizing this geometry is a $U(1)$ gauge theory with matter content and charges
\[
\begin{array}{ccccccc}
X_0 & X_1 & \cdots & X_n & P_1 & \cdots & P_k \\
1 & 1 & \cdots & 1 & -d_1 & \cdots & -d_k
\end{array}
\]
and superpotential
\begin{equation}\label{W_CI}
W = \sum _{l=1}^k P_l G_l(X),
\end{equation}
where each $G_l(X)$ is a degree $d_l$ polynomial and we assume $n+1 \geq \sum_{l=1}^k d_l$. The positive phase of this GLSM is a NLSM with target space $X$, while the negative phase is a Landau-Ginzburg model with target space
\begin{equation}\label{LG_CI}
\mathrm{Tot}\left( \mathcal{O}(-1)^{\oplus(n+1)} \rightarrow 
W\mathbb{P}[d_1,d_2,\cdots,d_k] \right)
\end{equation}
and superpotential \eqref{W_CI}, where $X_i$ are fiber coordinates and $P_l$ are 
base coordinates. If we denote the LG model \eqref{LG_CI} by 
$\mathrm{LG}(d_1,\cdots,d_k)$, then we have a semiorthogonal decomposition
\[
D(X) = \langle \mathrm{MF}(\mathrm{LG}(d_1,\cdots,d_k)), \mathcal{O},\mathcal{O}(1),\cdots,\mathcal{O}(n - \sum_{l=1}^k d_l) \rangle,
\]
and a Lefschetz decomposition
\begin{equation}\label{Lef_CI}
D(X) = \langle \mathcal{A}_0, \mathcal{A}_1(1),\cdots,\mathcal{A}_{n - \sum_{l=1}^k d_l}(n - \sum_{l=1}^k d_l) \rangle,
\end{equation}
where
\[
\mathcal{A}_0 = \langle \mathrm{MF}(\mathrm{LG}(d_1,\cdots,d_k)), \mathcal{O} \rangle, \quad
\mathcal{A}_i = \mathcal{O},i>0.
\]
According to our general discussion in section \ref{sec:proposal}, the HPD of $X$ with respect to the Lefschetz decomposition \eqref{Lef_CI} can be realized by a $U(1) \times U(1)$ GLSM with matter content
\begin{equation}\label{GLSM_HPD_CI}
\begin{array}{ccccccccccc}
X_0 & X_1 & \cdots & X_n & P_0 & P_1 & \cdots & P_k & Y_0 & \cdots & Y_n \\
1 & 1 & \cdots & 1 & -1 & -d_1 & \cdots & -d_k & 0 & \cdots & 0 \\
0 & 0 & \cdots & 0 & -1 &  0   & \cdots &  0 & 1 & \cdots & 1
\end{array}
\end{equation}
and superpotential
\[
\widehat{W} = \sum _{l=1}^k P_l G_l(X) + P_0 \sum_{i=0}^n X_i Y_i.
\]
\begin{figure}
 \centering
 \begin{tikzpicture}
 	\draw[thin,->] (0,0) -- (4,0) node[anchor=north west]{$\zeta_1$};
	\draw[thin,->] (0,0) -- (0,4)node[anchor=south east]{$\zeta_2$};
	\draw [ultra thick] (0,0) -- (0,3.5);
	\draw [ultra thick] (0,0) -- (3.5,0);
	\draw [ultra thick] (0,0) -- (-3.5,-3.5);
	\node at (-4.45,-3.65) {\footnotesize $\zeta_1-\zeta_2=0$};
	\draw [ultra thick] (0,0) -- (-4,0);
	\draw [dashed, thick] (0,0) -- (-3.75,2);
	\node at (-5.3,2) {\footnotesize $\zeta_1+(R-1)\zeta_2 =0$};
	\draw [dashed, thick] (0,0) -- (1.5, -3.75);
	\node at (1.5, -4) {\footnotesize $n\zeta_1 + \zeta_2=0$};
	\node at (3,2.5) {\small Geometric phase};
	\node at (-2.5,3)  {\small LG phase I };
	\node at (-3.75,0.75) {\small LG phase II};
	\node at (-3.75,- 1.55) {\small LG phase III};
	\node at (-0.7, -2.75) {\small Mixed phase II};
	\node at (3, -2) {\small Mixed phase I};
 \end{tikzpicture}	
 \quad \quad \quad \caption{Phase diagram of GLSM for HPD of complete intersection $\mathbb{P}^n [d_1, \cdots, d_k]$.}\label{fig:phase_CI}
	{\footnotesize \begin{flushleft} Here $R = n+1 - \sum_{l=1}^k d_l$. The geometric phase describes the universal hyperplane section. LG phase I has a Higgs branch described by LG model $\mathrm{LG}_{\mathrm{HPD}}(d_1,\cdots,d_k)$ (eq. \eqref{LG1_space}\eqref{LG1_potential}), and (R-1) mixed branches, each described by $\mathbb{P}^n$. LG phase II has the same Higgs branch as LG phase I and $((n+1)(R-1))$ Coulomb vacua. LG phase III contains a Higgs branch described by LG model $\mathrm{LG}'(d_1,\cdots,d_k)$ (eq. \eqref{LG2_space}\eqref{LG2_potential}), $n$ mixed branches (each described by $\mathrm{LG}(d_1,\cdots,d_k)$ defined in eq. \eqref{LG_CI}) and $((n+1)(R-1))$ Coulomb vacua. Mixed phase I contains $n$ mixed branches, each described by NLSM on $\mathbb{P}^n [d_1, \cdots, d_k]$. Mixed branch II has the same mixed branches as LG phase III, and $(nR)$ Coulomb vacua. \end{flushleft}}
\end{figure}
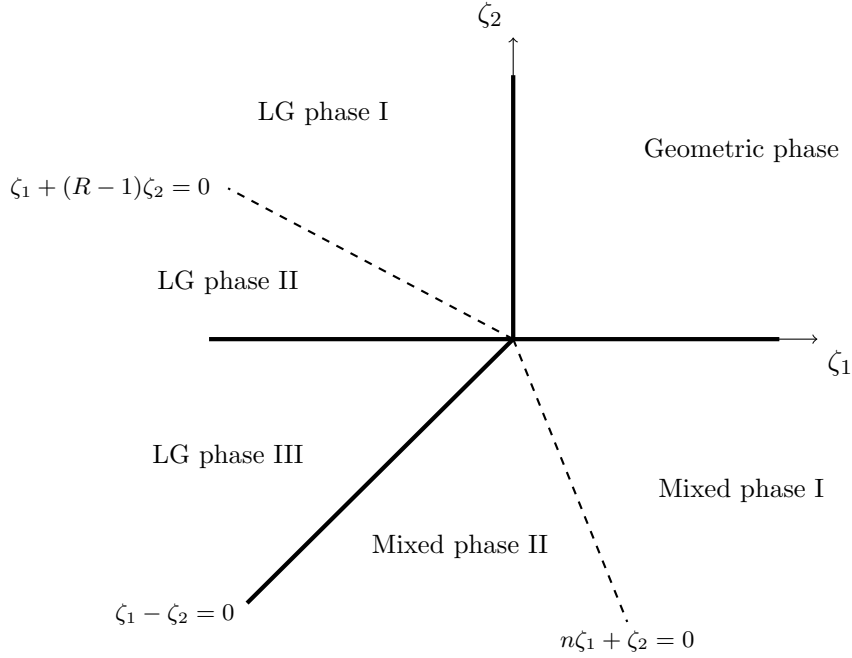
The phase diagram of this GLSM is shown in Figure \ref{fig:phase_CI}. The Higgs branches in various phases are as follows:\\
(i) $\zeta_1>0,\zeta_2>0$: NLSM with target space the universal hyperplane section
\[
\mathcal{X} = \{G_1=\cdots=G_k=0,\sum_{i=0}^n X_i Y_i = 0\} \subset \mathbb{P}^n \times \check{\mathbb{P}}^n,
\]
where the $X_i$'s and $Y_i$'s are homogeneous coordinates on $\mathbb{P}^n$ and $\check{\mathbb{P}}^n$ respectivlely.\\
(ii) $\zeta_1<0, \zeta_2>0$: Landau-Ginzburg model with target space
\begin{equation}\label{LG1_space}
\mathrm{Tot}\left( \mathcal{O}(-1,0)^{\oplus(n+1)} \oplus \mathcal{O}(1,-1) 
\rightarrow \mathrm{W}\mathbb{P}[d_1,\cdots,d_k] \times \check{\mathbb{P}}^n 
\right)
\end{equation}
and superpotential
\begin{equation}\label{LG1_potential}
\widetilde{W} = \sum _{l=1}^k P_l G_l(X) + P_0 \sum_{i=0}^n X_i Y_i,
\end{equation}
where $P_l$ are homogeneous coordinates of the weighted projective space 
$\mathrm{W}\mathbb{P}[d_1,\cdots,d_k]$, $Y_i$ are homogeneous coordinates of 
$\check{\mathbb{P}}^n$, $X_i$ and $P_0$ are coordinates along the fibers of 
$\mathcal{O}(-1,0)^{\oplus(n+1)}$ and $\mathcal{O}(1,-1)$ respectively. This LG 
model describes the HPD of the complete intersection $X = 
\mathbb{P}^n[d_1,d_2,\cdots,d_k]$ with respect to the Lefschetz decomposition 
\eqref{Lef_CI}, so let us denote it by 
$\mathrm{LG}_{\mathrm{HPD}}(d_1,\cdots,d_k)$. In the special case of 
hypersurfaces, $k=1$, $\mathrm{LG}_{\mathrm{HPD}}(d)$ has target space
\[
\mathrm{Tot} \left( \mathcal{O}^{\oplus(n+1)} \oplus \mathcal{O}(-1) \rightarrow 
\check{\mathbb{P}}^n \right) / \mathbb{Z}_d
\]
and superpotential
\[
\widetilde{W} = \sqrt{\frac{-\zeta_1}{d}} G(X) + P_0 \sum_{i=0}^n X_i Y_i.
\]\\
(iii) $\zeta_1 < \zeta_2 < 0$: LG model with target space
\begin{equation}\label{LG2_space}
\mathrm{Tot}\left( \mathcal{O}(-1)^{\oplus(n+1)} \oplus \mathcal{O}(1)^{\oplus 
(n+1)} \rightarrow \mathrm{W}\mathbb{P}[d_1,\cdots,d_k] \right)
\end{equation}
and superpotential
\begin{equation}\label{LG2_potential}
W '= \sum _{l=1}^k P_l G_l(X) + \sqrt{-\zeta_2} \sum_{i=0}^n X_i Y_i.
\end{equation}
Let's denote this LG model by $\mathrm{LG}'(d_1,\cdots,d_k)$.\\
(iv) $\zeta_2<0, \zeta_1 > \zeta_2$: No Higgs branch.

The Coulomb and mixed branches of different phases can be analyzed by studying the local models associated with wall crossing and the asymptotic behavior of the Coulomb vacua \footnote{See appendix \ref{app:CoulombEOM} for details.}.

\section{\label{sec:mutually orthogonal section}GLSMs for mutually orthogonal linear sections}

The main theorem of \cite{kuznetsov2007homological} reviewed in section \ref{sec:section2} tells us that, for a HPD pair $X \hookrightarrow \mathbb{P}(V)$ and $Z \hookrightarrow \mathbb{P}(V^\vee)$ and a subspace $L\subseteq V^\vee$, we have mutually orthogonal linear sections $X_L$ and $Z_{L}$. As discussed in section \ref{sec:linearsec}, we can take the $U(1)\times U(1)$ GLSM for HPD associated with embedding $X \hookrightarrow \mathbb{P}(V)$ and restrict it to the subspace $L$. In practice, this is done by deleting some of the fields corresponding to the homogeneous coordinates of $\mathbb{P}(V^\vee)$ and keeping only those of $L$. The resulting GLSM will have a phase with Higgs branch described by $X_L$ and another phase described by $Z_L$. This construction embeds $X_L$ and $Z_L$ into the same GLSM as long as $X_L$ and $Z_L$ are both nonempty. In general, $Z$ is described by a hybrid model with base $\mathbb{P}(V^\vee)$, therefore $Z_L$ can be described by a hybrid model with base $\mathbb{P}(L)$. Here we regard $Z$ and $Z_L$ as noncommutative spaces whose derived categories are equivalent to the category of matrix factorizations of the corresponding hybrid models.

In the case that $D(Y_{\zeta_{l}\ll -1},W_{\zeta_{l}\ll -1})$ is empty in the original GLSM describing $X$, such as in the case of Veronese embedding, we can go to the strong coupling limit of the $U(1)_{s+1}$ gauge group under which the deleted fields are charged, and the resulting GLSM also realizes $X_L$ and $Z_L$ in different phases.

\subsection{Veronese embedding}

Let's first consider the Veronese embedding $\mathbb{P}(V) \rightarrow \mathbb{P}(\mathrm{Sym}^d V)$. Suppose that $\dim V = n+1$, the linear space $L \subseteq \mathrm{Sym}^d V^\vee$ has dimension $m$ and the complete intersection $\mathbb{P}(V)_L$ is defined by
\[
\mathbb{P}(V)_L = \{X \in \mathbb{P}(V) | Q_1(X) = \cdots = Q_m(X) = 0\},
\]
where the $Q_a(X)$'s are independent degree-$d$ polynomials associated with $L$ and $m < n$.

Upon restricting to $L$, the GLSM \eqref{GLSM_Veronese_universal} reduces to the GLSM with the following matter content:
\begin{equation}\label{2veroL}
\begin{array}{cccccccc}
X_0 & X_1 & \cdots & X_n & P & S_1 & \cdots & S_m \\
1 & 1 & \cdots & 1 & -d & 0 & \cdots & 0 \\
0 & 0 & \cdots & 0 & -1 & 1 & \cdots & 1
\end{array}
\end{equation}
and superpotential
\[
\widehat{W} = P \sum_{a=1}^m S_a Q_a(X).
\]
As discussed in section \ref{sec:proposal}, by solving the D-term and F-term equations one can show:\\
(i)~$\zeta_1>0,\zeta_2>0$:~The Higgs branch is the subvariety defined by
\[
\sum_a S_a Q_a(X) = 0
\]
in $\mathbb{P}^n \times \mathbb{P}(L)$.\\
(ii)~$\zeta_2<0,\zeta_1 > d \zeta_2$:~The Higgs branch is $X_L$ provided that the transversality conditions are satisfied.\\
(iii)~$\zeta_1 <0, \zeta_1< d \zeta_2$:~The Higgs branch is $Z_{L}$, where $Z_L$ is the noncommutative space whose derived category is defined by the matrix factorizations of the Landau-Ginzburg model on
\[
\left[ \mathrm{Tot}(\mathcal{O}\left(-\frac{1}{d}\right)^{\oplus (n+1)} \rightarrow \mathbb{P}^{m-1})/\mathbb{Z}_d \right]
\]
with superpotential
\begin{equation}\label{potential2verL}
\widetilde{W} = \sqrt{-\frac{\zeta_1}{d}} \sum_{a=1}^m S_a Q_a(X),
\end{equation}
where the $\mathbb{Z}_d$ action acts on the fiber. When $d=2$, the category of the matrix factorization is equivalent to
\[
D(Z_{L}) = D(\mathbb{P}(L), \mathcal{C}l_0).
\]

In the strong coupling limit of the second $U(1)$, the GLSM describing $X_L$ and $Z_{L}$ is a $U(1)$ gauge theory with matter content
\begin{equation}\label{2veroLSim}
\begin{array}{ccccccc}
X_0 & X_1 & \cdots & X_n & S_1 & \cdots & S_m \\
1 & 1 & \cdots & 1 & -d & \cdots & -d
\end{array}
\end{equation}
and superpotential
\begin{equation}\label{W_YL_Veronese}
W_{0} = \sum_{a=1}^m S_a Q_a(X).
\end{equation}
From D-term and F-term equations, one can see that the Higgs branch of the positive phase is given by $X_L$. The Higgs branch of the negative phase is the LG model above describing $Z_L$.

Some examples of the GLSM \eqref{2veroLSim} were studied before in the literature \cite{Caldararu:2007tc, Addington:2012zv}. The geometric interpretation was also investigated by various means. Here we list a few examples:\\

$\bullet$ $X_L = \mathbb{P}^3[2,2]:$ Double Veronese embedding of $X=\mathbb{P}^3$, $\dim L = 2$, $Z_L=$ branched double cover of $\mathbb{P}^1$.

In the LG phase, the $X_i$'s can be integrated out due to their mass obtained from the superpotential, the theory becomes a non-linear sigma model on a branched double cover of $\mathbb{P}^1$ \cite{Caldararu:2007tc, Addington:2012zv}, which is an elliptic curve $C$. If we rewrite the superpotential \eqref{W_YL_Veronese} as $W_{0} = \sum_{i,j} X_i A_{ij}(S) X_j$, where the $4 \times 4$ matrix $A$ is linear in $S_a$, then the four branch points are given by the solutions of
\[
\det(A) = 0.
\]
Clearly, the matrix factorization
\begin{equation}\label{MF_P3_22}
\xymatrix{
\mathfrak{W}(q-4) \ar@<0.5ex>[r]^-{X\bar{\eta}} & \mathfrak{W}(q-3)^{\oplus 4} \ar@<0.5ex>[l]^-{AX\eta} \ar@<0.5ex>[r]^-{X\bar{\eta}} & \mathfrak{W}(q-2)^{\oplus 6} \ar@<0.5ex>[l]^-{AX\eta} \ar@<0.5ex>[r]^-{X\bar{\eta}} & \mathfrak{W}(q-1)^{\oplus 4} \ar@<0.5ex>[l]^-{AX\eta} \ar@<0.5ex>[r]^-{X\bar{\eta}} & \mathfrak{W}(q) \ar@<0.5ex>[l]^-{AX\eta}}
\end{equation}
is equivalent to\footnote{See appendix \ref{app:MFgerbe} for more details on matrix factorization on gerbes.}
\[
\mathcal{O}_{[\mathbb{P}^1/\mathbb{Z}_2]}\left(-\frac{q}{2}\right)
\]
in the IR limit because $W_{0}=0$ on the base $\mathbb{P}^1$.
Let $f$ be the projection from $C$ to $\mathbb{P}^1$. For the degree $m$ line bundle $\mathcal{O}_{\mathbb{P}^1}(m)$ on $\mathbb{P}^1$, the pullback $f^*(\mathcal{O}(m))$ is a degree $2m$ line bundle $\mathcal{E}_{2m}$ on $C$. From the fact that
\[
f^*(\mathcal{L}_1 \otimes \mathcal{L}_2) = f^*(\mathcal{L}_1) \otimes f^*(\mathcal{L}_2)
\]
for any pair of line bundles $\mathcal{L}_1$ and $\mathcal{L}_2$,
we conclude that the matrix factorization \eqref{MF_P3_22} projects to the sheaf of sections of $\mathcal{E}_{-q}$ on the elliptic curve $C$ in the IR limit.

The argument above shows the following equivalence between categories
\[
D(C) \cong D(Z_L) \cong D^b(\mathbb{P}^1,\mathcal{C}l_0),
\]
where $\mathcal{C}l_0$ consists of sheaves of modules of the even part of the Clifford algebra determined by $W_{0}$.

By comparing the categories of B-branes of the two phases, we also recover
\[
\mathcal{D}^b(C) \cong \mathcal{D}^b(M),
\]
which is consistent with the fact that $C$ and $M$ are both elliptic curves.

$\bullet$ $X_L = \mathbb{P}^5[2,2,2]:$ Double Veronese embedding of $X=\mathbb{P}^5$, $\dim L = 3$, $Z_L=$ branched double cover of $\mathbb{P}^2$.

The analysis of this example is essentially the same as that of the last example. The IR physics of the positive phase is described by a non-linear sigma model on the complete intersection $M=\mathbb{P}^5[2,2,2]$, which is a K3 surface. The IR physics of the negative phase is described by a non-linear sigma model on the branched double cover of $\mathbb{P}^2$, which is also a K3 surface $S$. The branch locus is the degree 6 curve in $\mathbb{P}^2$. Again we have the following equivalence between categories
\[
D(S) \cong D(Z_L) \cong D(\mathbb{P}^2,\mathcal{C}l_0) \cong D(M).
\]

$\bullet$ $X_L = \mathbb{P}^7[2,2,2,2]:$ Double Veronese embedding of $X = \mathbb{P}^7$, $\dim L = 4$, $Z_L=$ noncommutative resolution of branched double cover of $\mathbb{P}^3$.

This example is different from the last two because the double cover of $\mathbb{P}^3$ is ramified along a degree 8 surface, which is generically singular while the complete intersection $X_L=\mathbb{P}^7[2,2,2,2]$ is in general a smooth Calabi-Yau 3-fold. \cite{Addington:2012zv} shows that the moduli space of point-like B-branes in the negative phase is a small resolution of the double cover\footnote{See appendix \ref{app:D0} for more details on point-like branes.}. However, the low energy physics cannot be described by the small resolution because it is not K\"{a}hler.

The GLSM analysis is nevertheless valid, so we still have
{\footnote{A mathematical proof can be found in \cite{kuznetsov2008derived}.}}
\begin{equation}\label{hpdual1}
D(X_L) \cong D(Z_L) \cong D(\mathbb{P}^3,\mathcal{C}l_0),
\end{equation}
where $D(Z_L)$ is equivalent to the category of matrix factorizations of LG model on
\[
\left[\mathrm{Tot}(\mathcal{O}(-\frac{1}{2})^{\oplus 8} \rightarrow \mathbb{P}^3)/\mathbb{Z}_2\right].
\]
\eqref{hpdual1} justifies the statement that the negative phase of this GLSM is described by the noncommutative resolution $(\mathbb{P}^3,\mathcal{C}l_0)$.

$\bullet$ $X_L = \mathbb{P}^5[3,3]:$ Degree 3 Verondese embedding of $X=\mathbb{P}^5$, $\dim L=2$, $Z_L=$ noncommutative K3 surfaces fibered over $\mathbb{P}^1$.

The GLSM is a $U(1)$ gauge theory with chiral multiplets $X_i$, $i=1,\cdots,6$, and $P_j$, $j=1,2$ with the following charge assignment:
\[
\begin{array}{cccccccc}
X_1 & X_2 & X_3 & X_4 & X_5 & X_6 & P_1 & P_2 \\
1 & 1 & 1 & 1 & 1 & 1 & -3 & -3
\end{array}
\]
and a superpotential:
\begin{equation}\label{potential1}
W = P_1 G_1(X)+P_2 G_2(X),
\end{equation}
where $G_1, G_2$ are cubic polynomials in $X_i$. The positive phase is described by a NLSM whose target space is the complete intersection
\[
X_L = \{ G_1 = G_2 = 0 \} \subseteq \mathbb{P}^5,
\]
which is a Calabi-Yau 3-fold. The negative phase is described by a LG model on
\[
\left[\mathrm{Tot}\left(\mathcal{O}\left(-\frac{1}{3}\right)^{\oplus 6} \rightarrow \mathbb{P}^1\right)/\mathbb{Z}_3\right]
\]
with superpotential \eqref{potential1}. Thus the HPD category $D(Z_L)$ can be described by matrix factorizations of this LG model. Because the category of matrix factorizations of LG model with cubic superpotential in six variables is equivalent to the derived category of a noncommutative K3 surface, $Z_L$ can be thought of as a family of noncommutative K3 surfaces fibered over $\mathbb{P}^1$.

\subsection{Complete intersection}

For complete intersection $X = \mathbb{P}^n[d_1,d_2,\cdots,d_k]$ embedded in $\mathbb{P}^n$, the HPD associated with the Lefschetz decomposition \eqref{Lef_CI} is described by the GLSM \eqref{GLSM_HPD_CI}. Assume that $L$ is a $(m+1)$-dimensional subspace of $V^\vee$, $0 \leq m < n-k$, then the GLSM realizing the duality between $X_L$ and $Z_L$ has the following matter content and charges under $U(1) \times U(1)$ gauge symmetry:
\begin{equation}\label{GLSM_mutual_CI}
\begin{array}{ccccccccccc}
X_0 & X_1 & \cdots & X_n & P_0 & P_1 & \cdots & P_k & Y_0 & \cdots & Y_m \\
1 & 1 & \cdots & 1 & -1 & -d_1 & \cdots & -d_k & 0 & \cdots & 0 \\
0 & 0 & \cdots & 0 & -1 &  0   & \cdots &  0 & 1 & \cdots & 1
\end{array}
\end{equation}
and superpotential:
\[
W = \sum_{l=1}^k P_l G_l(X) + P_0 \sum_{j=0}^m Y_j L_j(X),
\]
where $\deg(G_l) = d_l$ and $L_i$'s are linear. D-term and F-term constraints give rise to the following Higgs branches:\\
(i)~$\zeta_1 \gg 1, \zeta_2 \gg 1$: Nonlinear sigma model with target space
\[
\mathcal{X}_L = \{ G_1(X) = \cdots =G_k(X)=0, \sum_{j=0}^m Y_j L_j(X) = 0\} \subset \mathbb{P}(V) \times \mathbb{P}(V^\vee).
\]
(ii)~$\zeta_1 \gg 1, \zeta_2 \ll -1$ or $\zeta_2 < \zeta_1 \ll -1$: Nonlinear sigma model with target space
\[
X_L = \{G_1(X)=\cdots=G_k(X)=0,L_0(X) = L_1(X) = \cdots = L_m(X) = 0\} \subset \mathbb{P}(V).
\]
(iii)~$\zeta_1 \ll -1, \zeta_2 \gg 1$: the Higgs branch is $Z_L$, which can be described by the Landau-Ginzburg model with target space
\[
\mathrm{Tot}\left( \mathcal{O}(-1,0)^{\oplus(n+1)} \oplus \mathcal{O}(1,-1) 
\rightarrow \mathrm{W}\mathbb{P}[d_1,\cdots,d_k] \times \mathbb{P}(L) \right)
\]
and superpotential
\[
W = \sum _{l=1}^k P_l G_l(X) + P_0 \sum_{j=0}^m Y_j L_j(X),
\]
where $P_l$ are homogeneous coordinates of the weighted projective space $\mathrm{W}\mathbb{P}[d_1,\cdots,d_k]$, $Y_i$ are homogeneous coordinates of $\mathbb{P}(L)$, $X_i$ and $P_0$ are coordinates along the fibers of $\mathcal{O}(-1,0)^{\oplus(n+1)}$ and $\mathcal{O}(1,-1)$ respectively.\\
(iv)~$\zeta_1 < \zeta_2 \ll -1$: LG model with target space
\begin{equation}\label{LG2_CI_mutual}
\mathrm{Tot}\left( \mathcal{O}(-1)^{\oplus(n+1)} \oplus 
\mathcal{O}(1)^{\oplus(m+1)} \rightarrow \mathrm{W}\mathbb{P}[d_1,\cdots,d_k] 
\right) 
\end{equation}
and superpotential
\[
W = \sum_{l=1}^k P_l G_l(X) + \sum_{i=0}^m Y_i L_i(X),
\]
where $P_l$ serve as homogeneous coordinates on the base, $X_i$, $Y_j$ are coordinates along the fiber.

We see $X_L$ and $Z_L$ are described by Higgs branches of phases (ii) and (iii) above. Note that in this case, because $\mathrm{LG}(d_1,d_2,\cdots,d_k)$ described in section \ref{subsec:CI_HPD} is nonempty, we cannot take the strong coupling limit of the second $U(1)$ gauge group to embed $X_L$ and $Z_L$ in a single $U(1)$ GLSM. Actually, the resulting $U(1)$ GLSM would have $X_L$ in the positive phase and the LG model \eqref{LG2_CI_mutual} in the negative phase, the matrix factorizations of the latter form a subcategory of the HPD category.

\section{\label{sec:nonabelian}Comments on nonabelian models: Pl\"ucker embedding}

Though our main concern in this paper is HPDs of abelian theories, in this section we briefly mention that our construction can also be applied to nonabelian theories as well by sketching its application to Pl\"ucker embedding. We leave a detailed analysis of nonabelian theories to future work.

The Grassmannian $G(k,V)$ with $\dim V = N$ can be implemented by the geometric phase of a nonabelian GLSM with $U(k)$ gauge group and $N$ chiral fields in the fundamental representation. From our general construction, the HPD of the Pl\"ucker embedding
\[
G(k,V) \rightarrow \mathbb{P}(\wedge^k V)
\]
can be described by a nonabelian GLSM with gauge group $U(k) \times U(1)$ and the following matter content
\begin{equation}\label{GLSM_Gr}
\begin{array}{cccccc}
& \Phi_1 & \cdots & \Phi_N & P & Y_{i_1\cdots i_k}\\
U(k) & \square & \cdots & \square & \det^{-1} & 0 \\
U(1) & 0 & \cdots & 0 & -1 &  1
\end{array}
\end{equation}
where $\square$ stands for fundamental representation of $U(k)$, and the indices $i_p$ satisfy $1 \leq i_1 < i_2 < \cdots < i_k \leq N$. The superpotential is
\[
W = P \sum_{1 \leq i_1 < i_2 < \cdots < i_k \leq N} B_{i_1\cdots i_k} Y_{i_1\cdots i_k},
\]
where the Pl\"ucker coordinates read
\[
B_{i_1\cdots i_k} = \epsilon_{a_1\cdots a_k} \Phi^{a_1}_{i_1} \cdots \Phi^{a_k}_{i_k}.
\]
When the FI parameters satisfy $\zeta_1 \gg 1, \zeta_2 \gg 1$, the IR theory is a NLSM with target space the universal hyperplane section
\[
\mathcal{X} = \left\{ \sum_{1 \leq i_1 < i_2 < \cdots < i_k \leq N} B_{i_1\cdots i_k} Y_{i_1\cdots i_k} = 0 \right\} \subset G(k,V) \times \mathbb{P}(\wedge^k V^\vee),
\]
where $Y_{i_1\cdots i_k}$ serve as homogeneous coordinates of $\mathbb{P}(\wedge^k V^\vee)$.

When $\zeta_1 \ll -1, \zeta_2 > \zeta_1$, the $P$ field receives a vev $\langle P \rangle = \sqrt{-\zeta_1}$, breaking the $U(k)$ gauge symmetry to $SU(k)$. Thus we get a family of $SU(k)$ gauge theories fibered over $\mathbb{P}(\wedge^k V^\vee)$, each has $N$ fundamentals and a superpotential
\[
W = \sqrt{-\zeta_1} \sum_{1 \leq i_1 < i_2 < \cdots < i_k \leq N} B_{i_1\cdots i_k} Y_{i_1\cdots i_k}.
\]
Note that every $B_{i_1\cdots i_k}$ is a section of $\mathcal{O}(-1)$ over $\mathbb{P}(\wedge^k V^\vee)$. The HPD category is expected to be equivalent to the category of matrix factorizations of this $SU(k)$ gauge theory.

Now we want to know which Lefschetz decomposition it corresponds to, this requires the knowledge of the twisting bundle $\mathcal{L}$, which is the pull-back of $\mathcal{O}(1)$ under the Pl\"ucker embedding, and $\mathcal{A}_0$ in the Lefschetz decomposition. Obviously,
\[
\mathcal{L} = \wedge^k \mathcal{S}^\vee,
\]
where $\mathcal{S}$ is the tautological bundle over $G(k,V)$. For a semiorthogonal decomposition of $G(k,V)$, every generator that is not in $\mathcal{A}_0$ will give rise to a mixed branch in the HPD phase above. Therefore, in order to determine the number of generators in $\mathcal{A}_0$, we only need to count the number of mixed branches in the HPD phase, which is equal to the number of Coulomb vacua of the local model corresponding to the wall crossing at the boundary between the geometric phase and the HPD phase. The local model is a $U(2)$ gauge theory with $N$ fundamentals and one chiral field in $\det^{-1}$ representation. The Coulomb vacua satisfy the equations
\[
\sigma_a^N = (-1)^{k} q \sum_{b=1}^k \sigma_b,\quad a=1,\cdots,k,
\]
and
\[
\sigma_a \neq \sigma_b \quad \mathrm{for} ~ a \neq b.
\]
In counting the number of solutions, we should take the residual $S_k$ gauge symmetry that swaps the $\sigma_a$'s into account. For $k=2$, the number of Coulomb vacua is $(N-1)(N-2)/2$, which is exactly the number of generators that is not in $\mathcal{A}_0$ of the Lefschetz decomposition with respect to Kapranov's collection, namely
\begin{equation}\label{Lef_Gr}
\begin{split}
D(G(2,N)) &= \langle \mathcal{A}_0, \mathcal{A}_1(1),\cdots,\mathcal{A}_{N-1}(N-1) \rangle,
\\
\mathcal{A}_0 = \langle \mathcal{O}, \mathcal{S}^\vee, \cdots, \mathrm{Sym}^{N-2}\mathcal{S}^\vee \rangle, \quad  &\mathcal{A}_1 = \langle \mathcal{O}, \mathcal{S}^\vee,\cdots, \mathrm{Sym}^{N-3}\mathcal{S}^\vee \rangle, \cdots, \mathcal{A}_{N-2} = \langle \mathcal{O} \rangle.
\end{split}
\end{equation}
Therefore, we predict that the HPD of Pl\"ucker embedding realized by the GLSM \eqref{GLSM_Gr} is with respect to the Lefschetz decomposition \eqref{Lef_Gr}.

If we restrict the theory \eqref{GLSM_Gr} to a subspace $L \subseteq \wedge^k V^\vee$ and $\dim L = m$, then we get a $U(k) \times U(1)$ GLSM with matter content
\begin{equation}\label{GLSM_Gr_mutual}
\begin{array}{cccccccc}
& \Phi_1 & \cdots & \Phi_N & P & Y_1 & \cdots & Y_m \\
U(k) & \square & \cdots & \square & \det^{-1} & 0 & \cdots & 0 \\
U(1) & 0 & \cdots & 0 & -1 & 1 & \cdots & 1
\end{array}
\end{equation}
and superpotential
\[
W = P \sum_{j=1}^m Y_j L_j(B),
\]
where $L_j(B)$'s are linear functions in the Pl\"ucker coordinates. If we take the strong coupling limit of the $U(1)$ gauge group, then we get a $U(k)$ GLSM with matter content
\begin{equation}\label{GLSM_Gr_mutual_sim}
\begin{array}{ccccccc}
& \Phi_1 & \cdots & \Phi_N & P_1 & \cdots & P_m \\
U(k) & \square & \cdots & \square & \det^{-1} & \cdots & \det^{-1}
\end{array}
\end{equation}
and superpotential
\[
W = \sum_{j=1}^m P_j L_j(B).
\]
This theory is of the type studied in \cite{Hori:2006dk}. The positive phase is described by a NLSM with target space $X_L$, which is a complete intersection in $G(k,N)$ defined by $L_j = 0$ for all $j$. The negative phase is described by a family of $SU(k)$ gauge theories fibered over $\mathbb{P}(L)$. The category of matrix factorizations of this theory is defined to be $D(Z_L)$. In some cases, $Z$ and $Z_L$ have geometric interpretations. For example, when $k=2, N=7$ and $\dim L = 7$, $X = G(2,7)$ and $Z$ is the Pfaffian variety $\mathrm{Pf}(4,\wedge^2 V^\vee)$, $X_L$ is the Calabi-Yau three-fold $G(2,7)[1,1,1,1,1,1,1]$ and $Z_L$ is the intersection $Z \cap \mathbb{P}(L)$ in $\mathbb{P}(\wedge^2 V^\vee)$ as shown in \cite{Hori:2006dk}.



\section{Future directions}

In this work, we proposed a construction for a GLSM $\mathcal{T}_{\mathcal{X}}$ that realizes the HPD of $(X, \mathcal{T}_{X})$, as well as its linear sections. This raises many questions as well as allows us to formulate some future directions:
\begin{itemize}
 \item \textbf{More general Lefschetz decompositions}. In the work of  \cite{rennemo2017fundamental} the Lefschetz decomposition of $D(X)$ is arbitrary. However, we are constrained by the Lefschetz decomposition induced by our pair $(\mathcal{T}_{X},\mathcal{T}_{\mathcal{X}})$. A natural question would be, given a Lefschetz decomposition of $D(X)$, how can we construct a GLSM that induces it or possibly refine the window categories in order to restrict uniquely to an $\mathcal{A}_{i}$ component.
 
 \item \textbf{Nonabelian theories}. An obvious extension of our work is to consider nonabelian GLSMs. We only sketched the simplest generalization of the projective space embedding, namely $G(k,N)\hookrightarrow \mathbb{P}^{r}$. Already this example is not completely solved in the mathematics literature. Even in this seemingly simple situation HPD is only known for certain Grassmannians \cite{kuznetsov2006homological,deliu2011homological} and conjectured for the rest. A deeper study of the GLSM presented in section  \ref{sec:nonabelian} may lead to new results on this subject. Let us mention that there are also results regarding HPD of other varieties that can be realized via nonabelian GLSMs, see for instance \cite{bernardara2016homological}. 
 
 \item \textbf{HPD of nongeometric phases}. In the examples, we studied $\mathrm{dim}_{\mathbb{C}}\mathcal{M}_{K}=1$ but it should be possible to generalize the construction to multiple parameter models, and define HPD for nongeometric phases, for instance, for LG orbifold models or hybrid models. On the other hand, as we saw in the quadric example \ref{sec:quadrics}, the hybrid HPD has a geometric interpretation, it will be interesting if this also is the case in other examples where the HPD obtained via $\mathcal{T}_{\mathcal{X}}$ is new.
 
\item \textbf{Mirror of HPD}. Having a GLSM interpretation $\mathcal{T}_{\mathcal{X}}$ for the HPD of a space, opens the possibility of using mirror symmetry for GLSMs \cite{Morrison:1995yh,Hori:2000kt,Gu:2018fpm,Gu:2019byn}. Even for the simple abelian models studied here, this would be interesting and provide new results.
\end{itemize}

\section*{Acknowledgements}

We would like to thank Will Donovan, Alexander Kuznetsov, David Favero, Daniel Pomerleano, Johanna Knapp, Richard Eager, Kentaro Hori and Eric Sharpe for useful discussions and comments. JG acknowledges support from the China Postdoctoral Science Foundation No. 2020T130353. MR thanks Harvard U. and IASM at Zhejiang U. for hospitality. MR acknowledges support from the National Key Research and Development Program of China, grant No. 2020YFA0713000, and the Research Fund for International Young Scientists, NSFC grant No. 11950410500.

\appendix

\section{Matrix factorization on gerbes}\label{app:MFgerbe}

As we have seen that the HPD category of a geometric embedding is usually described by matrix factorizations of orbifold of LG model defined on gerbes (quotient stack). In this appendix, we discuss the definition of matrix factorization on gerbes. The definition requires the notion of orbibundles. The reader may refer to \cite{adem_leida_ruan_2007} for more details.

\subsection{Orbibundle}

Let $X$ be a smooth manifold admitting a $G$-action, where $G$ is a group.
An orbibundle on the quotient stack $[X/G]$ is a fiber bundle $E \stackrel{\pi}{\rightarrow} X/G $ with each fiber an orbifold. Explicitly, let $V$ be a vector space admitting a representation of $G$:
\[
\rho:\quad G \rightarrow GL(V).
\]
The fibre of $E$ is $V/\rho(G)$. If $\{U_\alpha:~\alpha \in I\}$ is an open cover of $X/G$ and
\[
\phi_\alpha: U_\alpha \times V/\rho(G) \rightarrow \pi^{-1}(U_\alpha)
\]
are the corresponding local trivializations. Then the transition functions $g_{\alpha\beta} = \phi^{-1}_\alpha \circ \phi_\beta$ take values in $GL(V)/\rho(G)$. A local section of $E$ is given by a $\rho(G)$-invariant function $s_\alpha: U_\alpha \rightarrow V$ so the relation
\[
s_\alpha = g_{\alpha\beta}\cdot s_\beta
\]
is well defined on $U_\alpha \cap U_\beta$. Given a representation of $G$ as above, the orbibundles on $[X/G]$ are classified by $H^1(X,GL(V)/\rho(G))$. When the representation is trivial, the orbibundle is just an ordinary vector bundle. When $\dim_{\mathbb{C}}V=1$, we call it a line bundle.

A morphism between two orbibundles $E_1\stackrel{\pi_1}{\rightarrow} X/G$ and $E_2\stackrel{\pi_2}{\rightarrow} X/G$ is a bundle map $f: E_1 \rightarrow E_2$, i.e. $\pi_2 \circ f = \pi_1$. Given local trivializations of $E_1$ and $E_2$ in an open set $U$:
\[
\phi_1: U \times V_1/\rho_1(G) \rightarrow \pi_1^{-1}(U),
\]
\[
\phi_2: U \times V_2/\rho_2(G) \rightarrow \pi_2^{-1}(U),
\]
and for each $x \in U$,
$f_U(x) := \phi_2^{-1} \circ f \circ \phi_1|_x$ is a linear map from $V_1$ to $V_2$ satisfying
\[
f_U(x) \circ \rho_1(g) = \rho_2(g) \circ f_U(x)
\]
for all $g \in G$.

\noindent {\bf Example 1.} $[\mathbb{C}/\mathbb{Z}_k]$, where $\mathbb{Z}_k$ acts on $\mathbb{C}$ by
\[
z \mapsto \exp{\left( \frac{2 \pi i}{k} \right)} z.
\]
The representation $\rho_m$ of $\mathbb{Z}_k$ is defined by
\[
\rho_m \left(\exp{\frac{2 \pi i}{k}} \right) = \exp{\left( \frac{2 \pi m i}{k}\right)}
\]
for $m =0,\cdots,k-1$. Because $H^1(\mathbb{C},\mathbb{C}^*/\rho_m(\mathbb{Z}_k))=0$, the line bundles on $[\mathbb{C}/\mathbb{Z}_k]$ are in one to one correspondence with the representation of $\mathbb{Z}_k$. Let's denote by $\mathcal{L}_m$ the line bundle determined by $\rho_m$. It is easy to see that the complex
\[
\mathcal{L}_{m-1} \stackrel{z}{\rightarrow} \mathcal{L}_m
\]
is quasi-isomorphic to the skyscraper sheaf $\mathcal{O}_0(m)$ at the origin $z=0$, carrying the representation $\rho_m$ along its fiber.

\noindent {\bf Example 2.} $[\mathbb{P}^n/\mathbb{Z}_2]$, where $\mathbb{Z}_2$ acts as the identity map. There are two irreducible representations for $\mathbb{Z}_2$,
\[
\rho_0(1)=\rho_0(-1)=1,\quad \rho_1(1)=1,~\rho_1(-1)=-1.
\]
The line bundles defined by $\rho_0$ are just ordinary line bundles on $\mathbb{P}^n$, they are of the form $\mathcal{O}(m)$ for some integer $m$.

The line bundles defined by $\rho_1$ are twisted line bundles. Let's denote by $\mathcal{O}(m/2)$ for some odd integer $m$ the orbibundle whose transition functions are given by $g_{\alpha\beta}^{m/2}$, where $g_{\alpha\beta}$ are the transition functions of $\mathcal{O}(1)$. The square root makes sense because the transition functions take values in $\mathbb{C}^*/\mathbb{Z}_2$. For example, the line bundle $\mathcal{O}(1/2)$ on $\mathbb{P}^1$ has the transition function
\[
g_{12} = \underline{z_1^{\frac{1}{2}}} = \underline{z_2^{-\frac{1}{2}}}
\]
where $\underline{z}$ is the class of $\pm z$ in $\mathbb{C}^*/\mathbb{Z}_2$.

\subsection{Derived category}

We consider coherent sheaves on $[X/G]$ as sheaves with finite resolutions by orbibundles (see for example \cite{katz2003d} for a review of sheaves on stacks). A sheaf $\mathcal{F}$ is called coherent if there is an exact sequence
\[
0 \rightarrow \mathcal{E}_1 \rightarrow \mathcal{E}_2 \rightarrow \cdots \rightarrow \mathcal{E}_n \rightarrow \mathcal{F},
\]
where each $\mathcal{E}_i$ is the sheaf of sections of an orbibundle. Then the derived category of $[X/G]$ is defined to be the derived category of the category of coherent sheaves with morphisms being given by $G$-equivariant chain maps.

\subsection{Matrix factorization}

Let $W$ be a holomorphic $G$-invariant function on $X$, we can define a matrix factorization as being given by bundle maps $F_1 \in \mathrm{Mor}(\mathcal{E}_1,\mathcal{E}_2)$ and $F_2 \in \mathrm{Mor}(\mathcal{E}_2,\mathcal{E}_1)$, where $\mathcal{E}_1$ and $\mathcal{E}_2$ are orbibundles, such that
\[
F_2 \circ F_1 = W \cdot id_{\mathcal{E}_1},\quad F_1 \circ F_2 = W \cdot id_{\mathcal{E}_2}.
\]
Morphisms between two matrix factorizations are required to be $G$-equivariant.

\noindent {\bf Example.} Let $X$ be the space $\mathbb{C}^2_z \oplus \mathbb{C}^2_p - \{p_1=p_2=0\}$. $\mathbb{C}^*$ acts on $X$ with the following charges
\[
\begin{array}{cccc}
z_1 & z_2 & p_1 & p_2 \\
-1 & -1 & 2 & 2
\end{array}
\]
Let's consider matrix factorization for the function
\[
W = z_1^2 p_1 + z_2^2 p_2.
\]
Clearly, there is a projection
\[
[X/\mathbb{C}^*] \stackrel{\phi}{\rightarrow} [\mathbb{P}^1/\mathbb{Z}_2],
\]
where $\mathbb{Z}_2$ acts on $\mathbb{P}^1$ trivially. For any integer $Q$, define the representation $\rho_Q$ to be the one-dimensional representation with charge $Q$, i.e.
\[
\rho_{Q}(\lambda) = \lambda^Q
\]
for all $\lambda \in \mathbb{C}^*$, then the line bundle on $[X/\mathbb{C}^*]$ determined by the representation $\rho_Q$ is
\[
\mathcal{L}(Q) = \phi^* \mathcal{O}\left(\frac{Q}{2} \right).
\]
It is easy to check that
\[
\xymatrix{
{\begin{array}{c}
\mathcal{L}(Q+2) \\  \oplus \\ \mathcal{L}(Q)
\end{array}} \ar@<0.5ex>[rrr]^{\left(
\begin{array}{cc}
z_1 &  -p_2 z_2 \\
z_2 & p_1 z_1
\end{array}\right)} & & & \mathcal{L}(Q+1)^{\oplus 2} \ar@<0.5ex>[lll]^{\left(
\begin{array}{cc}
p_1 z_1 &  p_2 z_2 \\
-z_2 & z_1
\end{array}\right)}}
\]
is a matrix factorization for $W$.

\section{Analysis of Coulomb vacua}\label{app:CoulombEOM}

The phase diagrams in Figures \ref{fig:phase_Ver} and \ref{fig:phase_CI} can be determined by examining the asymptotic behavior of the equations of motion on the Coulomb branch. In this appendix, we perform this analysis in detail. For ease of notation, we present the computation for double Veronese embedding of $\mathbb{P}^2$ and quadric in $\mathbb{P}^3$. The general cases can be analyzed in the same way. In the following, $q_a = \exp(-t_a)$, where $t_a = \zeta_a - i \theta_a$ is the complexified FI parameter.

\subsection{$\mathbb{P}^2 \rightarrow \mathbb{P}^5$}
The equations of motion for the Coulomb vacua read
\begin{equation}\label{coulombeq2}
\left\{ \begin{array}{l}
\sigma_1^3 = q_1 (2 \sigma_1+ \sigma_2)^2, \\
\sigma_2^6 = -q_2 (2 \sigma_1+ \sigma_2).
\end{array} \right.
\end{equation}
Nonzero solutions satisfy
\begin{equation}\label{2sigma2}
\sigma_2^{15}+3 q_2 \sigma_2^{10} + 8 q_1 q_2 \sigma_2^9 + 3 q_2^2 \sigma_2^5 + q_2^3 = 0
\end{equation}
and
\begin{equation}\label{2sigma1}
\sigma_1^3 = q_1 q_2^{-2} \sigma_2^{12}.
\end{equation}
For generic $q_1$ and $q_2$, \eqref{2sigma2} cannot be solved exactly, but in order to investigate the asymptotic behavior of the solutions, we can use asymptotic approximations of the equation in different phases.

(i)~$\zeta_1 \gg 1,\zeta_2 \gg 1$:~$q_1 \rightarrow 0$ and $q_2 \rightarrow 0$, so \eqref{2sigma2} can be approximated by
\[
\sigma_2^{15} + 3 q_2 \sigma_2^{10} = 0.
\]
Nonzero solutions tend to zero as $\sigma_2 \sim q_2^{1/5}$, then $\sigma_1 \propto \sigma_2 (q_2^{-1} \sigma_2^5 + 1) \rightarrow 0$ in the asymptotic region. Therefore, all solutions of \eqref{coulombeq2} approach $(\sigma_1,\sigma_2)=(0,0)$ as $\zeta_1 \rightarrow \infty, \zeta_2 \rightarrow \infty$, which means that we only have Higgs branch in this phase, described by the universal quadric in $\mathbb{P}^2$.\\
(ii)~$\zeta_1 \ll -1, \zeta_2 \gg 1$:~$q_1 \rightarrow \infty$ and $q_2 \rightarrow 0$, so \eqref{2sigma2} can be approximated by
\[
\sigma_2^{15} + 8 q_1 q_2 \sigma_2^{9} = 0.
\]
Aside from the nine zero solutions, which correspond to the Higgs branch (the HPD), there are six solutions satisfying
\[
\sigma_2^6 = -8 q_1 q_2.
\]
Let's look at the line $\zeta_2 = -\lambda \zeta_1$ for $\lambda >0$. Then
\[
\sigma_2^6 \rightarrow \left\{ \begin{array}{ll} \infty, & \lambda<1 \\
 0, & \lambda>1 \end{array} \right.
\]
and
\[
\sigma_1^3 = q_1 q_2^{-2} \sigma_2^{12} \sim q_1^3 \rightarrow \infty.
\]
Therefore, other than the Higgs branch, there is a mixed branch described by $\mathbb{P}^5$ for $\lambda>1$, and there are six Coulomb vacua for $\lambda<1$.\\
(iii)~$\zeta_1 \ll -1,\zeta_2 \ll -1$:~$q_1 \rightarrow \infty$ and $q_2 \rightarrow \infty$. Again, let's look at the line $\zeta_2 = \lambda \zeta_1$, i.e. $q_2 = q_1^\lambda$. When $\lambda > 1/2$, \eqref{2sigma2} can be approximated by
\[
\sigma_2^{15} + q_1^{3 \lambda} = 0.
\]
Consequently, $\sigma_2 \sim - q_1^{\lambda/5} \rightarrow \infty$ and $\sigma_1^3 \sim q_1^{1+2 \lambda/5} \rightarrow \infty$. So all the fifteen solutions approach infinity and we have fifteen Coulomb vacua. On the other hand, when $\lambda < 1/2$, \eqref{2sigma2} can be approximated by
\[
\sigma_2^{15} + 8 q_1^{1+\lambda} \sigma_2^9 = 0.
\]
Nonzero solutions satisfy
\[
\sigma_2^6 \sim q_1^{1+\lambda} \rightarrow \infty,\quad \sigma_1^3 \sim q_1^3 \rightarrow \infty.
\]
Therefore we have a Higgs branch described by the HPD and six Coulomb vacua.\\
(iv)~$\zeta_1 \gg 1,\zeta_2 \ll -1$:~$q_1 \rightarrow 0, q_2 \rightarrow \infty$, then \eqref{2sigma2} can be approximated by
\[
\sigma_2^{15} + q_2^3 = 0,
\]
and all solutions satisfy $\sigma_2 \sim q_2^{1/5} \rightarrow \infty$. Assume that $\zeta_2 = -\lambda \zeta_1$, then \eqref{2sigma1} implies
\[
\sigma_1^3 \sim q_1^{1-2 \lambda/5} \rightarrow \left\{ \begin{array}{ll} 0, & \lambda<5/2 \\
 \infty, & \lambda>5/2 \end{array} \right.
\]
Thus there are five mixed branches, each of which is described by $\mathbb{P}^2$, when $\lambda<5/2$. When $\lambda>5/2$, each $\mathbb{P}^2$ split into three Coulomb vacua, and we are left with fifteen Coulomb vacua in total.

\subsection{$\mathbb{P}^3[2]$}

The equations of motion for the Coulomb vacua read
\begin{equation}\label{coulombeq3}
\left\{ \begin{array}{l}
\sigma_1^4 = -4 q_1 \sigma^2_1 (\sigma_1+ \sigma_2), \\
\sigma_2^4 = -q_2 (\sigma_1 + \sigma_2).
\end{array} \right.
\end{equation}
Nonzero solutions satisfy
\begin{equation}\label{EOMQ1}
(\sigma_2^3 + q_2)^2=0,
\end{equation}
or
\begin{equation}\label{EOMQ2}
\sigma_2^6 + 2 q_2 \sigma_2^3 + q_2^2 - 4 q_1 q_2 \sigma_2^2 = 0,
\end{equation}
and
\begin{equation}\label{EOMQ3}
\sigma_1 = -q_2^{-1} \sigma_2^4 - \sigma_2.
\end{equation}
As $q_2 \rightarrow 0$ for $\zeta_2 \gg 1$ and $q_2 \rightarrow \infty$ for $\zeta_2 \ll -1$, solutions to \eqref{EOMQ1} and \eqref{EOMQ3} contribute to Higgs branch on the upper half plane of the real FI-space, and contribute to mixed branches on the lower half plane. Now let us analyze the asymptotic behavior of solutions to equations \eqref{EOMQ2} and \eqref{EOMQ3}. \\
(i)~$\zeta_1 \gg 1,\zeta_2 \gg 1$:~$q_1 \rightarrow 0, q_2 \rightarrow 0$, all solutions approach $(0,0)$ as $\zeta_1 \rightarrow \infty, \zeta_2 \rightarrow \infty$. Therefore we only have Higgs branch in this phase, which is described by the universal hyperplane section in $\mathbb{P}^3[2] \times \mathbb{P}^3$. \\
(ii)~$\zeta_1 \ll -1,\zeta_2 \gg 1$:~$q_1 \rightarrow \infty$ and $q_2 \rightarrow 0$, so \eqref{EOMQ2} can be approximated by
\[
\sigma_2^6 - 4 q_1 q_2 \sigma_2^2 = 0.
\]
Other than the two zero solutions, which contribute to the Higgs branch (the HPD), there are four solutions satisfying
\[
\sigma_2^4 = 4 q_1 q_2.
\]
Consequently,
\[
\sigma_2^4 \rightarrow \left\{ \begin{array}{ll} \infty, & \zeta_1+\zeta_2<0 \\
 0, & \zeta_1+\zeta_2>0 \end{array} \right.
\]
and
\[
\sigma_1^2 = 4 q_1 q_2^{-1} \sigma_2^4 = 16 q_1^2 \rightarrow \infty.
\]
Therefore, other than the Higgs branch, there is a mixed branch described by $\mathbb{P}^3$ for $\zeta_2 > -\zeta_1$, and there are four Coulomb vacua for $\zeta_2 < -\zeta_1$.\\
(iii)~$\zeta_1 \ll -1,\zeta_2 \ll -1$:~$q_1 \rightarrow \infty$ and $q_2 \rightarrow \infty$. Let's take the semi-infinite line $\zeta_1 = \lambda \zeta_2$, i.e. $q_1 = q_2^\lambda$. When $\lambda < 1$, \eqref{EOMQ2} can be approximated by
\[
\sigma_2^6 + q_2^2 = 0.
\]
Thus $\sigma_2^6 = -q_2^2 \rightarrow \infty$ and $\sigma_1^2 = 4 q_1 q_2^{-1} \sigma_2^4 \sim q_1 q_2^{1/3} \rightarrow \infty$. Then we have six Coulomb vacua. When $\lambda > 1$, \eqref{EOMQ2} can be approximated by
\[
\sigma_2^6 - 4 q_2^{1+\lambda} \sigma_2^2 = 0.
\]
There are four nonzero solutions satisfying
\[
\sigma_2^4 = 4 q_2^{1+\lambda} \rightarrow \infty, \quad
\sigma_1^2 \sim 16 q_1 q_2^\lambda \rightarrow \infty,
\]
which contribute to four Coulomb vacua. The two zero solutions contribute to the Higgs branch.\\
(iv)~$\zeta_1 \gg 1,\zeta_2 \ll -1$:~$q_1 \rightarrow 0, q_2 \rightarrow \infty$, then \eqref{EOMQ2} can be approximated by
\[
\sigma_2^6 + q_2^2 = 0.
\]
Thus $\sigma_2 \rightarrow \infty$, and
\[
\sigma_1^2 \sim 4 q_1 q_2^{1/3} \rightarrow \left\{ \begin{array}{ll} 0, & \zeta_1+\zeta_2/3>0 \\
 \infty, & \zeta_1+\zeta_2/3<0 \end{array} \right.
\]
Therefore, we have mixed branch when $\zeta_2 > -3 \zeta_1$, and Coulomb vacua when $\zeta_2 < -3 \zeta_1$. The full phase diagram is given by figure \ref{fig:phase_CI} with $n=3, k=1, d_1=2$.

\section{D0-brane probes}\label{app:D0}

We have seen that the HPD category can be described by the category of matrix factorizations in general. However, the geometric meaning of the HPD is vague in this description. In some cases, we can completely or partially recover a geometric entity from the matrix factorizations by studying the so-called point-like branes or D0-brane probes. In this appendix, we provide a careful analysis of the point-like branes. A similar analysis was performed in \cite{Addington:2012zv} for LG models, here we take the GLSM point of view. We take the $\mathbb{P}^3[2,2]$ model as a demonstrative example explaining our construction of a GLSM D0-brane, where the moduli space of this D0-brane describes a branched double cover.

The corresponding GLSM has the matter content with the $U(1)$ charges
\[
\begin{array}{ccccccc}
X_0 & X_1 & X_2 & X_3 & P_0  & P_1 \\
1 & 1 &1 & 1 & -2 & -2
\end{array}
\]
and superpotential
\[
	W = P_0 G_0(X) + P_1 G_1(X),
\]
where $G_1(X)$ and $G_2(X)$ are quadratic polynomials of $X_i$. It is easy to see that the geometric phase $( \zeta \gg 1 )$ of this model is a NLSM on $\mathbb{P}^3[2,2]$. The LG phase $( \zeta \ll -1 )$ is better understood if we rewrite the superpotential as
\[
	W = \sum_{ij} X_i A_{ij}(P) X_j,
\]
where $A_{ij}$ is a $4 \times 4$ matrix with entries linear in $P_i$. In this phase, $p_1$ and $p_2$ expand a $\mathbb{Z}_2$-gerby $\mathbb{P}^1$ with branch points at
\[
	\det A_{ij} = 0,
\]
where some of $X_i$ become massless. In other words, it is a branched double cover ramified over four points, which is a torus.

Since $A_{ij}$ is a symmetric matrix, we can always diagonalize it with suitable field redefinitions of $X_i$. Therefore, without loss of generality, we can work with the superpotential
\begin{equation}
	W = f_0 X_0^2 + f_1 X_1^2 + f_2 X_2^2 + f_3 X_3^2,
\end{equation}
where $f_i(P_0,P_1)$ are linear polynomials and they determine the branch points. The superpotential can be written as
\begin{align}
	W =& (P_0 + \lambda P_1) G_0 + P_1(G_1 - \lambda G_0), \\
	  =& (P_0 + \lambda P_1) G_0 + P_1( L_1 F_1 + L_2 F_2),
\end{align}
with
\begin{align*}
	L_1 = \sqrt{f_0'} X_0 + \sqrt{-f_1'} X_1, \quad F_1 = \sqrt{f_0'} X_0 - \sqrt{-f_1'} X_1,\\
	L_2 = \sqrt{f_2'} X_2 + \sqrt{-f_3'} X_3, \quad F_2 = \sqrt{f_2'} X_2 - \sqrt{-f_3'} X_3,\\
\end{align*}
where
\[
	f_i' = f_i(-\lambda, 1).
\]

Our proposal of the matrix factorization is
\begin{equation}
	Q = (P_0 + \lambda P_1) \bar{\eta}_0 + G_0 \eta_0 + P_1 F_1 \bar{\eta}_1 + L_1 \eta_1 + P_1 F_2 \bar{\eta}_2 + L_2 \eta_2. \label{mf-proposal}
\end{equation}
This brane is supported on
\begin{align}
	P_0 + \lambda P_1 = 0, \\
	L_1 =L_2 = 0,\\
	G_0 =0, \\
	P_1 F_1 = P_1 F_2 = 0.
\end{align}
The first constraint restricts the brane to a specific point on the base $\mathbb{P}^1$; the second equation describes the isotropic submanifold of the fiber as in \cite{Addington:2012zv}; the last equation restricts the brane support to the origin of the fiber, so indeed we get a D0-brane. The corresponding B-brane is described by a $\mathbb{Z}_2$ complex,
\begin{multline}
{\cal B}(q): \\
\begingroup
\footnotesize
	\xymatrix{
	W(q)_r \oplus  W(q-2)_{r-2} \oplus W(q-3)_{r-2}^{\oplus 2}  
\ar@<0.5ex>[r]^<<<<<f & \ar@<0.5ex>[l]^<<<<<{g} 
  W(q-1)_{r-1}^{\oplus 2} \oplus W(q-2)_{r-1} \oplus  W(q-4)_{r-3}} , \label{z2-b-brane}
  \endgroup
\end{multline}
with the morphisms
\begin{align*}
	f &= \begin{pmatrix}
		P_1 F_1 & -L_2 & G_0 & 0 \\
		P_1 F_2 & L_1 & 0 & G_0 \\
		P_0 + \lambda P_1 & 0 & -L_1 & - L_2 \\
		0 & P_0 + \lambda P_1 & P_1 F_2 & -P_1 F_1
	\end{pmatrix}, \\ \\
	g &= \begin{pmatrix}
		L_1 & L_2 & G_0 & 0 \\
		-P_1 F_2 & P_1 F_1 & 0 & G_0 \\
		P_0 + \lambda P_1 & 0 & - P_1 F_1 & L_2\\
		0 & P_0 + \lambda P_1 & -P_1 F_2 & -L_1 
	\end{pmatrix}.
\end{align*}

First, notice that ${\cal B}(n)$ is quasi-isomorphic to ${\cal B}(n+2)$ which can be shown by the cone 
\[
\xymatrix{
{\cal E}_0(n+2) \ar[d]_{\phi_0} \ar@<0.5ex>[r]^-{f} & \ar@<0.5ex>[l]^-{g}    {\cal E}_1(n+2) \ar[d]^{\phi_1}\\
{\cal E}_0(n) \ar@<0.5ex>[r]^-f& \ar@<0.5ex>[l]^-{g}            {\cal E}_1(n)}
\]
where we denote the branes ${\cal B}(n)$ and ${\cal B}(n+2)$ as
\[
\xymatrix{
{\cal E}_0(n)\ar@<0.5ex>[r]^<<<<<f & \ar@<0.5ex>[l]^<<<<<{g} 
 {\cal E}_1(n)}, \quad 	\xymatrix{{\cal E}_0(n+2) \ar@<0.5ex>[r]^<<<<<f & \ar@<0.5ex>[l]^<<<<<{g} {\cal E}_1(n+2)}.
 \]
Notice that the chain maps $\phi_0$ and $\phi_1$ satisfying the following properties:
\begin{enumerate}
	\item All entries are holomorphic;
	\item All entries should have the correct gauge charges;
	\item The commutative relations: $ \phi_0 g = g \phi_1 $, $f \phi_0 =  \phi_1 f$.
\end{enumerate}
In order to show the quasi-isomorphism, one need to find morphisms $\phi_0$ and $\phi_1$ such that at every point the cone potential $\{ Q_c, Q_c^{\dagger} \} > 0$ in the Landau-Ginzburg phase where $P_0$ and $P_1$ can not vanish simultaneously. The matrix factorization here is given by
\begin{equation}
	Q_c = \begin{pmatrix}
		0 & \tilde{g} \\
		\tilde{f} & 0
	\end{pmatrix},
\end{equation}
with
\begin{equation}
	\tilde{f} = \begin{pmatrix}
		f & 0 \\
		\phi_0 & -g
	\end{pmatrix}, \quad
	\tilde{g} = \begin{pmatrix}
		g & 0 \\
		\phi_1 & -f
	\end{pmatrix}.
\end{equation}
The potential for the cone is   
\begin{align*}
	\{&Q_c, Q_c^{\dagger} \} =  \\
	&\begin{pmatrix}
		gg^{\dagger} + f^{\dagger}f+ \phi_0^{\dagger} \phi_0 & g \phi_1^{\dagger} - \phi_0^{\dagger} g & 0 & 0 \\
		\phi_1 g^{\dagger} - g^{\dagger} \phi_0 & gg^{\dagger} + f^{\dagger} f + \phi_1 \phi_1^{\dagger} & 0 & 0 \\
		0 & 0 &  f f^{\dagger}+ g^{\dagger} g + \phi_1^{\dagger} \phi_1 & f \phi_0^{\dagger} - \phi_1^{\dagger} f \\
		0 & 0 & \phi_0 f^{\dagger} - f^{\dagger} \phi_1 & f f^{\dagger} + g^{\dagger} g + \phi_0 \phi_0^{\dagger}
	\end{pmatrix}.
\end{align*}
Taking 
\begin{equation}
	\phi_0 = \phi_1 = P_1 \, \mathrm{Id}_{4 \times 4},
\end{equation}
one can easily see the cone potential only vanishes at $P_0 = P_1 = 0$ which means that the cone potential always greater than zero at Landau-Ginzburg phase. Therefore, there are two equivalent families of branes ${\cal B}(n)$ with $n$ even or $n$ odd. WLOG, one can focus on ${\cal B}(0)$ and ${\cal B}(1)$. Let's denote ${\cal B}(0)$ and ${\cal B}(1)$ as
\[
	\xymatrix{
{\cal E}_0\ar@<0.5ex>[r]^<<<<<f & \ar@<0.5ex>[l]^<<<<<{g} 
 {\cal E}_1}, \quad 	\xymatrix{{\cal F}_0 \ar@<0.5ex>[r]^<<<<<f & \ar@<0.5ex>[l]^<<<<<{g} {\cal F}_1}.
 \]
To study the relations between ${\cal B}(0)$ and ${\cal B}(1)$, we take the cone  
\[
\xymatrix{
{\cal E}_0 \ar[d]_{\tilde{\phi}_0} \ar@<0.5ex>[r]^<<<<<f & \ar@<0.5ex>[l]^<<<<<{g}    {\cal E}_1 \ar[d]^{\tilde{\phi}_1}\\
{\cal F}_1 \ar@<0.5ex>[r]^<<<<<g& \ar@<0.5ex>[l]^<<<<<{f}            {\cal F}_0}
\]
with the chain maps $\tilde{\phi}_0$ and $\tilde{\phi}_1$. Notice that the brane ${\cal B}(1)$ has been shifted by 1, and the commutative relations for the diagram becomes $g \phi_0 = \phi_1 f$, $\phi_0 g = f \phi_1$. The matrix factorization for the cone is given by
\begin{equation}
	\tilde{Q}_c = \begin{pmatrix}
		0 & \tilde{g} \\
		\tilde{f} & 0
	\end{pmatrix},
\end{equation}
with
\begin{equation}
	\tilde{f} = \begin{pmatrix}
		f & 0 \\
		\tilde{\phi}_0 & -f
	\end{pmatrix}, \quad
	\tilde{g} = \begin{pmatrix}
		g & 0 \\
		\tilde{\phi}_1 & -g
	\end{pmatrix}.
\end{equation}
The potential for the cone is   
\begin{align*}
	\{& \tilde{Q}_c, \tilde{Q}_c^{\dagger} \} =  \\
	&\begin{pmatrix}
		gg^{\dagger} + f^{\dagger}f+ \tilde{\phi}_0^{\dagger} \tilde{\phi}_0 & g \tilde{\phi}_1^{\dagger} - \tilde{\phi}_0^{\dagger} f & 0 & 0 \\
		\tilde{\phi}_1 g^{\dagger} - f^{\dagger} \tilde{\phi}_0 & gg^{\dagger} + f^{\dagger} f + \tilde{\phi}_1 \tilde{\phi}_1^{\dagger} & 0 & 0 \\
		0 & 0 &  f f^{\dagger}+ g^{\dagger} g + \tilde{\phi}_1^{\dagger} \tilde{\phi}_1 & f \tilde{\phi}_0^{\dagger} - \tilde{\phi}_1^{\dagger} g \\
		0 & 0 & \tilde{\phi}_0 f^{\dagger} - g^{\dagger} \tilde{\phi}_1 & f f^{\dagger} + g^{\dagger} g + \tilde{\phi}_0 \tilde{\phi}_0^{\dagger}
	\end{pmatrix}.
\end{align*}
Away from the branch points, it is not hard to find morphisms $\phi_0$ and $\phi_1$ such that the cone potential vanishes at some points. Thus, the brane ${\cal B}(0)$ and brane ${\cal B}(1)$ are not quasi-isomorphic to each other. We have two different sets of branes away from branch points corresponding to two copies of $\mathbb{P}^1$. 

To show that ${\cal B}(0)$ and ${\cal B}(1)$ are quasi-isomorphic at branch points, one needs to find a special pair of morphisms $\phi_0$ and $\phi_1$ such that at every point the potential $\{Q_c, Q_c^{\dagger} \} > 0$. 
What is special about the branch points is that some of the polynomials defining the matrix factorization become the same,
\[
 L_1 = \pm F_1 , \quad L_2 = \pm F_2.
\]
If $L_1 = F_1$, we find the chain maps can be
\begin{equation}
	\phi_0 = c \cdot \mathrm{diag}\{ -P_1, 1, -1, P_1 \}, \quad \phi_1 = c \cdot \mathrm{diag} \{ -1, P_1, -P_1, 1\},
\end{equation}
where $c$ is a constant. 
The determinant of the potential is
\begin{equation}
	\det \{Q_c, Q_c^{\dagger}\} = \Big( (A+|c P_1|^2)(A+|c|^2) - |c|^2(|P_1|^2-1)^2 |L_1|^2\Big)^8,
\end{equation}
with 
\[
 A =|P_0 + \lambda P_1|^2 + |L_1|^2+ |L_2|^2 + |G_0|^2 + |P_1 F_1|^2 + |P_1 F_2|^2.
\]
Vanishing of the determinant implies that $P_0 = P_1 = 0$, which means the potential is always greater than zero at Landau-Ginzburg phase. Therefore, we showed that the branes ${\cal B}(0)$ and ${\cal B}(1)$ are quasi-isomorphic at the branch point corresponding to $L_1 =F_1$.

Similarly, when $L_1 = - F_1$, we find
\begin{equation}
	\phi_0 = c \cdot \mathrm{diag}\{ P_1, 1, -1, -P_1 \}, \quad \phi_1 = c \cdot \mathrm{diag} \{ -1, -P_1, P_1, 1\},
\end{equation}
such that the cone potential is always greater than zero on the Landau-Ginzburg phase. For the branch points corresponding to $L_2 = \pm F_2$, the branes are also quasi-isomorphic since $L_2, F_2$ and $L_1, F_1$ are on the same footing. One can see this by constructing ${\cal B}'(0)$ and ${\cal B}'(1)$ with $L_2, F_2$ and $L_1, F_1$ exchanged which are quasi-isomorphic to ${\cal B}(0)$ and ${\cal B}(1)$.


In extreme case that $L_1 = F_1 = L_2 = F_2 = 0$ at branch locus, the morphisms of matrix factorization become
\begin{equation}
	f = g = \begin{pmatrix}
		0 & 0 & G_0 & 0 \\
		0 & 0 & 0 & G_0 \\
		P_0 + \lambda P_1 & 0 & 0 & 0 \\
		0 & P_0 + \lambda P_1 & 0 & 0
	\end{pmatrix}.
\end{equation}
The brane ${\cal B}(0)$ and ${\cal B}(1)$ are obviously quasi-isomorphic.
 
It is also possible that $L_1 = F_1 = 0$ or $L_2 = F_2 = 0$ at the branch point which means that two of the chiral fields $X_i$ are always massive and can be integrated out. In this case, the matrix factorization can be reduced to 
\begin{equation}
	Q = (P_0 + \lambda P_1) \bar{\eta}_0 + G_0 \eta_0 + P_1 F_1 \bar{\eta}_1 + L_1 \eta_1, 
	\end{equation}
with the assumption that $X_2$ and $X_3$ are massive. The brane complex is 
\begin{equation}
{\cal B}'(q):	\xymatrix{
	W(q)_r \oplus  W(q-3)_{r-2}   
\ar@<0.5ex>[r]^<<<<<{f'} & \ar@<0.5ex>[l]^<<<<<{g'} 
  W(q-1)_{r-1}\oplus W(q-2)_{r-1} } , 
\end{equation} 
with 
\begin{equation}
	f' = \begin{pmatrix}
		P_1F_1 & G_0 \\
		P_0 + \lambda P_1 & -L_1 
	\end{pmatrix}, \quad
	g' = \begin{pmatrix}
		L_1 & G_0 \\
		P_1 + \lambda P_1 & -P_1 F_1
	\end{pmatrix}.
\end{equation}
Following the same procedure above, one can easily show that the brane ${\cal B}'(0)$ and ${\cal B}'(1)$ are quasi-isomorphic. On the LG phase, the Wilson line brane descends to orbibundle supporting on the $[\mathbb{P}^1]_{\mathbb{Z}_2}$, 
\begin{equation}
	W(q) \rightarrow \cO(q/2).
\end{equation}
Then, ${\cal B}'(0)$ reduces to 
\begin{equation}
\xymatrix{
	{\cal O}_r \oplus  {\cal O}(-3/2)_{r-2}   
\ar@<0.5ex>[r]^<<<<<{f'} & \ar@<0.5ex>[l]^<<<<<{g'} 
  {\cal O}(-1/2)_{r-1}\oplus {\cal O}(-1)_{r-1} } , \label{mf-eric1}	
\end{equation}
and ${\cal B}'(1)$ reduces to
\begin{equation}
\xymatrix{
	{\cal O}(1/2)_r \oplus  {\cal O}(-1)_{r-2}   
\ar@<0.5ex>[r]^<<<<<{f'} & \ar@<0.5ex>[l]^<<<<<{g'} 
  {\cal O}_{r-1}\oplus {\cal O}(-1/2)_{r-1} } .  \label{mf-eric2}
\end{equation}
Notice that on the local patch of $\mathbb{P}^1$ matrix factorization (\ref{mf-eric1}) and (\ref{mf-eric2}) reduce to the local matrix factorization studied in \cite{Addington:2012zv}. Those two local matrix factorizations are obviously quasi-isomorphic.

In summary, we have found morphisms $\phi_0$, $\phi_1$ such that the cone potential $\{ Q_c, Q_c^{\dagger} \} > 0$ for all possible scenarios. Also, our proposal recovers the local model studied in the literature. In this way,  we have shown that the LG phase of $\mathbb{P}^3[2,2]$ model is indeed a branched double cover and our proposed matrix factorization (\ref{mf-proposal}) gives a global description of the D0-brane on the branched double cover geometry.

\bibliography{ref.bib}
\bibliographystyle{fullsort}

\end{document}